\documentclass[a4paper, 11pt]{article}
\pdfoutput=1
\usepackage{jcappub}
\usepackage[utf8]{inputenc}
\usepackage{physics}
\usepackage{amsmath, amssymb, amsthm}
\usepackage{siunitx}
\usepackage{xcolor}
\usepackage{graphicx}
\usepackage{cancel}
\usepackage{caption}
\usepackage{subcaption}
\usepackage{mathrsfs}
\usepackage{numprint}
\usepackage{multirow,multicol, makecell, booktabs}

\renewcommand{\tilde}{\widetilde}
\newcommand{\mpl}{M_{\rm pl}}

\newcommand{\sinc}{\operatorname{sinc}}

\numberwithin{equation}{section}

\newcommand{\decay}{f} 
\newcommand{\dimps}{P} 
\newcommand{\dimlessps}{\mathcal{P}}
\newcommand{\trap}{\ast} 
\newcommand{\present}{0} 
\newcommand{\critical}{c} 
\newcommand{\weak}{\rm frag} 
\newcommand{\strong}{\rm s} 
\newcommand{\ini}{i} 
\newcommand{\kin}{\rm kin} 
\newcommand{\fend}{\rm end} 
\newcommand{\osc}{\rm osc} 
\newcommand{\eq}{\rm eq} 
\newcommand{\eps}{\varrho} 
\newcommand{\ellipticE}{\textrm{E}} 
\newcommand{\ellipticK}{\textrm{K}} 
\newcommand{\barscale}{\gamma} 
\newcommand{\kap}{\kappa} 
\newcommand{\amp}{N} 
\newcommand{\eff}{\Delta} 

\title{ALP Dark Matter from Kinetic Fragmentation: Opening up the Parameter Window}

\author[a]{Cem Eröncel,}
\author[b]{Ryosuke Sato,}
\author[a,c]{Géraldine Servant,}
\author[a,c]{Philip S\o{}rensen}

\affiliation[a]{Deutsches Elektronen-Synchrotron DESY, Notkestr. 85, 22607 Hamburg, Germany}
\affiliation[b]{Department of Physics, Osaka University, Toyonaka, Osaka 560-0043, Japan}
\affiliation[c]{II. Institute of Theoretical Physics, Universit\"{a}t Hamburg D-22761, Germany}

\emailAdd{cemeroncel@gmail.com}
\emailAdd{rsato@het.phys.sci.osaka-u.ac.jp}
\emailAdd{geraldine.servant@desy.de}
\emailAdd{philip.soerensen@pd.infn.it}

\abstract{The main mechanism responsible for Axion-Like-Particle (ALP) production in the early universe is the so-called misalignment mechanism. Three regimes have been investigated in this context: 
standard misalignment, large misalignment and kinetic misalignment. The latter applies if the axion inherits a large initial velocity in the early universe, such that the field rolls through many wiggles during its evolution, before it gets trapped in one minimum.  This largely opens the region of parameter space for ALP dark matter towards higher values for the axion-photon coupling, which can be probed by the whole set of next decade's upcoming experiments. In fact, almost the entire parameter space in the [mass, decay constant] plane can now accommodate dark matter.
In this paper,  we  show that in kinetic misalignment, the axion field is almost always entirely fragmented, meaning that the energy density of the homogeneous field is redistributed over higher-mode axions. 
We present a general model-independent analytical description of kinetic fragmentation, including discussion of the  modified initial conditions  for the mode functions due to the axion's initial velocity, and how they impact the growth of the adiabatic fluctuations. We calculate precisely  
the parameter regions corresponding respectively to standard misalignment, kinetic misalignment with weak fragmentation, fragmentation after trapping and fragmentation before trapping. While axion fragmentation can impact the precise determination of the relic abundance,  another main observational implication is the formation of much denser compact axion halos, that is described in a companion paper. We also point out a new gravitational-wave signature that arises in the large misalignment regime with complete fragmentation and could be seen in measurements of $\mu$ distortions in the Cosmic Microwave Background.}

\begin{document}

\begin{flushright}
\footnotesize
DESY 22-106\\
OU-HET-1148\\
\end{flushright}
\color{black}

\maketitle
\flushbottom

\section{Introduction}
\label{sec:introduction}

Axion-Like-Particles (ALPs) appear in many extensions of the Standard Model.  Their physics case has been extensively scrutinised in the last decades, especially in the last few years which have seen an increased interest in novel detection strategies and new experimental proposals \cite{Irastorza:2018dyq,Adams:2022pbo}.
ALPs are pseudo-Nambu-Goldstone bosons, arising from the spontaneous breaking of a global  $U(1)$ symmetry. As such, they are naturally light particles, with a mass $m$ much smaller than the abelian symmetry breaking scale  $f$, also referred to as the ALP decay constant. ALPs are
characterised by their small mass $m \ll f$, and by their small coupling to Standard Model particles which scales as  $1/f$. 
The axion mass is generated at some lower scale $\Lambda_b \ll f$ typically related to some new strong dynamics that break the $U(1)$ symmetry explicitly, leading to $m \propto \Lambda_b^2/f$.  For a generic ALP, $m$ and $f$ are independent parameters. In the case of the QCD axion, they are related as $m f \propto \Lambda_{QCD}^2$.
The best-known example of an axion is in fact the QCD axion, which arises as a consequence of the Peccei-Quinn solution to the strong-CP problem \cite{Peccei:1977hh,Weinberg:1977ma,Wilczek:1977pj}.
ALPs have a characteristic two-photon vertex. This coupling is at the basis of most search strategies which rely on axion-photon conversion in external magnetic fields.

\begin{figure}[tbp]
	\centering
	\includegraphics[width=\textwidth]{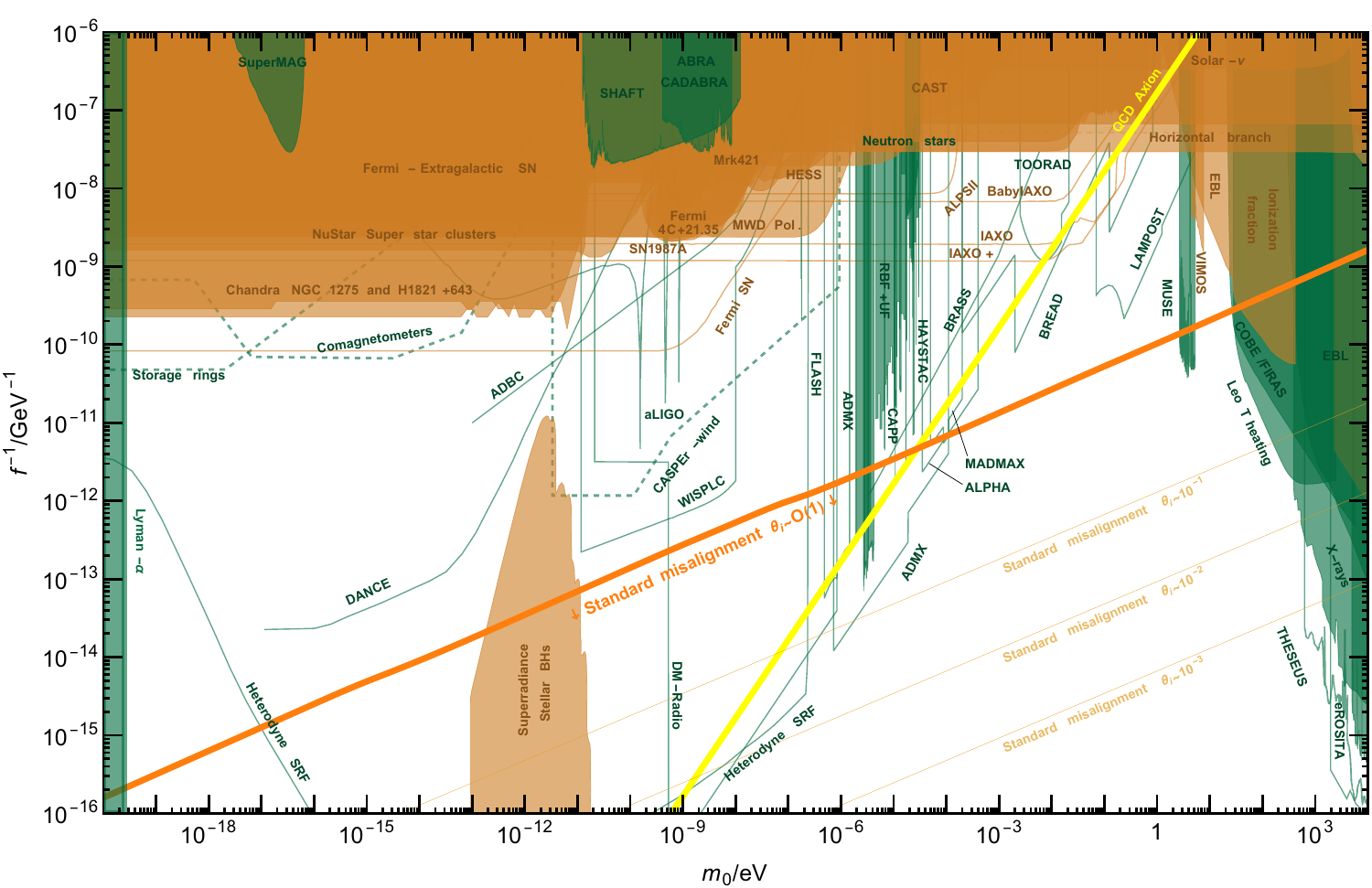}
	\caption{ \it \small The experimental landscape in the hunt for ALPs assuming a KSVZ-like  axion-photon coupling $g_{\theta\gamma\gamma}^{\rm KSVZ}$ given in \eqref{eq:282}. 
		Coloured regions are excluded. The thin lines indicate the sensitivities of future experiments. Used data is listed in Appendix \ref{sec:experimental-surveys}.
		Orange constraints apply to any ALP while the green ones assume the ALP is DM.
		The yellow thick line corresponds to the QCD axion. The four other parallel straight lines indicate the correct dark matter relic abundance contours for different assumptions of the initial misalignment angle. Above the thick orange line, the axion produced from the standard misalignment mechanism is under-abundant to explain DM. }
	\label{fig:experimentallandscape}
\end{figure}

Particularly well-motivated is the case that ALPs compose the dark matter of the Universe \cite{Preskill:1982cy,Abbott:1982af,Dine:1982ah}.  Because of their small mass and small couplings, ALPs are cosmologically stable.
In most of the literature, ALPs are  produced as extremely cold relics in the early universe through the {\it misalignment mechanism}. The energy density of the axion field is stored into the  oscillations of the zero-mode homogeneous axion field $f \theta(t)$. The scalar field equation takes the form
$\ddot{\theta}+3H(t)\dot{\theta}+m^2(t) \sin \theta =0$. The initial conditions are usually taken to be $\theta(0)= \theta_i $ where $\theta_i \in [-\pi, \pi]$ is referred to as the initial misalignment angle,
and $\dot{\theta}(0)=0$. 
At early times, the field is frozen due to Hubble friction.  
Oscillations start as soon as the expansion rate of the universe $H(t)$ drops below the axion mass.  From that time, the field behaves effectively as pressureless cold matter.
The axion contribution to the dark matter density today 
crucially depends on the initial displacement of the axion field with respect to the minimum of the axion potential. 
We show the corresponding predictions for different assumptions of this initial angle in Fig.~\ref{fig:experimentallandscape}.
Constraints can be represented in the $[m, 1/f]$ plane.
The testable region of parameter space of both the QCD axion and of generic ALPs is still restricted. However, there are many plans for testing further the parameter space.
Another ALP production mechanism is from the decay of cosmic strings or domain walls into axion particles, which roughly predicts comparable regions of parameter space as those from the misalignment mechanism. Experiments such as ADMX can probe some parts of the parameter space of the QCD axion, although not in the region where it is Dark Matter,  while other regions of generic ALP DM are out of reach of most future experiments, see Fig.~\ref{fig:experimentallandscape}.

In this paper, we consider a new mechanism for ALP production in the early universe, so far overlooked, in which ALPs are first produced relativistically while the energy density initially stored in the homogeneous zero mode of the axion field gets transferred into relativistic axion particles. This happens generically if the axion field has initially sufficiently large velocity to roll over many wiggles of the periodic axion potential until the expansion rate of the universe is sufficiently small to enable axion fragmentation.  
This can happen at relatively low temperatures and leads to a viable scenario of Cold Dark Matter if the fragmentation temperature is below a  TeV. These axion particles quickly cool down and dominate the energy density.
The corresponding axion periodic potential emerges from some new confining dynamics happening in a dark sector at energies between a keV and a TeV. 

Recently,  the usual assumption that the axion field is initially static in the early universe has been questioned \cite{Co:2019jts,Chang:2019tvx}. 
In particular, it was shown that such initial large velocity can delay the onset of the usual misalignment mechanism for axion DM production.
The so-called axion kinetic misalignment has been put forward to predict axion dark matter with a  larger coupling than in the original setup, which is excellent news for experiments which now have stronger motivations to search for ALP DM\footnote{Another recent proposal to expand the ALP DM parameter space to larger couplings is the \emph{frictional misalignment} \cite{Papageorgiou:2022prc}.}.
We show that there is an additional effect which takes place in the axion kinetic misalignment framework, called axion fragmentation \cite{Fonseca:2019ypl}.
Axion fragmentation was studied in the context of the relaxion in \cite{Fonseca:2019lmc} and for  dark matter in axion monodromy models in \cite{Jaeckel:2016qjp,Berges:2019dgr}.

While most axion cosmology literature usually assumes that the axion field does not have any initial velocity, 
such initially large velocity is actually a natural outcome in a class of UV completions extensively discussed in 
\cite{Co:2019jts,Co:2019wyp,Co:2020jtv,Co:2020xlh,Gouttenoire:2021jhk}.
Considering the early evolution of the full complex Peccei-Quinn scalar field,  if the radial mode has a large initial VEV, it can induce a kick for the angular (axion) mode that starts rotating for a long time (similar to what happens in Affleck-Dine baryogenesis \cite{Affleck:1984fy,Dine:1995kz}). A large kinetic energy can also arise in the so-called \emph{trapped misalignment} mechanism \cite{DiLuzio:2021gos,DiLuzio:2021pxd}, and as a result of level crossing in the axiverse models \cite{Daido:2015cba,Daido:2015bva}.
In this paper, we remain agnostic about the UV completion. We work in full generality, only considering the late-stage evolution when the complex scalar field has reached the bottom of the potential. The radial mode of the complex scalar field can be ignored, we only consider the axion degree of freedom after it has acquired a mass. Because of its large initial velocity, it can overshoot the potential barrier and keep rotating until it eventually fragments entirely.

It is the main goal of this paper to investigate in detail and in full generality (model-independently) the properties of ALP DM produced by fragmentation in the kinetic misalignment mechanism.  We generalise the analysis of  Ref.\cite{Fonseca:2019ypl} in two ways. First, we focus on the axion being DM (while \cite{Fonseca:2019ypl} did not impose that the axion is DM). Second, we not only investigate axion fragmentation before the axion is trapped by the barriers but also show that fragmentation can take place after trapping.
We provide a discussion of the initial conditions for the mode functions and derive the field power spectrum. 
The distinctive feature in this scenario is the momentum distribution of the axion particles. 
Ultimately, our analysis sets the basis to derive the 
phenomenological implications of this modified power spectrum, which will be presented in \cite{Eroncel:2022efc}.
We will provide predictions in the explicit UV completions in another companion publication \cite{Eroncel:2022a}, deriving the precise model parameter space regions for ALP dark matter through kinetic fragmentation.

The plan of this article is as follows.
In Section \ref{sec:kin-mis} we provide an analytical derivation of the ALP relic abundance and trapping temperature in the kinetic misalignment framework. The latter is crucial when calculating the power spectrum. The analytical theory of parametric resonance in kinetic misalignment is derived in section \ref{sec:analyt-theory-param}.
The determination of the various regions in the  [axion mass, decay constant] parameter space are determined in Section \ref{sec:alp-dark-matter} where the impact of fragmentation on the relic abundance prediction is also discussed. All phenomenological constraints are compiled in Section \ref{sec:constr-alp-param}. Section \ref{sec:gravitational-waves} presents the prediction of the gravitational-wave spectrum induced by fragmentation. 
We summarise our results by listing the important equations and figures in Section \ref {sec:conclusion}.
Technical details are gathered in appendices \ref{sec:deta-disc-param}, \ref{sec:calc-adiab-init} and \ref{sec:fragm-before-trapp}. Appendix \ref{sec:experimental-surveys} lists all the experimental bounds and projections that are reported in our plots together with the corresponding references.

\section{ALP dark matter from kinetic misalignment}
\label{sec:kin-mis}

We consider the cosmological evolution of an ALP field $\theta$ whose Lagrangian is given by
\begin{equation}
  \label{eq:1}
  \mathcal{L}=-\frac{\decay^2}{2}g^{\mu\nu}\partial_{\mu}\theta\partial_{\nu}\theta-V(\theta)=-\frac{\decay^2}{2}g^{\mu\nu}\partial_{\mu}\theta\partial_{\nu}\theta-m^2(T)\decay^2\qty[1-\cos\qty(\theta)],
\end{equation}
where $\decay$ is the vacuum expectation value of the complex scalar field radial component. The metric is taken to be the flat Friedmann-Lemaitre-Robertson-Walker metric\footnote{In general, the metric does also have curvature perturbation terms. These will determine the initial conditions for the mode functions as we will demonstrade in Section \ref{sec:initial-conditions}.}
\begin{equation}
  \label{eq:2}
  \dd{s}^2=-\dd{t}^2+a^2(t)\delta_{ij}\dd{x}^i\dd{x}^j.
\end{equation}
We decompose the ALP field $\theta(t,\vb{x})$ into a homogeneous mode $\Theta(t)$, and small fluctuations $\delta \theta(t,\vb{x})$, where the latter can be expanded into Fourier modes as
\begin{equation}
  \label{eq:3}
  \delta\theta(t,\vb{x})=\int \frac{\dd[3]{k}}{(2\pi)^3}\theta_{\vb{k}}(t)e^{-i\vb{k}\cdot\vb{x}}.
\end{equation}
In this work we assume that the Fourier modes $\theta_{\vb{k}}(t)$ evolve independently. We also separate their time evolution  from their statistical properties by defining
\begin{equation}
  \label{eq:4}
  \theta_{\vb{k}}(t)=\hat{\theta}_{\vb{k}}\theta_k(t),
\end{equation}
where $\hat{\theta}_{\vb{k}}$'s carry the statistical properties, and $\theta_k$'s are $c$-number functions which depend only on the magnitude of the momentum $k\equiv \abs{\vb{k}}$ due to the isotropy of the equations of motions in the linear approximation. We also assume that $\hat{\theta}_{\vb{k}}$'s obey statistical homogeneity and isotropy:
\begin{equation}
  \label{eq:5}
  \expval{\hat{\theta}_{\vb{k}}\hat{\theta}^{\ast}_{\vb{k}'}}=(2\pi)^3\delta^{(3)}\qty(\vb{k}-\vb{k}')
\end{equation}
The initial conditions for the mode function $\theta_k$ will be determined by the initial field power spectrum $\dimps_{\theta}$ which depends on the physical process that creates the initial conditions for the Fourier modes. We study the initial conditions in Section \ref{sec:initial-conditions}. 

If we neglect the back-reaction,  the equations of motion of the homogeneous mode $\Theta$ and of the fluctuations $\theta_k$ are decoupled from each other. These are respectively given by
\begin{equation}
  \label{eq:6}
  \ddot{\Theta}+3H\dot{\Theta}+m^2(T)\sin \Theta=0,
\end{equation}
and
\begin{equation}
  \label{eq:7}
  \ddot{\theta}_k+3H\dot{\theta}_k+\qty[\frac{k^2}{a^2}+m^2(T)\cos\Theta]\theta_k=0.
\end{equation}
We start by studying the evolution of the homogeneous mode, and derive today's value of the relic density  in Sections \ref{sec:adiab-invar-relic} and \ref{sec:trapping-temperature}. Then we discuss the implications of the parametric resonance in Section \ref{sec:parametric-resonance}. We conclude this section by providing a semi-analytical  estimate of the back-reaction of the fluctuations on the homogeneous mode. We comment on the implications in Section \ref{sec:an-estim-backr}.

\subsection{Derivation of  the relic density today from the adiabatic invariant}
\label{sec:adiab-invar-relic}

In the Kinetic Misalignment mechanism \cite{Co:2019jts}, we work under the assumption that the ALP field has a large initial velocity such that its initial kinetic energy is much larger than its potential energy. Therefore it travels over many barriers without being trapped. As the universe expands, the kinetic energy will redshift while the potential energy will grow since it is proportional to the axion mass. At some  temperature $T_{\trap}$, the energy in the homogeneous mode $\rho_{\Theta}$  becomes smaller than the height of the barriers. After this time the ALP field cannot continue rolling, as it is \emph{trapped} by the barrier. The trapping temperature $T_{\trap}$ is defined by the condition\footnote{From now on the $\trap$-subscript denotes the quantities evaluated at $T=T_{\trap}$.}
\begin{equation}
  \label{eq:8}
  \rho_{\Theta}(T_{\trap})=\frac{1}{2}\decay^2\dot{\Theta}^2(T_{\trap})+m^2(T_{\trap})\decay^2\qty[1-\cos(\Theta(T_{\trap}))] =2 m^2(T_{\trap})\decay^2\qty[1-\cos(\Theta(T_{\trap}))].
\end{equation}
We introduce the dimensionless parameter $\eps(t)$ as the ratio between the total ALP energy density and its potential energy\footnote{We will use the time and temperature interchangebly depending on the nature of the quantity that we want to calculate. We always assume radiation domination so $t=(2H)^{-1}$, and $H(T)\propto g_{\ast}(T)^{1/2}T^2$ where $g_{\ast}$ is the effective number of relativistic degrees of freedom. We assume that $g_{\ast}$ is constant throughout the parametric resonance so we can change from temperature to time and vice versa by $T^2\propto t^{-1}$. }
\begin{equation}
  \label{eq:9}
  \eps(t)\equiv \frac{\rho_{\Theta}}{2 m^2(t) \decay^2}=\frac{1}{4}\frac{\dot{\Theta}^2}{m^2(t)}+\sin^2\qty(\frac{\Theta}{2}).
\end{equation}
The field is trapped at $\eps(t_{\ast})=1$. If $\eps>1$ it keeps rolling, while for $\eps<1$ it is oscillating around its minimum. 

The evolution of $\eps(t)$ can be studied analytically by using \emph{action-angle formalism} \cite{goldsteinClassicalMechanics2001,landauMechanicsVolume1976}, see also \cite{Masso:2005zg}. Neglecting the back-reaction, the evolution of the homogeneous mode is governed by the Lagrangian density
\begin{equation}
  \label{eq:10}
  \mathcal{L}_0=\decay^2a^3(t)\qty[\frac{1}{2}\dot{\Theta}^2-m^2(t)\qty(1-\cos\Theta)].
\end{equation}
This Lagrangian has a discrete shift symmetry given by $\Theta \rightarrow \Theta + 2\pi$, so the motion in $\Theta$-space can be considered periodic. In the following, we define the period to be the time it takes for the field to roll from one maximum to the other.


Now we assume that the parameters carrying explicit time dependence, namely $a$ and $m$, change \emph{adiabatically}, in other words they remain approximately constant during a single oscillation. This assumption is justified if the Hubble time scale is larger compared to the period of the motion. We will justify this assumption later. In the case of periodic motion where the Lagrangian changes adiabatically, the \emph{action variable} $J$ defined by
\begin{equation}
  \label{eq:11}
  J\equiv \oint \Pi_q\dd{q},
\end{equation}
is a constant of motion, where the integral is over a single period, and
\begin{equation}
  \label{eq:12}
  q=\decay\Theta\qand \Pi_q=\pdv{\mathcal{L}_0}{\dot{q}}=a^3(t) f\dot{\Theta}.
\end{equation}
Assuming that the scale factor and the axion mass are constant during one period, $J$ becomes
\begin{equation}
  \label{eq:13}
  J\approx \decay^2a^3(t)\oint \dd{\Theta}\dot{\Theta}=2m(t)\decay^2a^3\oint\sqrt{\eps-\sin^2\qty(\frac{\Theta}{2})}.
\end{equation}
The integral can be performed analytically and one gets
\begin{equation}
  \label{eq:14}
  J\approx m(t)\decay^2a^3(t)\times
  \begin{cases}
    8\sqrt{\eps}\,\ellipticE\qty(1/\sqrt{\eps}),& \eps>1\\
    8 \qty[(\eps - 1)\ellipticK(\sqrt{\eps})+\ellipticE(\sqrt{\eps})],&\eps <1
  \end{cases},
\end{equation}
where $\ellipticK$ and $\ellipticE$ are complete elliptic integrals of the first and second kind respectively\footnote{We use the following definitions for the elliptic integrals:
  \begin{equation*}
    \ellipticK(k)=\int_0^{\pi/2}\frac{\dd{\varphi}}{\sqrt{1-k^2\sin^2\varphi}}\qand \ellipticE(k)=\int_0^{\pi/2}\dd{\varphi}\sqrt{1-k^2\sin^2\varphi}
  \end{equation*}
  Note that most software packages such as \texttt{Mathematica} and \texttt{scipy} \cite{2020SciPy-NMeth} uses $m=k^2$ instead of $k$ as their argument when defining elliptic integrals.
}.  By using the limiting behavior of these as $\eps \rightarrow 1$ we can show that
\begin{equation}
  \label{eq:16}
  \lim_{\eps \rightarrow 1^+}J(\eps)=\lim_{\eps \rightarrow 1^-}J(\eps)=8m(t)f^2a^3(t).
\end{equation}
The adiabatic invariant $J$ remains approximately constant throughout the whole evolution. This makes $J(\eps)$ a perfect quantity to compare the early and late time behaviors. At early times when $\eps \gg 1$ we have
\begin{equation}
  \label{eq:18}
  J(\eps \gg 1)\approx 2\pi \decay^2\dot{\Theta}a^3,
\end{equation}
whereas at late times we get
\begin{equation}
  \label{eq:19}
  J(\eps \ll 1)\approx \pi \frac{\rho_{\Theta}a^3}{m_{\present}},
\end{equation}
where $m_{\present}$ is the zero-temperature axion mass. Let us introduce the \emph{yield} quantity defined by 
\begin{equation}
Y\equiv n_{\Theta}/s, 
\end{equation}
where $s$ is the entropy density of the universe, and 
\begin{equation}
n_{\Theta}=\decay^2\dot{\Theta}
\end{equation}
 is the Noether charge of the axion shift symmetry. This quantity is conserved during the early evolution when the ALP mass is negligible so the equation of motion of the ALP field is just $\ddot{\Theta}+3H\dot{\Theta}\approx 0$. The conservation of $J(\eps)$ implies that (\ref{eq:18}) and (\ref{eq:19}) should be equal which gives
\begin{equation}
  \label{eq:20}
  \frac{\rho_{\Theta,\present}}{s_{\present}}\approx 2m_{\present}Y.
\end{equation}
where $\rho_{\Theta,\present}$ and $s_{\present}$ are respectively the relic density and entropy density today. This equation is also derived in \cite{Co:2019jts}, however the numerical factor there was obtained numerically. Here we did present an analytical derivation.

From this result we can obtain a simple formula for the fractional energy density of the ALP field today as
\begin{equation}
  \label{eq:21}
  h^2\Omega_{\Theta,\present}\approx h^2\Omega_{\rm DM}\qty(\frac{m_{\present}}{5\times 10^{-3}\,\si{\electronvolt}})\qty(\frac{Y}{40}),
\end{equation}
where we took $h^2\Omega_{\rm DM}=0.12$ \cite{Aghanim:2018eyx}. Note that the yield does only depend on the zero temperature axion mass, not on the high-temperature behavior. The latter will be important in determining the trapping temperature $T_{\trap}$. For the QCD axion the zero-temperature mass and the axion decay constant are related by $m_{\present}\decay \approx (75.6\si{\mega\electronvolt})^2$ \cite{Borsanyi:2016ksw}. Then the above relation implies
\begin{equation}
  \label{eq:22}
  h^2\Omega^{\rm QCD axion}_{\phi,\present}\approx h^2\Omega_{\rm DM}\qty(\frac{10^9\,\si{\giga\electronvolt}}{\decay})\qty(\frac{Y}{40}).
  \end{equation}

Before closing this subsection, we mention in which conditions the adiabacity assumption is justified. To do this we need to calculate the periods of the motion before and after trapping. By neglecting the energy loss due to the Hubble expansion during a single oscillation, the periods $\mathcal{T}_{>}$ before  and  $\mathcal{T}_{<}$ after trapping can be derived as
\begin{equation}
  \label{eq:341}
  \mathcal{T}_{>}(\eps)=\frac{2}{m\sqrt{\eps}}\ellipticK(1/\sqrt{\eps})\qand \mathcal{T}_{<}(\eps)=\frac{2}{m}\ellipticK(\sqrt{\eps}).
\end{equation}
At early times when $\eps\gg 1$ we have
\begin{equation}
  \label{eq:342}
  \mathcal{T}_{>}(\eps\gg 1)\approx \frac{\pi}{m}\sqrt{\frac{1}{\eps}}\approx \frac{2\pi}{\dot{\Theta}},
\end{equation}
so the adiabacity condition reads $\dot{\Theta}> 2\pi H$. At early times $\dot{\Theta}\propto a^{-3}$ and $H\propto a^{-2}$ so it is sufficient if this condition is satisfied at trapping $\dot{\Theta}=2m_{\trap}$. This yields $m_{\trap}>\pi H_{\trap}$ as the adiabacity condition. On the other hand, at late times when $\eps\ll 1$ we have
\begin{equation}
  \label{eq:343}
  \mathcal{T}_{<}(\eps \ll 1)\approx \frac{\pi}{m},
\end{equation}
so the adiabacity condition becomes $m>\pi H$. Again, it is sufficient that this condition is satisfied at trapping which yields
\begin{equation}
  \label{eq:344}
  \frac{m_{\trap}}{H_{\trap}}\gtrsim \pi,
\end{equation}
as the adiabacity condition. As we will show explicity in Section \ref{sec:alp-dark-matter}, the most interesting region of the parameter space is where $m_{\trap}/H_{\trap}\gg 1$ which justifies the adiabacity assumption. We note that even for  $m_{\trap}/H_{\trap}\gg 1$, the adiabacity is broken for a period around $\eps= 1$, but this period is shorter for larger $m_{\trap}/H_{\trap}$.

\subsection{Trapping temperature}
\label{sec:trapping-temperature}

\begin{figure}[tbp]
  \centering
  \includegraphics[width=\textwidth]{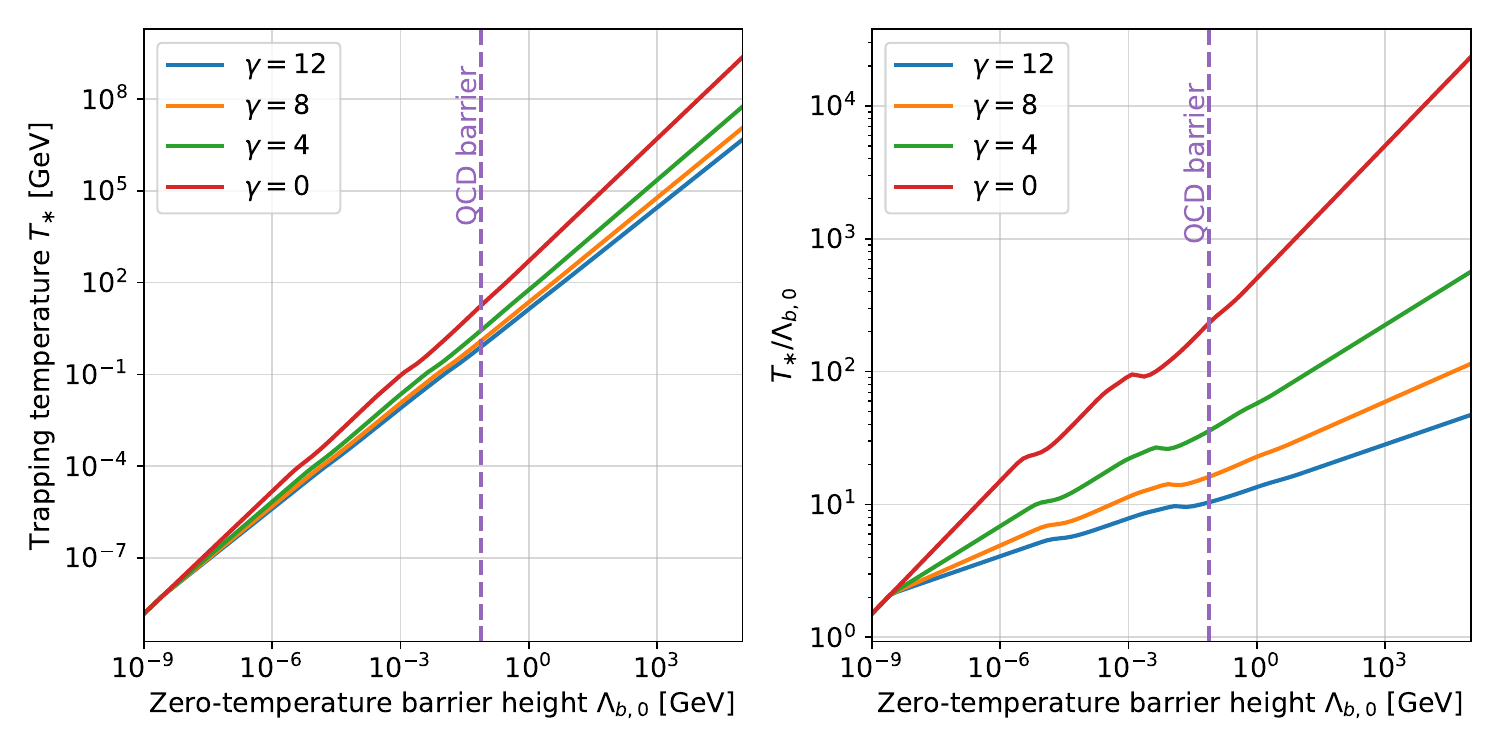}
  \caption{\it \small Trapping temperature $T_{\trap}$ given in Eq.~ \ref{eq:26}
(left figure), and the ratio between $T_{\trap}$ and the zero-temperature barrier height $\Lambda_{b,\present}$ (right plot) as a function of  $\Lambda_{b,\present}$ for different choices of $\barscale$ which parametrizes the early-time scaling of the barrier height, and thus of the axion mass, with temperature as defined in (\ref{eq:25}). We see that this ratio increases for larger $\Lambda_{b,\present}$, and the trapping temperature becomes more sensitive to the early time behavior of the axion potential. The dashed vertical purple line shows the zero-temperature QCD barrier height $75.6\,\si{\mega\electronvolt}$.}
  \label{fig:trapping-temperature}
\end{figure}

 The trapping temperature is an important quantity since fragmentation happens around the same time as trapping.
We recall that the trapping happens at $\eps=1$ for which
\begin{equation}
  \label{eq:23}
  J(\eps=1)=8 m_{\trap}\decay^2a_{\trap}^3.
\end{equation}
By comparing this with (\ref{eq:19}), and using conservation of $J(\eps)$ we obtain
\begin{equation}
  \label{eq:24}
  \frac{m_{\trap}}{m_{\present}}\qty(\frac{a_{\trap}}{a_{\present}})^3=\frac{\pi}{8}\frac{\rho_{\Theta,\present}}{\Lambda_{b,\present}^4},
\end{equation}
where $\Lambda_{b,\present}^4=m_{\present}^2\decay^2$ is the zero-temperature barrier height. In order to calculate the trapping temperature we need to specify the scaling of the axion mass with respect to the temperature. We consider the following ansatz:
\begin{equation}
  \label{eq:25}
  m^2(T)=m_{\present}^2\times
  \begin{cases}
    \qty(T/T_{\critical})^{-\barscale},&T>T_c\\
    1,&T\leq T_c
  \end{cases},\qquad T_{\critical}\equiv 2.12\times \Lambda_{b,\present}.
\end{equation}
The idea behind this choice is for $\Lambda_{b,\present}=75.6\,\si{\mega\electronvolt}$ and $\barscale=8.16$, this reproduces the lattice result \cite{Borsanyi:2016ksw} for the QCD axion quite well at high temperatures $T \gtrsim 1\,\si{\giga\electronvolt}$. By using this ansatz we can calculate the trapping temperature via (\ref{eq:24}). The result is
\begin{equation}
  \boxed{
    \label{eq:26}
  \frac{T_{\trap}}{\Lambda_{b,\present}}\approx (2.12)^{\frac{\barscale}{6+\barscale}}\qty(2\times 10^8)^{\frac{2}{6+\barscale}}\qty(\frac{g_s(T_{\trap})}{72})^{-\frac{2}{6+\barscale}}\qty(\frac{\Lambda_{b,\present}}{\rm{GeV}})^{\frac{2}{6+\barscale}}\qty(\frac{h^2 \Omega_{\phi,\present}}{h^2\Omega_{\rm DM}})^{-\frac{2}{6+\barscale}}.
  }
\end{equation}
We can observe that the trapping temperature does depend only on the zero temperature barrier height, and on the scaling of the axion mass at high temperature. For the QCD axion, the trapping temperature does not depend on the axion mass, and it is given by
\begin{equation}
  \label{eq:27}
  T_{\trap}^{\rm QCD}\approx \qty(1.21\,\si{\giga\electronvolt})\qty(\frac{h^2 \Omega_{\phi,\present}}{h^2\Omega_{\rm DM}})^{-0.141}.
\end{equation}
We show a plot of the trapping temperature as a function of the zero-temperature barrier height for various choices of $\gamma$ in Figure \ref{fig:trapping-temperature}. 
 Equation (\ref{eq:26}) is a key quantity that determines  the different fragmentation regimes.


\section{Analytical theory of parametric resonance in kinetic misalignment}
\label{sec:analyt-theory-param}

\subsection{Parametric resonance}
\label{sec:parametric-resonance}

After the discussion of the evolution of the homogeneous mode $\Theta$, now we turn our attention to fluctuations. We generalise the results of \cite{Fonseca:2019ypl} by including the regime of fragmentation after trapping. We also use different initial conditions for the fluctuations. If one neglects the expansion of the universe, the equation of motion for axion mode functions \eqref{eq:7} takes the form
\begin{equation}
  \label{eq:43}
  \ddot{\theta}_k+\qty[\frac{k^2}{a^2}+m^2\cos\Theta]\theta_k=0,
\end{equation}
where $a$ and $m$ are constants. This is a second-order differential equation with periodic coefficients, so it has the form of the \emph{Hill's equation} \cite{Hill1966}. According to the Floquet theorem, the solutions should be of the form
\begin{equation}
  \label{eq:44}
  \theta_k(t)=\theta_+(t;k) e^{\mu_k t} + \theta_-(t;k) e^{-\mu_k t},
\end{equation}
where $\theta_{\pm}(t;k)$ are periodic functions in time, and $\mu_k$'s are in general complex numbers known as \emph{Floquet exponents}. If $\Re{\mu_k}>0$, then the mode does grow exponentially during one oscillation, and one says that the mode experiences instability via parametric resonance. The primary goal of the \emph{Floquet analysis} is to determine the \emph{instability bands} of the system, which are the regions where $\Re{\mu_k}>0$.

In the case of oscillations after trapping, a closed-form expression for the Floquet exponents has been obtained in \cite{Greene:1998pb}. We have found that this method can easily be extended to derive an analogous expression for Floquet exponents during the evolution before trapping. In this section we will just state the result, and discuss the consequences. The detailed derivation in presented in Appendix \ref{sec:deta-disc-param}.

Let $\eps$ be the dimensionless energy of the homogeneous mode as defined in \eqref{eq:9}, and let us  introduce a dimensionless momentum 
\begin{equation}
\kap\equiv k/(ma).
\label{eq:def-kappa}
\end{equation} 
Then the Floquet exponents have the following analytical solution:
\begin{equation}
  \label{eq:55}
  \mu_k=\sqrt{8\kap^2\qty(\eps-\kap)\qty(1-\eps+\kap^2)}\times
  \begin{cases}
    \mathcal{T}_{>}^{-1}(\eps)\,\mathcal{I}\qty(\arcsin\qty(1/\sqrt{2\eps-1})),& \eps>1\\
    \mathcal{T}_{<}^{-1}(\eps)\,\mathcal{I}\qty(\pi/2),&\eps<1
  \end{cases},
\end{equation}
 The first line in (\ref{eq:55}) is new while the second line was derived in the literature.
 $\mathcal{T}_{>}$ and $\mathcal{T}_{<}$ are the periods before and after trapping respectively, and $\mathcal{I}(\varphi)$ is the following integral:
\begin{equation}
  \label{eq:56}
  \mathcal{I}(\varphi)=\int_0^{\varphi}\frac{\dd{\vartheta'}}{1+\qty(1-2\eps+2\kap^2)\sin^2\vartheta'}\frac{2\sin^2\vartheta'}{\sqrt{\qty(1+\sin^2\vartheta')\qty[1+(1-2\eps)\sin^2\vartheta']}}.
\end{equation}
The periods $\mathcal{T}$ can be derived from the conservation of energy. They read
\begin{equation}
  \label{eq:57}
  \mathcal{T}_{>}(\eps)=\frac{2}{m\sqrt{\eps}}\ellipticK(1/\sqrt{\eps})\qand \mathcal{T}_{<}(\eps)=\frac{2}{m}\ellipticK(\sqrt{\eps}).
\end{equation}
From the square root term in \eqref{eq:55} we can immediately obtain the instability band. Before trapping it is given by
\begin{equation}
  \label{eq:58}
  \eps-1<\kap^2<\eps,
\end{equation}
while its form after trapping is
\begin{equation}
  \label{eq:59}
  0<\kap^2<\eps.
\end{equation}
If one neglects the Hubble expansion, both $\eps$ and $\kap$ do not change, so these modes keep growing. Therefore, in order to get realistic results we need to incorporate the Hubble expansion, which we do next.

The Hubble expansion and the temperature dependence of the axion mass can be incorporated by restoring the time dependence of $\eps$ and $\kap$ in the expression for the Floquet exponent \eqref{eq:55}:
\begin{equation}
  \label{eq:60}
  \kap \rightarrow \kap(t)=\frac{k/a(t)}{m(t)}\qand \eps \rightarrow \eps(t),
\end{equation}
where $\eps(t)$ is given by \eqref{eq:9}. As a result, the Floquet exponents now become time-dependent:
\begin{equation}
  \label{eq:61}
  \mu_k\rightarrow \mu_k\qty(\kap(t),\eps(t)).
\end{equation}
Then we can obtain the total amplification factor $\amp_k$ of a given mode by integrating the time-dependent Floquet exponents over time:
\begin{equation}
  \label{eq:62}
  \amp_k(t)=\exp(\int_{t_i}^t\dd{t'}\mu\qty(\kappa(t'),\eps(t'))),
\end{equation}
where $t_i$ is some initial time before which the parametric resonance is not effective. At this point it is  convenient to introduce the dimensionless quantities via
\begin{equation}
  \label{eq:63}
  \kappa_{\trap}\equiv \frac{k/a_{\trap}}{m_{\trap}}\qq{,}\tau\equiv 2H_{\trap}t\qq{,}\tilde{\mu}\equiv \frac{\mu}{m_{\trap}},
\end{equation}
where $\trap$-subscript implies that the quantities are evaluated at $T_{\trap}$ defined by \eqref{eq:8}. In terms of these quantities \eqref{eq:62} becomes
\begin{equation}
\boxed{
  \label{eq:64}
  \amp_k\qty(\tau)=\exp(\frac{m_{\trap}}{2H_{\trap}}\int_{\tau_i}^{\tau}\dd{\tau'}\tilde{\mu}\qty(\kap(\tau'),\eps(\tau')))\equiv \exp(\frac{m_{\trap}}{2H_{\trap}}\mathcal{B}_k(\tau)).
  }
\end{equation}
The growth factor $\mathcal{B}_k$ does depend on the temperature scaling $\gamma$ of the axion mass, but not on the model parameters such as $m_{\trap}$ and $H_{\trap}$. For the modes which amplify most efficiently its value is $\sim \mathcal{O}(0.5)$. Therefore the efficiency of the fragmentation is effectively determined by the hiearchy between the axion mass and Hubble rate at trapping. This result can be understood physically. If the Hubble is much smaller than the axion mass at the beginning of oscillations, the redshifting  of the homogenous mode becomes very slow allowing the axion to probe non-quadratic parts of its potential for a longer time. In the Standard Misalignment Mechanism, $m_{\trap}/H_{\trap}\sim 3$ so the expansion quickly redshifts the amplitude of the oscillations which makes the parametric resonance ineffective. In a nutshell, efficient parametric resonance requires a mechanism which delays the onset of oscillations. The Kinetic Misalignment Mechanism provides this via large initial kinetic energy. In the Large Misalignment Mechanism \cite{Zhang:2017dpp,Arvanitaki:2019rax} this is achieved by tuning\footnote{It is also possible to dynamically generate an initial angle very close to the top \cite{Co:2018mho} and bottom \cite{Co:2018phi} of the potential.} the initial angle to the top of the axion potential such that the onset of oscillations is delayed due to the small potential gradient at the top. 

\begin{figure}[tbp]
  \centering
  \includegraphics[width=\textwidth]{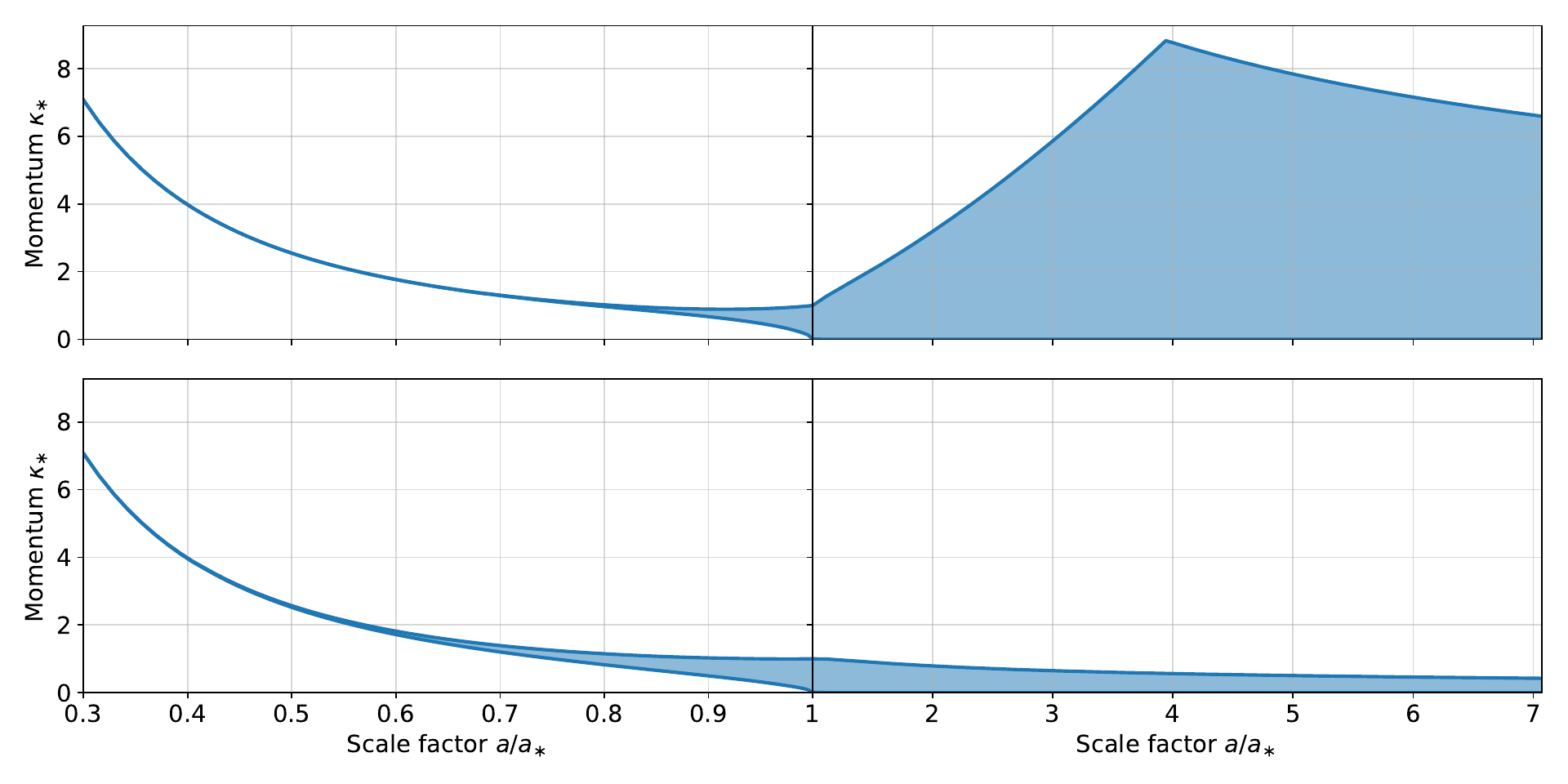}
  \caption{\it \small Time evolution of the instability bands, Eq.~\eqref{eq:70} and \eqref{eq:71}, as a function of the scale factor for two benchmark points with $\gamma=8$ (upper plot), and constant mass $\gamma=0$ (lower plot). In both cases we set $m_0=10^{-11}\,\rm{eV}$, and choose the decay constant $f$ such that $m_{\ast}/H_{\ast}=10^2$. }
  \label{fig:instability-bands}
\end{figure}

It is also instructive to study how the shape of the instability bands changes with time. For this, let us introduce the following quantity:
\begin{equation}
  \label{eq:65}
  \tilde{J}\qty(t)=\frac{m_{\trap}}{m(t)}\qty(\frac{a_{\trap}}{a(t)})^3.
\end{equation}
Then we can write $\kap(\tau)$ as
\begin{equation}
  \label{eq:66}
  \kap(\tau)=\kappa_{\trap}\tilde{J}(\tau)\tau.
\end{equation}
To obtain $\eps(\tau)$ we note that the adiabatic invariant \eqref{eq:14} can be expressed in terms of $\tilde{J}$ by
\begin{equation}
  \label{eq:67}
  J(\eps(\tau))=8\decay^2m(t)a^3(t)\tilde{J}(\tau),
\end{equation}
so that
\begin{equation}
  \label{eq:68}
  \tilde{J}(\eps)=
  \begin{cases}
    \sqrt{\eps}\,\ellipticE(1/\sqrt{\eps}),&\eps>1\\
    \qty(\eps-1)\ellipticK(\sqrt{\eps})+\ellipticE(\sqrt{\eps}),&\eps<1
  \end{cases}.
\end{equation}
At a given time this function can be inverted to get $\eps(\tau)$. In general this needs to be done numerically, however in asymptotic regimes we can use the following approximations:
\begin{equation}
  \label{eq:69}
  \tilde{J}(\eps\gg 1)\approx \frac{\pi}{2}\sqrt{\eps}\qand \tilde{J}(\eps \ll 1)\approx \frac{\pi}{4}\eps.
\end{equation}
We are now ready to discuss the behavior of the instability bands. Before trapping, the expression for the instability band \eqref{eq:58} can be written as
\begin{equation}
  \label{eq:70}
  \frac{\eps-1}{\qty(\tilde{J}(\tau)\tau)^2}<\kappa_{\trap}^2<\frac{\eps}{\qty(\tilde{J}(\tau)\tau)^2}.
\end{equation}
Initially $\eps$ is large, so the width of the instability band is narrow. As the energy of the homogeneous mode decreases, the instability band gradually widens. The behavior after trapping is slightly more involved. The expression for the instability band \eqref{eq:59} becomes
\begin{equation}
  \label{eq:71}
  0<\kappa_{\trap}^2<\frac{\eps}{\qty(\tilde{J}(\tau)\tau)^2}.
\end{equation}
At late times when $\eps\ll 1$ we have $\eps\approx 4\tilde{J}/\pi$. Therefore the upper limit of the instability band can be approximated as
\begin{equation}
  \label{eq:72}
  \frac{\eps}{\qty(\tilde{J}(\tau)\tau)^2}\approx \frac{4}{\pi}\tau^{-1/2}\times
  \begin{cases}
    \tau^{\gamma/4},&T>T_c\\
    m_{\present}/m_{\trap},&T<T_c
  \end{cases}.
\end{equation}
We see that for $\gamma>2$, the upper limit of the instability band grows with time until the axion reaches  its zero-temperature mass, and then decreases with time. This means that the modes with $\kappa_{\trap}>1$ which did enter and exit the instability band at the rolling stage will re-enter the instability band after trapping. On the other hand, for $\gamma<2$, and also for constant axion mass, the width of the instability band will shrink after trapping. We show the evolution of the instability bands, Equations \eqref{eq:70} and \eqref{eq:71}, as a function of the scale factor for these two cases in Figure \ref{fig:instability-bands}. 

\begin{figure}[tbp]
  \centering
  \includegraphics[width=1.1\textwidth]{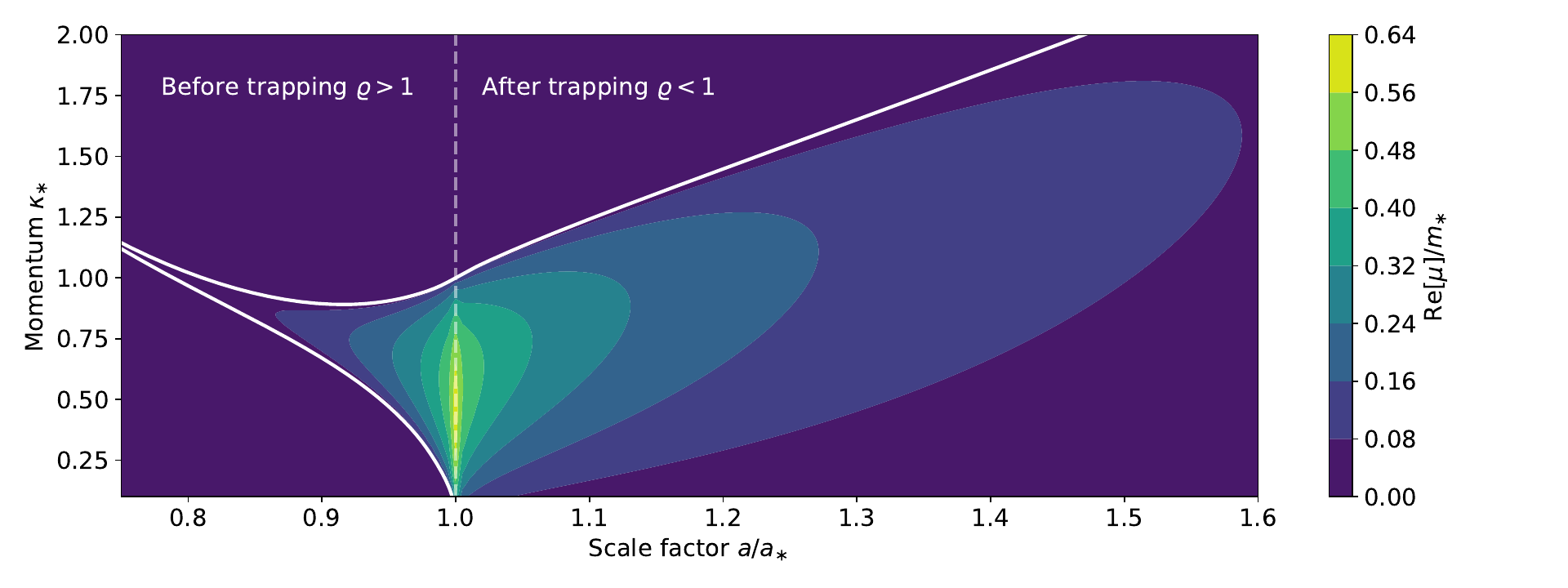}
  \caption{\it \small Evolution of the instability bands together with the value of the Floquet exponents by zooming into the region around trapping. The white lines denote the boundaries of the instability band. Fragmentation is most efficient around trapping $a\approx a_{\trap}$. The model parameters are identical to the ones used in Figure \ref{fig:instability-bands} with $\gamma=8$.}
  \label{fig:instability-bands-zoom}
\end{figure}

Even though the Floquet analysis predicts a very wide instability band, the fragmentation is most efficient around trapping as can be observed in Figure \ref{fig:instability-bands-zoom}. In this plot, we have used the parameters of Figure \ref{fig:instability-bands} with $\gamma=8$, but we also show the value of the Floquet exponents, and zoom-in into the region around trapping. The white lines denote the boundaries of the instability band. 

\begin{figure}[tbp]
  \centering
  \includegraphics[width=\textwidth]{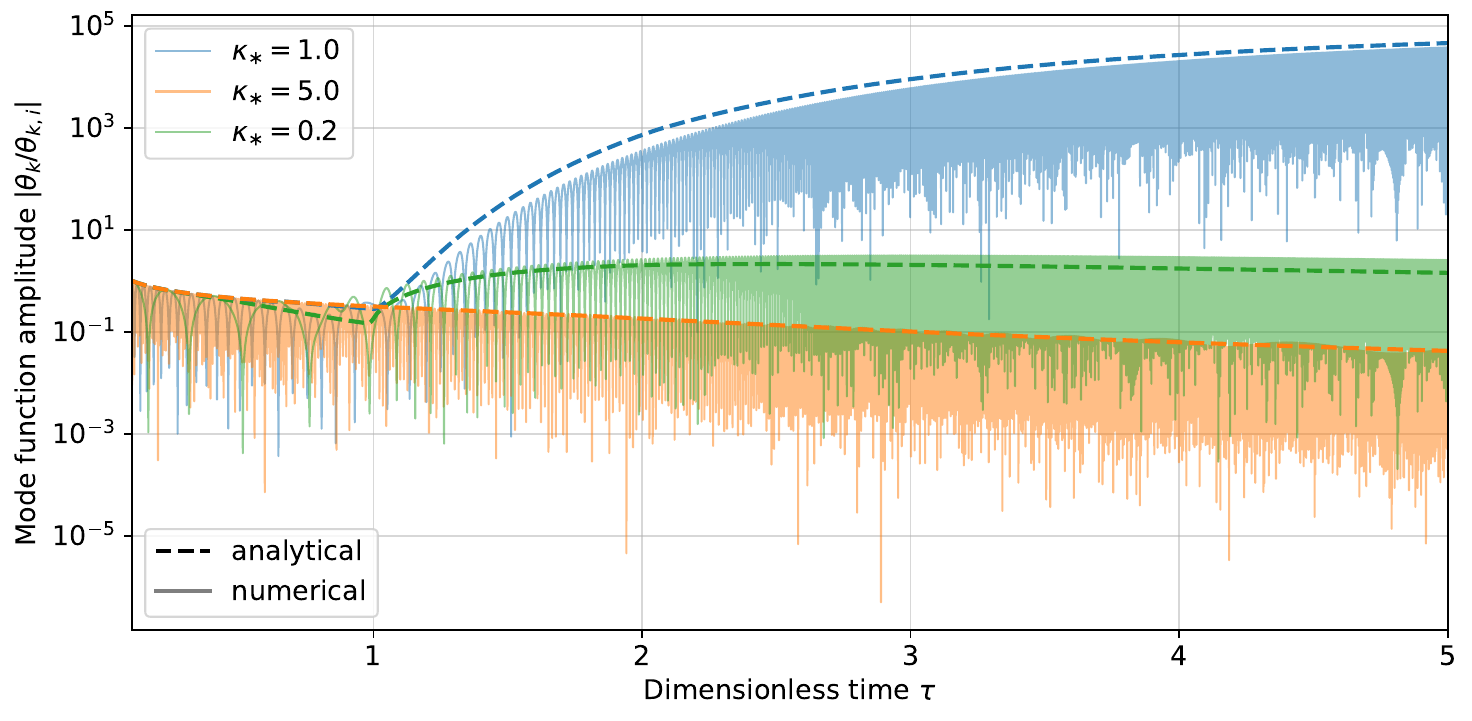}
  \caption{\it \small Comparison between the numerical solution (thin solid lines) and our analytical ansatz \eqref{eq:74} (thick dashed lines) for three benchmark momentum modes, showing a very reasonable agreement. The parameters are the same as in Figures \ref{fig:instability-bands} and \ref{fig:instability-bands-zoom}.  }
  \label{fig:ansatz-compare}
\end{figure}

Our approximation so far sizably  overpredicts the amplification factor. The reason is that the Floquet analysis gives the amplification factor during one oscillation by neglecting the damping of the mode due to the redshift. More specifically, if the Floquet exponents for a mode are imaginary at all times, our approximation would predict that the amplitude of this mode will stay constant which is of course not correct. Instead, the amplitude of this mode will get redshifted, and becomes smaller. To account for this, we assume that on top of the amplification given by \eqref{eq:64}, all modes redshift like a free particle. This decay factor can be calculated using WKB approximation and it reads
\begin{equation}
  \label{eq:73}
  A_k(t)\propto \omega_k^{-1/2}(t)a^{-3/2}(t)\qq{where} \omega_k(t)=\sqrt{\frac{k^2}{a^2(t)}+m^2(T)}.
\end{equation}
So our final ansatz for the mode functions is
\begin{equation}
  \label{eq:74}
  \abs{\theta_k(t)}=\theta_{k,i}A_k(t)\amp_k(t),
\end{equation}
where $\theta_{k,i}$ is related to the initial field power spectrum \eqref{eq:5} by $\theta_{k,i}=\sqrt{\dimps_{\theta}(k)}$, $A_k(t)$ is normalized such that it is unity initially, and we have omitted the oscillatory term. We provide a comparison between this ansatz and the full numerical solutions of the mode functions in Figure \ref{fig:ansatz-compare}. The parameters are the same as in Figures \ref{fig:instability-bands} and \ref{fig:instability-bands-zoom}. The thin solid lines show the numerical solutions, while the thick dashed lines are calculated via \eqref{eq:74}. We started the numerical solution at $\tau=0.1$ with an initial amplitude $\theta_{k,i}=1$ for all the modes. We confirm that our ansatz \eqref{eq:74} provides a very reasonable approximation to the numerical solution. 

\subsection{An estimate of the back-reaction}
\label{sec:an-estim-backr}

In our discussion so far we have neglected the back-reaction of the fluctuations on the homogeneous mode. However this approximation breaks down when the fragmentation is efficient so that the fluctuations cannot be considered small compared to the homogeneous mode \cite{Fonseca:2019ypl}. If the only thing we want to know is whether at a given point on the ALP parameter space the relic density is dominated by fragmented axions or by the homogeneous mode, then the back-reaction effects might not be important.  However,  ultimately we are interested  in observational consequences of fragmentation such as miniclusters \cite{Eroncel:2022efc}. For this we need to calculate the density power spectrum of the ALP field after fragmentation, and in order to get accurate results we need to take the back-reaction into the account.

Precise study of the back-reaction should be performed by using non-perturbative methods such as lattice. However, we can obtain a simple estimate semi-analytically. For this, we only need to assume that fragmentation steals energy from the homogeneous mode, and gives it to the fluctuations. This implies that the energy gained by the fluctuations during the fragmentation is equal to the energy lost by the homogeneous mode during the same period:
\begin{equation}
  \label{eq:75}
  \Delta \rho_{\rm fluct}=-\Delta \rho_{\Theta},
\end{equation}
where $\rho_{\Theta}$ is the energy density in the homogeneous mode. This idea has also been used in \cite{Fonseca:2019ypl} for an extensive discussion of the fragmentation before trapping. As we will show later, our calculations reproduce these results under appropriate limits.

\begin{figure}[tbp!]
  \centering
  \includegraphics[width=0.8\textwidth]{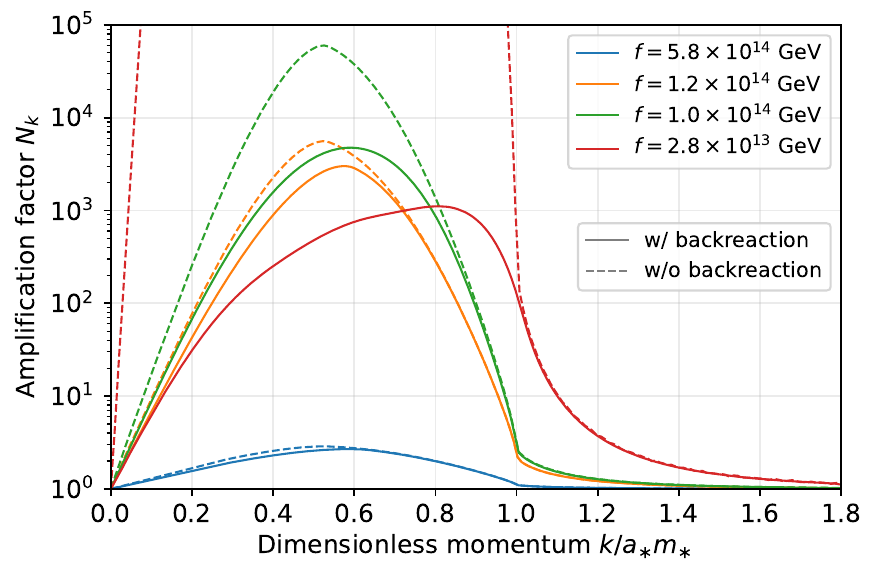}
  \caption{\it \small This plot compares the estimates of the amplification factor $N_k$ performed by including/not including the back-reaction effects through the procedure  \eqref{eq:81}. It shows the importance of back-reaction effects as $f$ is decreased. This plot is made for a constant mass $m_*=m_0=10^{-15}$ eV. }
  \label{fig:back-reaction}
\end{figure}

The energy density in the fluctuations is
\begin{equation}
  \label{eq:76}
  \rho_{\rm fluct}= \frac{\decay^2}{2}\int \frac{\dd[3]{k}}{(2\pi)^3}\qty[\abs{\dot{\theta}_k(t)}^2+\qty(\frac{k^2}{a^2}+m^2\cos\Theta)\abs{\theta_k(t)}^2],
\end{equation}
where we have neglected higher-order terms in $\theta_k$. Since the mode functions oscillate very rapidly, we can assume that the kinetic and potential terms contribute equally. We also average over the oscillations of the homogeneous mode. With these assumptions and our ansatz \eqref{eq:74}, we approximate the energy density in fluctuations as
\begin{equation}
  \label{eq:77}
  \rho_{\rm fluct}(t)\approx \frac{f^2}{4\pi^2}\int \dd{k}k^2\qty(\frac{k^2}{a^2}+m^2\overline{\cos\Theta})\dimps_{\theta}(k)A_k^2(t)\amp_k^2(t),
\end{equation}
where
\begin{equation}
  \label{eq:78}
  \overline{\cos\Theta}\equiv \frac{1}{\Theta_{\rm max}-\Theta_{\rm min}}\int_{\Theta_{\rm min}}^{\Theta_{\rm max}}\dd{\Theta}\cos\Theta=
  \begin{cases}
    0,&\eps>1,\\
    \sinc(2\arcsin \sqrt{\eps}),&\eps<1
  \end{cases}.
\end{equation}
The extra $1/2$ factor in \eqref{eq:77} arises from averaging over the mode function oscillations. From this result, we can estimate the energy lost by the homogeneous mode during a short time period $\Delta t$ as
\begin{equation}
  \label{eq:79}
  -\frac{\Delta \rho_{\Theta}}{\Delta t}=\frac{\Delta \rho_{\rm fluct}}{\Delta t}\approx \frac{f^2}{4\pi^2}\int \dd{k}k^2\qty(\frac{k^2}{a^2}+m^2\overline{\cos\Theta})\dimps_{\theta}(k)A_k^2(t)\frac{\amp_k^2(t)-\amp_k^2(t-\Delta t)}{\Delta t}.
\end{equation}
Note that we have only varied the amplification factor $\mathcal{N}_k$ since we are only interested in the change in the energy due to the fragmentation. In the $\Delta t\rightarrow 0$ limit we find
\begin{equation}
  \label{eq:80}
  -\dv{\rho_{\Theta}}{t}=\dv{\rho_{\rm fluct}}{t}\approx \frac{f^2}{4\pi^2}\int \dd{k}k^2\qty(\frac{k^2}{a^2}+m^2\overline{\cos\Theta})\dimps_{\theta}(k)A_k^2(t)\qty[2 N_k^2(t)\mu\qty(\kap(t),\eps(t))],
\end{equation}
where we have used \eqref{eq:62} when taking the time derivative.

With this result, we can employ the following procedure to calculate the amplification factors including back-reaction. Choose an initial time $\tau_i$ at which $\dimps_{\theta}(k)$ is known and $\amp_k(\tau_i)=1$ for all modes. Pick a sufficiently small time-step $\Delta \tau$, and at each time interval calculate the amplification factors by
\begin{equation}
  \label{eq:81}
  \ln \amp_k(\tau+\Delta \tau)=\ln \amp_k(\tau)+\frac{m_{\trap}}{2 H_{\trap}}\int_{\tau}^{\tau + \Delta \tau}\dd{\tau}'\tilde{\mu}_k(\tau')\approx \ln \amp_k(\tau)+\frac{m_{\trap}}{2 H_{\trap}}\Delta \tau\,\tilde{\mu}_k(\tau).
\end{equation}
For each time step, find $\eps$ via  \eqref{eq:65} and \eqref{eq:68} and 
therefore the time-dependent Floquet exponent $\mu_k$. Take into account the decrease in $\eps$ due to the back-reaction via  \eqref{eq:80}. Then repeat the steps either until fragmentation is no longer efficient, or when $\eps$ drops to zero indicating complete fragmentation. At this stage, the dynamics of the system becomes non-perturbative, so our calculation should be considered an estimate.
We illustrate the importance of back-reaction effects in Figure \ref{fig:back-reaction}.
We will use this procedure in the following, especially to determine the boundary between the `after-trapping' and `before-trapping' fragmentation regimes defined in Section \ref{fig:back-reaction}.

\subsection{Initial conditions for the mode functions}
\label{sec:initial-conditions}

In order to study the consequences of fragmentation for ALP dark matter, we need to specify the initial conditions for the mode functions prior to fragmentation. A scalar field, like any component in the universe, can have adiabatic and isocurvature fluctuations. Adiabatic ones arise solely due to the temperature fluctuations in the universe, and are unavoidable for any cosmological fluid including scalar fields. On the other hand, the isocurvature fluctuations are the fluctuations on constant density slices, and they describe any other kind of fluctuation which is not adiabatic.


In the case of standard misalignment mechanism, adiabatic fluctuations of an ALP field are negligible during the early evolution when it is Hubble frozen \cite{Marsh:2015xka}. They start to grow once the homogeneous mode starts oscillating. If the ALP field is present during inflation, for example if the Peccei-Quinn (PQ) symmetry is broken during the inflation in the case of QCD axion, then it will pick up quantum fluctuations given by $\delta \phi=H_I/2\pi$, where $H_I$ is the inflation scale. This causes fluctuations in the initial axion angle $\Theta_i$ given by $\delta \Theta_i\simeq H_I/(2\pi S_I)$ where $S_I$ is the effective axion decay constant during inflation which can be different than its value at low temperature \cite{Linde:1991km}.

The situation changes drastically in the case of rotating axions. In this case, the ALP field has a much more complicated cosmological history which is highly model-dependent. A common property in all of these histories is that the ALP field starts moving much earlier compared to the standard case, and it has a large velocity long before the ALP potential turns on. As we show below, this velocity together with curvature perturbations acts as a source term for adiabatic fluctuations and make them grow. In addition to these, the kick mechanism can induce isocurvature fluctuations from the quantum fluctuations of the field responsible for producing the kick. These can constrain the parameter space due to their isocurvature nature, and can also lead to domain wall problems in some cases \cite{Co:2020dya}.

An extensive discussion of model realizations of the kinetic misalignment mechanism including the possible cosmological histories as well as the implications of isocurvature fluctuations was presented 
in \cite{Gouttenoire:2021jhk}. The detailed consequences of ALP dark matter  in these UV completions will be presented 
in \cite{Eroncel:2022a}. For simplicity, and to remain agnostic about the UV completion, we consider a standard cosmological history in this work, and assume that the modes that are relevant for fragmentation do enter the horizon when the ALP field scales as kination. We will present the necessary condition for the validity of this assumption towards the end of this sub-section.


To get the evolution, we start by writing the FRLW metric with curvature perturbations included. In conformal time $\eta$ and Newtonian (conformal) gauge, the metric reads\footnote{A brief review of cosmological perturbation theory and our notation convention can be found in Appendix \ref{sec:calc-adiab-init}}. 
\begin{equation}
  \label{eq:148}
  \dd{s}^2=a^2(\eta)\qty{-\qty[1+2\Psi(\eta,\vb{x})]\dd{\eta}^2+\qty[1+2\Phi(\eta,\vb{x})]\delta_{ij}\dd{x}^i\dd{x}^j}.
\end{equation}
Here $\Psi$ and $\Phi$ are Bardeen potentials \cite{Bardeen:1980kt}, and they represent the curvature perturbations. In Appendix \ref{sec:deriv-adiab-init}, we show that the mode functions have the following equations of motion at early times:
\begin{equation}
  \label{eq:149}
  \theta_k''+2\mathcal{H}\theta_k'+k^2\theta_k=-4\Phi_k'\Theta',
\end{equation}
where $\prime$ denotes the derivative with respect to conformal time, and $\mathcal{H}=aH$ is the conformal Hubble parameter. The right side of this equation represents the source term due to the curvature fluctuations. This term clearly represents the difference between the standard and kinetic misalignment mechanisms. 
In the standard case, the ALP field is frozen due to the Hubble friction, so $\Theta'=0$. Therefore, the source term is absent, and the adiabatic fluctuations remain zero until the oscillations start. However, in the case of kinetic misalignment mechanism, the ALP field receives a kick at a much earlier time, so the right side is non-zero for a much longer time. As a result, the standard and kinetic misalignment mechanisms predict different initial conditions for mode functions.
To our knowledge, this is the first time that
this is pointed out. We will discuss more thoroughly further implications of this source term in 
\cite{Eroncel:2025qlk}

 By an explicit calculation of the mode functions which we describe in Appendix \ref{sec:deriv-adiab-init}, see (Eq.~\ref{eq:197}), we obtain the following result for the power spectrum:
\begin{equation}
\boxed{
  \label{eq:150}
  \dimps_{\theta}(k;a)=\abs{\theta_k(a)}^2\approx \frac{2\pi^2}{k^3}\qty(\frac{1}{3})^2A_s\qty(\frac{\dot{\Theta}}{H})^2
  }
\end{equation}
where $A_s=2.1\times 10^{-9}$ is the amplitude of the primordial power spectrum at the pivot scale $k=0.05\,\textrm{Mpc}^{-1}$ as measured by Planck 2018 (TT,TE,EE+lowE+lensing 68\%) \cite{Aghanim:2018eyx}. This result is valid for both super- and sub-horizon modes, but assumes that the mode is super-horizon when the ALP field starts its kination-like scaling which we denote by $a_{\kin}$. The behavior of the modes which were sub-horizon at $a_{\rm kin}$ cannot be determined without specifying the cosmological history before $a_{\rm kin}$.

From this result alone, we can put a bound on the duration of the kination-like scaling. For this, we evaluate the variance of the axion velocity:
\begin{equation}
  \label{eq:151}
  \expval{\qty(\delta \dot{\theta})^2}=\frac{1}{2\pi^2}\int_0^{\infty}\dd{k}k^2\abs{\dot{\theta}_k}^2=\frac{1}{2\pi^2a^2}\int_0^{\infty}\dd{k}k^2\abs{\theta_k'(\eta)}^2.
\end{equation}
The integral is dominated by the modes which are sub-horizon, but were super-horizon at $a_{\kin}$. Then by approximating $\abs{\theta_k'(\eta)}^2\approx k^2\abs{\theta_k(\eta)}^2$, and using \eqref{eq:150} we find at late times
\begin{equation}
  \label{eq:152}
  \expval{\qty(\delta \dot{\theta})^2}\approx \frac{1}{2}\qty(\frac{1}{3})^2A_s\dot{\Theta}^2\qty(\frac{a}{a_{\kin}})^2.
\end{equation}
Then we can estimate the density contrast by
\begin{equation}
  \label{eq:153}
  \delta_{\phi}\sim \frac{\delta\qty(\dot{\theta}^2/2)}{\dot{\Theta}^2/2}\sim \frac{2}{\dot{\Theta}}\sqrt{\expval{\qty(\delta \dot{\theta})^2}}\approx \sqrt{2 A_s}\qty(\frac{1}{3})\frac{a}{a_{\kin}}.
\end{equation}
This becomes $\mathcal{O}(1)$ when 
\begin{equation}
\boxed{
\label{eq:kination-bound}
a/a_{\kin} \gtrsim 10^5 .}
\end{equation}
 This implies that if the ALP field scales as kination more than $\ln(10^5)\sim 10$ number of e-folds, then the ALP field cannot be considered homogeneous anymore. A similar bound applies if the  ALP drives a kination era. The bound (\ref{eq:kination-bound}) is an important constraint  when considering UV completions \cite{Eroncel:2022a}, in particular models where the ALP drives temporarily a kination era in the early universe, enhancing primordial GW signals. Such bound on the total duration of a kination era was overlooked in previous literature on kination, as discussed in \cite{Gouttenoire:2021jhk}. 
 We will investigate this issue further in an upcoming work 
 \cite{Eroncel:2025qlk}.
 In this paper, we do assume that this bound is not violated, so that the ALP field can be considered homogeneous at the onset of fragmentation.

As we have stated previously, in this work we will assume that all the modes relevant for fragmentation are super-horizon at $a_{\rm kin}$. This assumption requires that
\begin{equation}
  \label{eq:273}
  \frac{k}{a_{\kin}H_{\kin}}< 1\quad\Rightarrow\quad \kappa_{\trap}\frac{m_{\trap}}{H_{\trap}}\frac{a_{\kin}}{a_{\trap}}<1.
\end{equation}
If we demand that this condition is satisfied for the relevant modes for fragmentation, i.e. $\kappa_{\trap}\sim \mathcal{O}(1)$, then we need to demand that
\begin{equation}
  \label{eq:274}
  \frac{m_{\trap}}{H_{\trap}}\lesssim \frac{a_{\trap}}{a_{\rm kin}}.
\end{equation}
The quantity $a_{\trap}/a_{\kin}$ depends on the specifics of the UV completion, however we have shown that the homogeneity of the ALP field prior to fragmentation requires $a_{\trap}/a_{\kin}\gtrsim 10^5$ which puts the bound $m_{\trap}/H_{\trap}\lesssim 10^5$. In the next section, we will show that this bound is satisfied in the region of the ALP parameter space where our analytical approximation is under control.

With all these assumptions above, we can fix the field power spectrum at the onset of fragmentation by
\begin{equation}
  \label{eq:154}
  \dimps_{\theta}(k;a_{\ini})=\frac{2\pi^2}{k^3}\qty(\frac{1}{3})^2A_s\qty(\frac{\dot{\Theta}_{\ini}^2}{H_{\ini}^2}).
\end{equation}
As we will see explicitly in the coming sections, the choice of $a_{\ini}$ is not relevant for the final result, as long as it is early enough that $\dot{\Theta}\propto a^{-3}$, and any fragmentation prior to $a_{\ini}$ can be neglected. 
\newpage

\section{ALP dark matter from fragmentation}
\label{sec:alp-dark-matter}

\subsection{Overview of the fragmentation regions}
\label{sec:overv-fragm-regi}

One can split the ALP parameter space into different regions depending whether the fragmentation happens and when. There are four different scenarios:
\begin{enumerate}
\item \textbf{Standard misalignment: }In this case the onset of oscillations is not delayed from its conventional value $m(T_{\osc})=3H(T_{\osc})$, and the standard misalignment mechanism is at play.
\item \textbf{Kinetic misalignment with weak fragmentation: }The ALP field has a non-zero initial velocity such that the onset of oscillations is delayed, but the particle production is not strong enough so that the energy density in fluctuations is always subdominant compared to the zero mode. 
\item \textbf{Fragmentation after trapping: }The ALP field is completely fragmented, but the fragmentation ends after it would have been trapped by the potential in the absence of fragmentation, i.e. $T_{\fend}<T_{\trap}$.
\item \textbf{Fragmentation before trapping: }The ALP field is completely fragmented, and the fragmentation ends before it would have been trapped by the potential in the absence of fragmentation. It other words the fragmentation is complete at $T_{\fend}>T_{\trap}$ where $T_{\trap}$ is given by \eqref{eq:27}.
\end{enumerate}
The boundaries between these regions depend strongly on the hierarchy between the axion mass and Hubble at trapping, and have a very mild dependence on the other model parameters.  In the rest of this section, we will calculate these boundaries while giving details on the properties of the fragmentation in each region.

\begin{figure}[tbp]
  \centering
  \includegraphics[width=\textwidth]{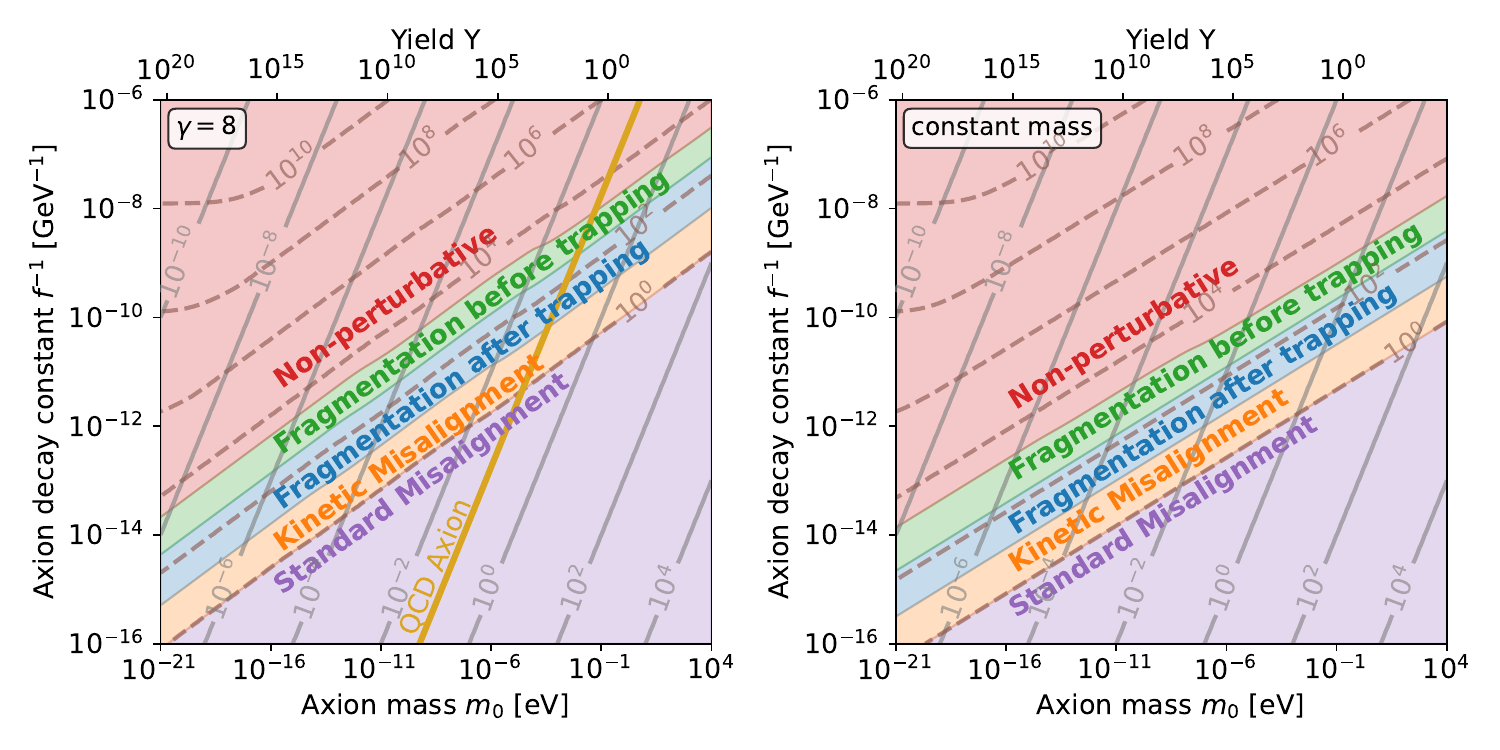}
  \caption{\it \small Overview of the fragmentation regions for temperature-dependent axion mass with $\gamma=8$ (left plot), and for constant axion mass (right plot). The solid contours denote the zero-temperature barrier heights, while the dashed ones are the $m_{\trap}/3H_{\trap}$ contours, where $T_{\trap}$  is given in Eq.~\ref{eq:26}.}
  \label{fig:mastoverHast}
\end{figure}

\begin{figure}[tbp]
  \centering
  \includegraphics[width=\textwidth]{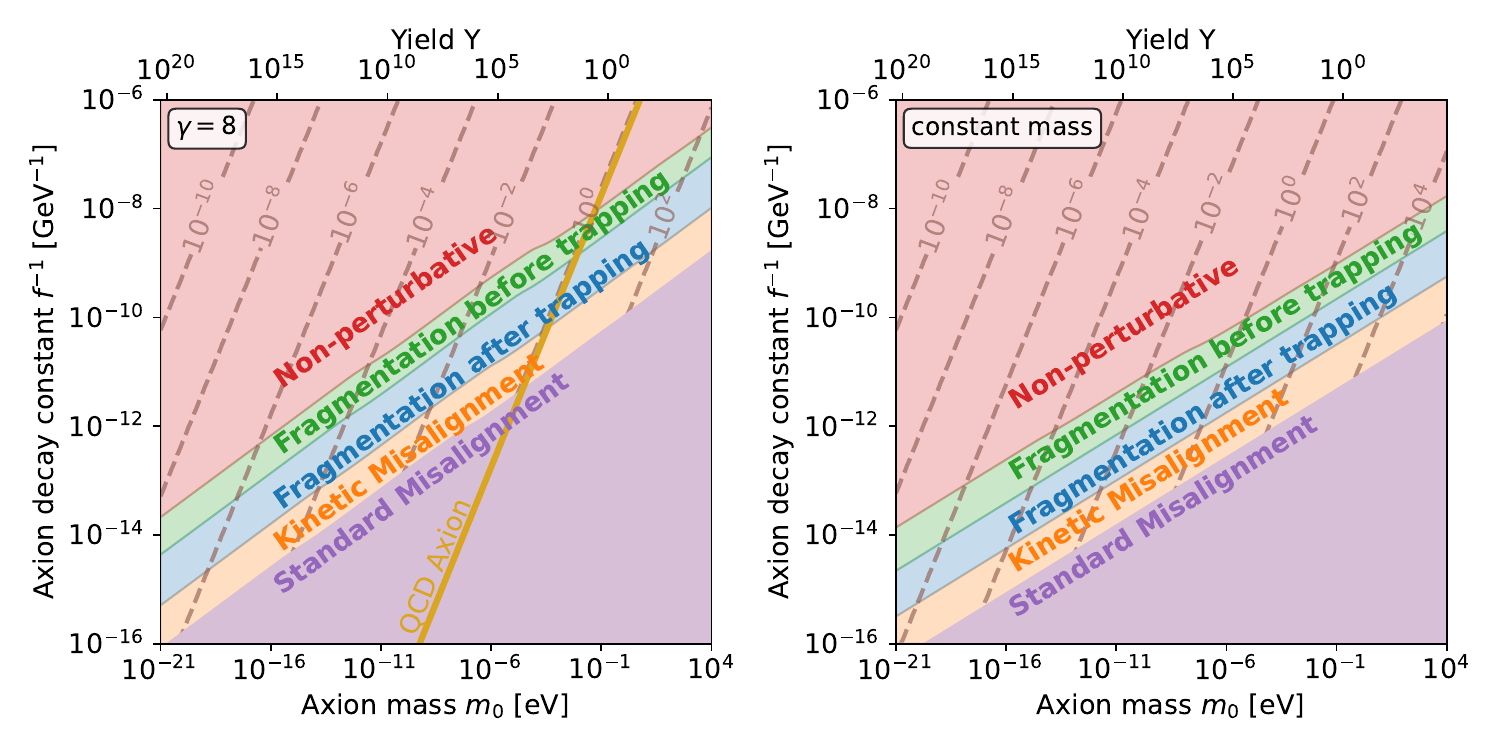}
  \caption{\it \small Same as Figure \ref{fig:mastoverHast}, but the dashed contours now show the trapping temperature $T_*$  in GeV, as given in Eq.~\ref{eq:26}.}
  \label{fig:trapping}
\end{figure}

An overview of these regions on the $m_0$--$f$ plane along with various model parameters can be seen in Figures \ref{fig:mastoverHast} and \ref{fig:trapping}.  

\subsubsection{Boundary between regions $1$ and $2$:}

In order to have kinetic misalignment, the onset of oscillations needs to be delayed so $m(T_{\trap})>3 H(T_{\trap})$. Therefore the boundary between the regions $1$ and $2$ is given by the condition
\begin{equation}
  \label{eq:584}
  m(T_{\trap})=3 H(T_{\trap}).
\end{equation}
In a generic ALP model, the critical value of the axion decay constant $\decay_{\rm crit}$ for a given zero-temperature axion mass $m_0$ can be calculated via \eqref{eq:24} and \eqref{eq:25}. It can approximately be expressed as
\begin{equation}
\boxed{
  \label{eq:377}
  \decay_{\rm crit}^{1\text{ -- }2}\approx 10^{12}\,\si{\giga\electronvolt} \qty[\qty(3\pi)^{-12-2\gamma}\qty(\frac{g_s(T_{\trap})}{72})^{2+\gamma}\qty(\frac{m_0}{\si{\electronvolt}})^{-4-\gamma}\qty(\frac{h^2\Omega_{\Theta,\present}}{h^2\Omega_{\rm DM}})^{8+2\gamma}]^{\frac{1}{16+3\gamma}}
  }
\end{equation}
For the QCD axion this value is 
\begin{equation}
  \label{eq:600}
  \decay_{\rm crit}^{1\text{ -- }2}\approx 2.15\times 10^{11}\;\si{\giga\electronvolt},\qq{QCD Axion.}
\end{equation}
This is in agreement with the result obtained in \cite{Co:2019jts}.

\subsubsection{Boundary between regions $2$ and $3$:}

To calculate this boundary, we need to find the region where the energy density in the fluctuation remains subdominant compared to the energy density in the homogeneous mode. In this region, the back-reaction can be neglected. The energy density in the homogeneous mode is
\begin{equation}
  \label{eq:118}
  \rho_{\Theta}=2\decay^2m^2\eps,
\end{equation}
while the energy density in the fluctuations is \eqref{eq:77}
\begin{equation}
  \label{eq:138}
  \rho_{\rm fluct}=\frac{f^2}{4\pi^2}\int \dd{k}k^2\qty(\frac{k^2}{a^2}+m^2\overline{\cos\Theta})\dimps_{\theta}(k)A_k^2(t)\amp_k^2(t).
\end{equation}
Let us define a quantity $\eff\equiv \lim_{t\rightarrow \infty}\rho_{\rm fluct}/\rho_{\Theta}$ as the ratio between the two energy densities at late time limit. We refer this quantity as the ``efficiency'' of the fragmentation. Then the transition from weak fragmentation to complete fragmentation should occur approximately when $\eff$ reaches to unity. This point will determine the boundary between the regions $2$ and $3$.

We are interested in the late time limit of $\eff$. In this limit $\overline{\cos\Theta}=1$, and we can assume that all the modes which did grow during the parametric resonance become non-relativistic so that the momentum term in \eqref{eq:138} can be ignored. The redshift factor $A_k^2(t)$ at late times can be approximated by
\begin{equation}
  \label{eq:139}
  A_k^2(t)=\frac{\omega_k(t_i)}{\omega_k(t)}\qty(\frac{a_i}{a})^3\approx \frac{k/a_i}{m_0}\qty(\frac{a_i}{a})^3=\frac{k a_i^2}{m_0a^3}.
\end{equation}
By plugging this result into \eqref{eq:138} and using the expression for the initial power spectrum, we find that the energy density in the fluctuations approach an asymptotic value given by
\begin{equation}
  \label{eq:141}
  \rho_{\rm fluct}\rightarrow \frac{m_0m_{\trap}\decay^2}{2}\qty(\frac{1}{3})^2A_s\qty(\frac{\dot{\Theta}_{\ini}}{H_{\ini}}\frac{a_{\ini}}{a_{\trap}})^2\qty(\frac{a_{\trap}}{a})^3\int\dd{\kappa_{\trap}}\exp(\frac{m_{\trap}}{H_{\trap}}\mathcal{B}_{\kappa}^{\infty}),
\end{equation}
where $\mathcal{B}_{\kappa}^{\infty}$ is the asymptotic value of the growth factor defined in \eqref{eq:64}. The late time limit of the energy density of the homogeneous mode can be found by using the fact that for $\eps\ll 1$ we have
\begin{equation}
  \label{eq:142}
  \eps\approx \frac{4}{\pi}\tilde{I}=\frac{4}{\pi}\frac{m_{\trap}}{m}\qty(\frac{a_{\trap}}{a})^3.
\end{equation}
With this result, the efficiency factor takes the form
\begin{equation}
  \label{eq:143}
  \Delta=\frac{\pi}{16}\qty(\frac{1}{3})^2A_s\qty(\frac{\dot{\Theta}_{\ini}}{H_{\ini}}\frac{a_{\ini}}{a_{\trap}})^2\int\dd{\kappa_{\trap}}\exp(\frac{m_{\trap}}{H_{\trap}}\mathcal{B}_{\kappa}^{\infty}).
\end{equation}
Note that at early time $\dot{\Theta}\propto a^{-3}$, therefore the pre-factor in this expression does not depend on the choice of the initial time provided that it is early enough so that $\dot{\Theta}\propto a^{-3}$, and any fragmentation prior to this time can be neglected. By expressing the axion velocity $\dot{\Theta}_i$ in terms of the yield parameter introduced in Section \ref{sec:adiab-invar-relic} as $\dot{\Theta}_i=s(T_i)Y/\decay^2$ where $s(T_i)$ is the entropy density at $T_i$, we can derive the following result:
\begin{equation}
  \label{eq:144}
  \frac{\dot{\Theta}_{\ini}}{H_{\ini}}\frac{a_{\ini}}{a_{\trap}}\approx \frac{4\pi}{3}\sqrt{\frac{g_s(T_{\trap})}{10}}\frac{Y T_{\ast}\mpl}{f^2}.
\end{equation}
When arriving this, we have neglected the difference between the effective degrees of freedom at $T_i$ and $T_{\ast}$, and also assume that the effective degrees of freedoms in the entropy and the energy density are the same. So our final expression for the efficiency factor is
\begin{equation}
  \label{eq:145}
  \Delta=\frac{\pi^3}{10}\qty(\frac{1}{9})^2A_sg_s(T_{\trap})\qty(\frac{Y T_{\trap}\mpl}{f^2})^2\int\dd{\kappa_{\trap}}\exp(\frac{m_{\trap}}{H_{\trap}}\mathcal{B}_{\kappa}^{\infty}).
\end{equation}
With this factor, we can calculate the critical point at which the efficiency becomes unity. We find that across the parameter space the transition happens around $m_{\trap}/H_{\trap}\approx 42$ for $\gamma=8$, and around $m_{\trap}/H_{\trap}\approx 38$ for constant axion mass ($\gamma=0$). This leads to a critical axion decay constant given by
\begin{equation}
\boxed{
  \label{eq:378}
  \decay_{\rm crit}^{2\text{ -- }3}\approx 10^{12}\,\si{\giga\electronvolt} \qty[\qty(40\pi)^{-12-2\gamma}\qty(\frac{g_s(T_{\trap})}{72})^{2+\gamma}\qty(\frac{m_0}{\si{\electronvolt}})^{-4-\gamma}\qty(\frac{h^2\Omega_{\Theta,\present}}{h^2\Omega_{\rm DM}})^{8+2\gamma}]^{\frac{1}{16+3\gamma}}}
\end{equation}
For the QCD-Axion, this boundary corresponds to a decay constant of
\begin{equation}
  \label{eq:146}
    \decay_{\rm crit}^{2\text{ -- }3}\approx 1.5\times 10^{10}\;\si{\giga\electronvolt},\qq{QCD Axion.}
  \end{equation}

\subsubsection{Boundary between regions 3 and 4:}
In both regions the parametric resonance is efficient enough for complete fragmentation. We want to determine whether the fragmentation ends before the ALP field would have been trapped by the potential in the absence of fragmentation. At early times when $\eps\gg 1$, we can simplify the fragmentation calculation to understand qualitative features of the fragmentation. In this regime, we can also show that the procedure we have outlined in Section \ref{sec:analyt-theory-param} reproduces the results of \cite{Fonseca:2019ypl}. We present this calculation in Appendix \ref{sec:fragm-before-trapp}. Here we will give a quick summary.

In the $\eps\gg 1$ regime, the exponential particle production happens when
\begin{equation}
  \label{eq:158}
  H(T)\lesssim \frac{\pi m^4(T)}{4 \dot{\Theta}^3(T)}.
\end{equation}
The temperature at which this inequality is satisfied can be defined as the onset of fragmentation provided that it is larger than the trapping temperature $T_{\trap}$. We denote this temperature by $T_{\weak}$. Since the back-reaction is not efficient in the beginning of fragmentation, the onset of fragmentation does not depend on the initial conditions of the mode functions.

The back-reaction becomes relevant when
\begin{equation}
  \label{eq:159}
  H(T)\lesssim \frac{\pi m^4(T)}{4 \dot{\Theta}^3(T)}\frac{1}{\ln \alpha^{-1}(T)},
\end{equation}
where for adiabatic initial conditions
\begin{equation}
  \label{eq:160}
  \alpha(T)=\qty(\frac{4\pi}{9})^2\qty(\frac{g_s(T_{\trap})}{80})^2\qty(\frac{Y T_{\trap}\mpl}{f^2})^2A_s \qty(\frac{T}{T_{\trap}})^2.
\end{equation}
If the inequality \eqref{eq:159} gets saturated before trapping, then the energy density of the homogeneous mode gets diluted quite rapidly, and the fragmentation is completed in a short amount of time. To estimate the temperature at which the back-reaction becomes prominent, we can replace $\dot{\Theta}(T)\approx s(T)Y/f^2$ and $m(T)=m_{\trap}(T_{\trap}/T)^{\gamma/2}$ in \eqref{eq:159} to get an equation in terms of the temperature $T$ and other model parameters. Let $T_{\strong}$ denote the solution of this equation. For fragmentation to be completed before trapping, this temperature $T_{\strong}$ should be larger than the trapping temperature $T_{\trap}$. Therefore, the Region 4 can be approximately defined by the condition $T_{\strong}\gtrsim T_{\trap}$. Across the parameter space, the transition happens around $m_{\trap}/H_{\trap}\sim 9(5)\times 10^2$ for $\gamma=8(0)$. The expression for the critical decay constant can be approximated by
\begin{equation}
\boxed{
  \label{eq:379}
  \decay_{\rm crit}^{3\text{ -- }4}\approx 10^{12}\,\si{\giga\electronvolt} \qty[\qty(\mathcal{O}(1)\pi \times 10^{2})^{-12-2\gamma}\qty(\frac{g_s(T_{\trap})}{72})^{2+\gamma}\qty(\frac{m_0}{\si{\electronvolt}})^{-4-\gamma}\qty(\frac{h^2\Omega_{\Theta,\present}}{h^2\Omega_{\rm DM}})^{8+2\gamma}]^{\frac{1}{16+3\gamma}}
  }
\end{equation}
where the $\mathcal{O}(1)$ factor is $\gamma$-dependent. For the QCD axion, the critical decay constant at this boundary is
\begin{equation}
  \label{eq:161}
  \decay_{\rm crit}^{1\text{ -- }2}\approx 7.3\times 10^{8}\;\si{\giga\electronvolt},\qq{QCD Axion.}
\end{equation}

\subsubsection{Breakdown of perturbativity for larger $m_{\trap}/H_{\trap}$}
\label{sec:breakd-pert-larg}

Finally we discuss what happens in the parameter space where $m_{\trap}/H_{\trap}$ is even larger. For this we revisit the equation of motion for the homogeneous mode \eqref{eq:6}, but we also include the back-reaction at leading order. Then the equation of motion is modified to \cite{Fonseca:2019ypl}
\begin{equation}
  \label{eq:162}
  \ddot{\Theta}+3H\dot{\Theta}+m^2\sin\Theta-\frac{1}{2}m^2\sin\Theta \expval{\qty(\delta\theta)^2}=0.
\end{equation}
The last term is responsible for the back-reaction of fluctuations, and as a result for the completion of the fragmentation. All of our analysis depends on the assumption that this term is negligible before the onset of fragmentation. Therefore we should check whether the adiabatic initial conditions for the mode functions \eqref{eq:150} do not violate this assumption at the beginning of fragmentation.

By using \eqref{eq:150}, the variance can be calculated by
\begin{equation}
  \label{eq:163}
  \expval{\qty(\delta\theta)^2}=\int \frac{\dd[3]{k}}{(2\pi)^3}\abs{\theta_k}^2\sim \qty(\frac{1}{3})^2A_s\qty(\frac{\dot{\Theta}}{H})^2,
\end{equation}
where we have neglected the modes with $k/a_{\kin}H_{\kin}>1$, and assumed $\ln(a/a_{\kin})\sim\mathcal{O}(1)$. At early times, the background evolution is dominated by the first two terms in \eqref{eq:162} since the mass is negligible. However, if the variance becomes much larger than unity at these times, it can affect the slow-roll evolution much earlier than the fragmentation does. To avoid this case, we demand that
\begin{equation}
  \label{eq:164}
  3H(T)\dot{\Theta}(T)>\frac{m^2(T)}{2}\qty(\frac{1}{3})^2A_s \qty(\frac{\dot{\Theta}(T)}{H(T)})^2
\end{equation}
until the onset of fragmentation $T_{\weak}$ given by the solution of \eqref{eq:158}. This puts a strong bound on the parameter space which can roughly be approximated by
\begin{equation}
\boxed{
  \label{eq:165}
  \frac{m_{\trap}}{H_{\trap}}\lesssim \mathcal{O}(1\text{---}10)\times 10^3,
  }
\end{equation}
where the factor depends on $\gamma$. We want the stress that this result does not exclude the parameter space where this bound is violated. It just implies that the analysis we describe might not be reliable, and a dedicated study is needed. In this paper, we concentrate on the region of the parameter space where \eqref{eq:165} is satisfied\footnote{We have also checked whether the variance term can dominate the mass term in \eqref{eq:162} before the onset of fragmentation. This yields a weaker bound given by $m_{\trap}/H_{\trap}\lesssim \mathcal{O}(1)\times 10^5$.}.

\subsection{ALP relic density with fragmentation}
\label{sec:alp-relic-density}

If the ALP field is completely fragmented, then all the energy density in the homogeneous mode gets transferred into the fluctuations. This will have an effect on the relic density today, since the redshift of the ALP energy density is not necessarily the same with and without fragmentation. Naive expectation is that the energy density will be diluted slightly since the modes that are enhanced exponentially are mildly relativistic right after the fragmentation. Therefore it is natural to ask whether this effect is significant or not.

 \begin{figure}[tbp]
   \centering
    \includegraphics[width=\textwidth]{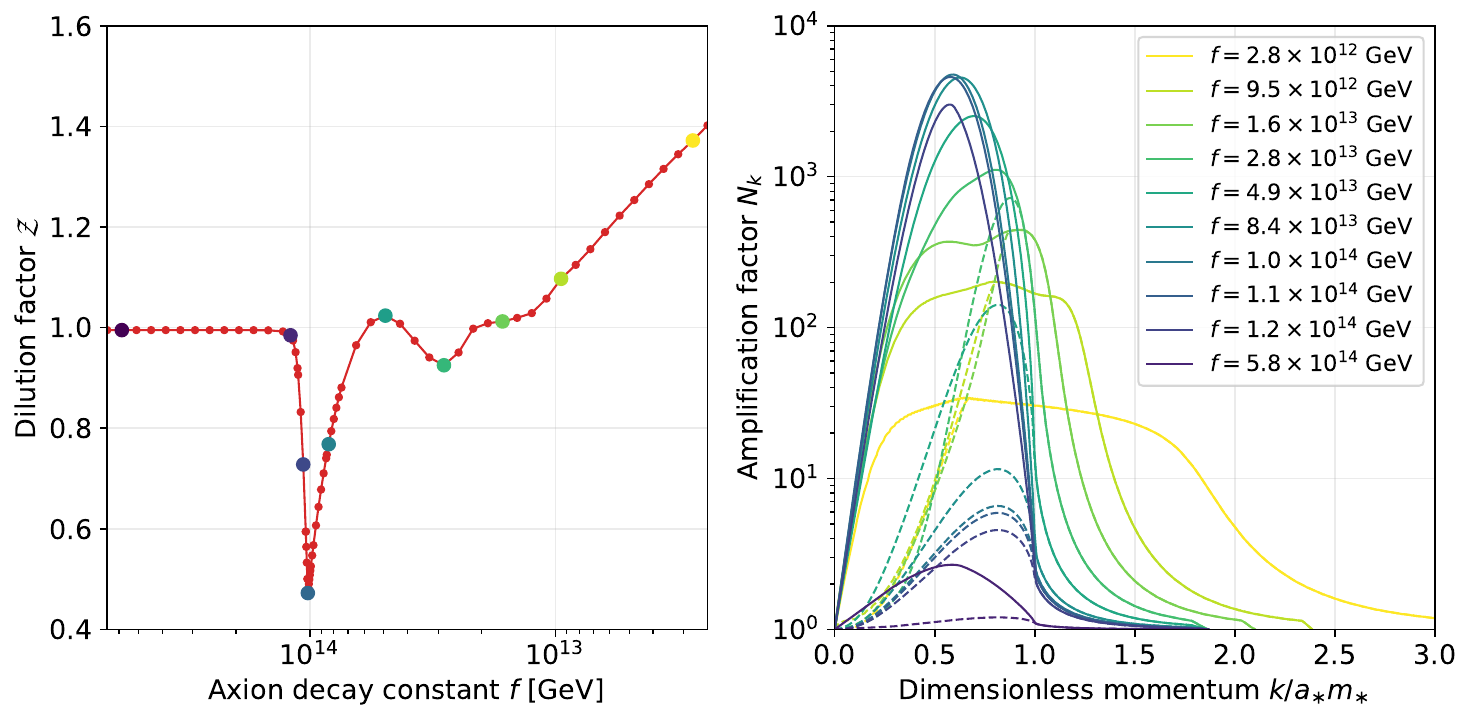}
   \caption{\it \small {\bf Left:}
  Dilution factor  (\ref{eq:169}) as a function of the axion decay constant for an ALP with constant mass $m=10^{-15}\,\si{\electronvolt}$. {\bf Right:} Amplification factor $N_k$ (\ref{eq:64}) as a function of the dimensionless momentum $k/a_{\trap}m_{\trap}$. The solid lines show the spectrum at the end of fragmentation, while the thin dashed lines together with the shaded regions show the spectrum when the ALP is trapped. The line colors match with the colored dots on the left plot.}
   \label{fig:dilution}
 \end{figure}

 At late times, the ALP energy density is given by
 \begin{equation}
   \label{eq:168}
   \rho_{\theta}=\overline{\rho}_{\theta} + \frac{f^2}{2}\int \frac{\dd[3]{k}}{(2\pi)^3}\qty[\dot{\theta}_k^2+\qty(\frac{k^2}{a^2}+m^2_{\present})\theta_k^2].
 \end{equation}
where $\overline{\rho}_{\theta}$ is the energy density remaining in the homogeneous mode which is negligible in the case of complete fragmentation. Let $\rho_{\Theta}$ be the energy density without fragmentation. We define the dilution factor as
 \begin{equation}
   \label{eq:169}
   \mathcal{Z}\equiv \frac{\rho_{\theta}}{\rho_{\Theta}}.
 \end{equation}
 which will be relevant for the section on the gravitational-wave signal. 
 Unfortunately, it is not possible to calculate this factor precisely  without a proper lattice simulation. In the non-linear regime after the fragmentation, the self-interactions between the enhanced momentum modes can modify the momentum spectrum of the fluctuations which can cause $\mathcal{O}(1)$ modifications in the dilution factor. We leave the careful study of this dilution factor with lattice simulations for future work.
 



 \begin{figure}[tbp]
   \centering
   \includegraphics[width=\textwidth]{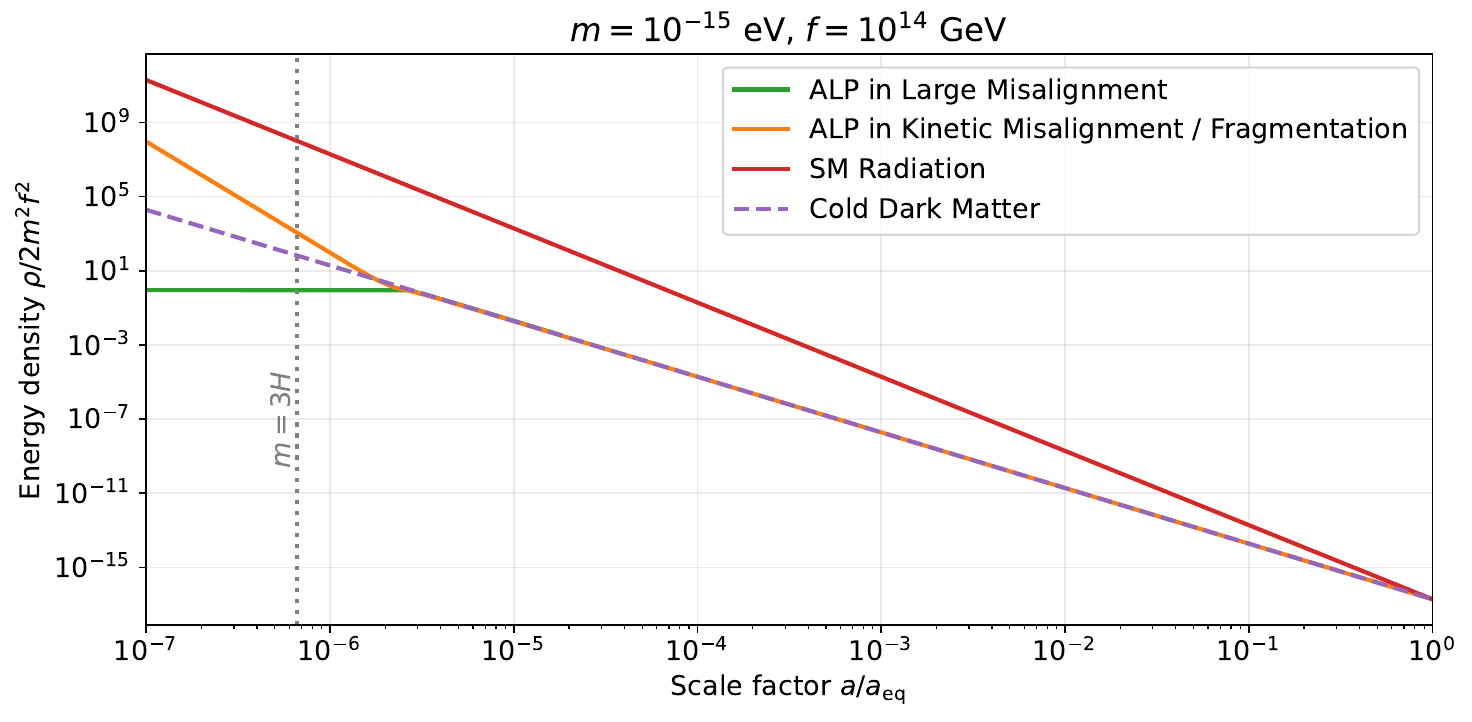}
   \caption{\it \small Evolution of the energy density of the axion in the large misalignment case (where the field is initially frozen and then redshifts as matter once oscillations turn on) compared to the kinetic misalignment/fragmentation case ($a^{-6}$ redshifting preceeding the matter era, with almost no time for a radiation-like $a^{-4}$ behaviour in between). }
   \label{fig:EoS}
 \end{figure}

 At least we show in  Fig.~\ref{fig:dilution}  our estimate for $\mathcal{Z}$. Here, we assume an ALP  with constant mass, fixed it to be $m=10^{-15}\,\si{\electronvolt}$, and show the behavior of the dilution factor as a function of the decay constant $\decay$ on the left plot. For large decay constants, the fragmentation is inefficient, therefore we have $\mathcal{Z}\approx 1$. As the fragmentation becomes more efficient, i.e. for smaller decay constants, more and more energy is transferred to the modes that are mildly relativistic after the fragmentation. As a result, the dilution factor decreases confirming our expectations.
 However, we observe the opposite trend at even lower decay constants where our calculation predicts dilution factors larger than one, meaning that the relic density is \emph{enhanced} by the fragmentation. This might be seen counter-intuitive at first, but it is not. For small decay constants, most of the fragmentation occurs before trapping as we can see from the right plot of  Fig.~\ref{fig:dilution}. Recall that the energy density of the homogeneous mode scales as $\rho_{\Theta}\propto a^{-6}$ in this regime. During fragmentation, the energy density is transferred to the fluctuations that redshift slower compared to $a^{-6}$. This yields to an enhancement in the relic density.
 Even though the behavior of the dilution factor can be understood for large and small decay constants, we cannot derive conclusions about the behavior in between with our simplified calculation.  The self-interactions between the excited modes can modify the momentum distribution and therefore the dilution factor. This effect cannot be captured by our approximation and require a lattice calculation.
 
We also show the evolution of the energy density of the axion in figure \ref{fig:EoS}.
This clearly shows that there is almost no time for a radiation-like equation of state. The produced axions are not much relativistic when they are produced and very quickly cool down so we transit quickly from a kination-like equation of state to a matter-like equation of state for the axion.

\subsection{Other initial conditions }
\label{sec:other-init-cond}

Before concluding this section, we briefly comment on the modifications of our results, if we had taken different initial conditions for the mode functions. In particular, we consider the case where the mode functions are in the Bunch-Davies vacuum initially. This implies that
\begin{equation}
  \label{eq:170}
  \eval{\delta\phi_k}_{\ini}=f\eval{\theta_k}_{\ini}=\frac{e^{-i(k/a_{\ini})t}}{a_{\ini}\sqrt{2k}} \quad\Rightarrow\quad \dimps(k)=\abs{\theta_k}^2=\frac{1}{2k a_{\ini}^2f^2}.
\end{equation}
This is the only modification we need to make. Then we can rederive all the results we obtained in this section by using this power spectrum instead of \eqref{eq:154}. We give a summary of modifications by skipping the details of the calculation.

The efficiency factor $\Delta$ with which we have determined the boundary between the regions 3 and 4 takes the form
\begin{equation}
  \label{eq:171}
  \Delta=\frac{1}{64\pi}\qty(\frac{m_{\trap}}{\decay})^2\int\dd{\kappa}_{\trap}\kappa_{\trap}^2\exp(\frac{m_{\trap}}{H_{\trap}}\mathcal{B}_{\kappa}^{\infty}),\qq{Bunch-Davies}
\end{equation}
Note that the factor in front of the integral is much more suppressed compared to factor with adiabatic initial conditions \eqref{eq:145}. Therefore larger $m_{\trap}/H_{\trap}$ values are needed for complete fragmentation. We find that $\Delta$ exceeds unity when $m_{\trap}/H_{\trap}\sim \mathcal{O}(1)\times 10^2$ where exact values show larger variations across the parameter space compared to the adiabatic case. For the QCD axion, the critical decay constant is
\begin{equation}
  \label{eq:172}
  \decay_{\rm crit}^{2\text{ -- }3}\approx 2.1\times 10^{9}\;\si{\giga\electronvolt},\qq{QCD Axion, Bunch-Davies vacuum},
\end{equation}
which is almost an order of magnitude larger than the critical value in the adiabatic case \eqref{eq:146}.

In the $\eps\gg 1$ regime, the fragmentation condition
\begin{equation}
  \label{eq:173}
  H(T)\lesssim \frac{\pi m^4(T)}{4\dot{\Theta}^3(T)}
\end{equation}
is not modified since it is independent of the initial conditions. The condition for strong back-reaction does have a similar form
\begin{equation}
  \label{eq:174}
  H(T)\lesssim \frac{\pi m^4(T)}{4\dot{\Theta}^3(T)}\frac{1}{\ln \alpha^{-1}_{\rm BD}(T)},
\end{equation}
except the expression for $\alpha(T)$ is modified to
\begin{equation}
  \label{eq:175}
  \alpha_{\rm BD}(T)=\frac{\dot{\Theta}^2(T)}{128\pi^2\decay^2}.
\end{equation}

\newpage
\section{Constraints on the ALP parameter space due to fragmentation}
\label{sec:constr-alp-param}

We need to impose a number of constraints to ensure that the fragmentation process does not spoil existing cosmological observations. These are the following\footnote{Note that axion fragmentation does not generate a domain wall problem~\cite{Morgante:2021bks}.}:

\subsection{Structure formation}
If all of the dark matter is made up of fragmented axions, they will need to be sufficiently cold at matter-radiation equality in order to be consistent with structure formation. This puts a constraint on the axion velocity $v_{\eq}$ at matter-radiation equality $a_{\eq}$. In Section \ref{sec:analyt-theory-param}, we learned that the momentum modes which grow most efficiently due to the parametric resonance are those with $k\sim a_{\trap}m_{\trap}$. Then we can estimate the axion velocity at matter-radiation equality as
  \begin{equation}
    \label{eq:166}
    v_{\eq}\sim \frac{k/a_{\eq}}{m_{\present}}\sim \qty(\frac{a_{\trap}}{a_{\eq}})\qty(\frac{m_{\trap}}{m_{\present}}).
  \end{equation}
  To get precise constraints from structure formation would require involved numerical simulations,  which is beyond the scope of this work. Instead, we will use the bound $v_{\eq}<10^{-3}$ which is commonly considered in the literature \cite{Berges:2019dgr}. So we demand
  \begin{equation}
    \label{eq:167}
    \qty(\frac{a_{\trap}}{a_{\eq}})\qty(\frac{m_{\trap}}{m_{\present}})\lesssim 10^{-3}.
  \end{equation}

\subsection{Big Bang Nucleosynthesis (BBN)}
The presence of an additional energy component at the time of BBN will modify the expansion rate and therefore be subject to constraints from observations of primordial helium-4 and deuterium abundances. Conventionally, the energy densities of new, dark and relativistic particle species are recast in terms of a neutrino density to be constrained through the effective number of neutrino species $N_{\rm eff}$. However, this recast is only possible when the new energy density redshifts like radiation. Because our axions can either behave as cold dark matter, radiation or be in a phase with a kination-like scaling, $N_{\rm eff}$ does not provide a convenient language to cast our BBN constraints in. Instead, we work directly in terms of energy densities, and run a full numerical BBN calculation in the presence of such an additional dark density component. Such a calculation can be done with the built-in routines of the numerical code \texttt{AlterBBN 2.2} \cite{Arbey:2018zfh}.

The routine \texttt{alter\_standmod} in \texttt{AlterBBN 2.2} permits the addition of an additional energy density component of the form
\begin{equation}
\label{eq:277}
\rho_D(T)=\zeta \rho_{\gamma}(T_{\rm BBN})\qty(\frac{T}{T_{\rm BBN}})^n,
\end{equation}
where $\zeta=\rho_D(T_{\rm BBN})/\rho_{\gamma}(T_{\rm BBN})$ is the ratio of the additional energy density to the photon density as measured at the reference temperature $T_{\rm BBN}=1\,\si{\mega\electronvolt}$ and $n$ is the scaling exponent such that $\rho_D\propto a^{-n}$. Expressed in terms of the total radiation density the parameter $\zeta$ is
\begin{equation}
\label{eq:280}
\zeta=\eval{\frac{\rho_{\theta}}{\rho_{\gamma}}}_{\si{\mega\electronvolt}}=\frac{g_{\rho}(T_{\rm BBN})}{2}\eval{\frac{\rho_{\theta}}{\rho_{\rm rad}}}_{\si{\mega\electronvolt}}\approx 5.29\eval{\frac{\rho_{\theta}}{\rho_{\rm rad}}}_{\si{\mega\electronvolt}}.
\end{equation}
If we assume that the ALP energy density scales cleanly with either $n=3$, $4$ or $6$ during BBN, we can then use this routine to calculate the produced helium-4 and deuterium abundances. These abundances are then compared to the most current PDG constraints \cite{Zyla:2020zbs}, which at $1\sigma$ uncertainty are
\begin{align}
\label{eq:281}
Y_p&=0.245\pm 0.003,\\
\eval{D/H}_p&=\qty(2.547\pm 0.025)\times 10^{-5},
\end{align}
where $Y_p$ and $D/H\vert_p$ are the helium-4 and deuterium fractions respectively. Using these constraints with the \texttt{AlterBBN} routine \texttt{alter\_standmod} we obtain $2\sigma$ contstraints on ALP densities. Depending on the scaling of the axion relic at $T\sim 1\,\si{\mega\electronvolt}$, the constraints are the following:
\begin{itemize}
	\item $\rho_{\theta}\propto a^{-3}$: Cold dark matter is not constrainted by BBN.
	\item $\rho_{\theta}\propto a^{-4}$: Constrained by BBN if $\rho_{\theta}/\rho_{\rm rad}\vert_{1\,\si{\mega\electronvolt}}\gtrsim 3.2\times 10^{-2}$.
	\item $\rho_{\theta}\propto a^{-6}$: Constrained by BBN if $\rho_{\theta}/\rho_{\rm rad}\vert_{1\,\si{\mega\electronvolt}}\gtrsim 1.9\times 10^{-1}$.
\end{itemize}
Only the last constraint is relevant for us. This corresponds to a lower bound of 
\begin{equation}
T_* \gtrsim 20 \mbox{ keV}.
\end{equation}
By using \eqref{eq:144} and assuming that the field is still rolling at $T_{\rm BBN}$ the last condition above becomes equivalent to
\begin{equation}
\label{eq:278}
\frac{4\pi^2}{135}\qty(\frac{g_s(T_{\trap})}{g_s(T_{\rm BBN})})^{1/3}g_s(T_{\rm BBN})\qty(\frac{Y T_{\rm BBN}}{f})^2\lesssim 0.19.
\end{equation}
Expressing the yield in terms of the zero-temperature axion mass by using \eqref{eq:21} we find that the BBN constraint implies a bound on the zero-temperature barrier height:
\begin{equation}
\boxed{
\label{eq:279}
\Lambda_{b,0}\gtrsim 9\times 10^{-7}\,\si{\giga\electronvolt}\times \qty(\frac{h^2\Omega_{\Theta,0}}{h^2 \Omega_{\rm DM}})^{1/2}.
}
\end{equation}
This result is independent of all other model parameters including the temperature-dependence of the axion mass. We observe that this bound is always stronger than constraint from structure formation.

\begin{figure}[tbp]
	\centering
	\includegraphics[width=\textwidth]{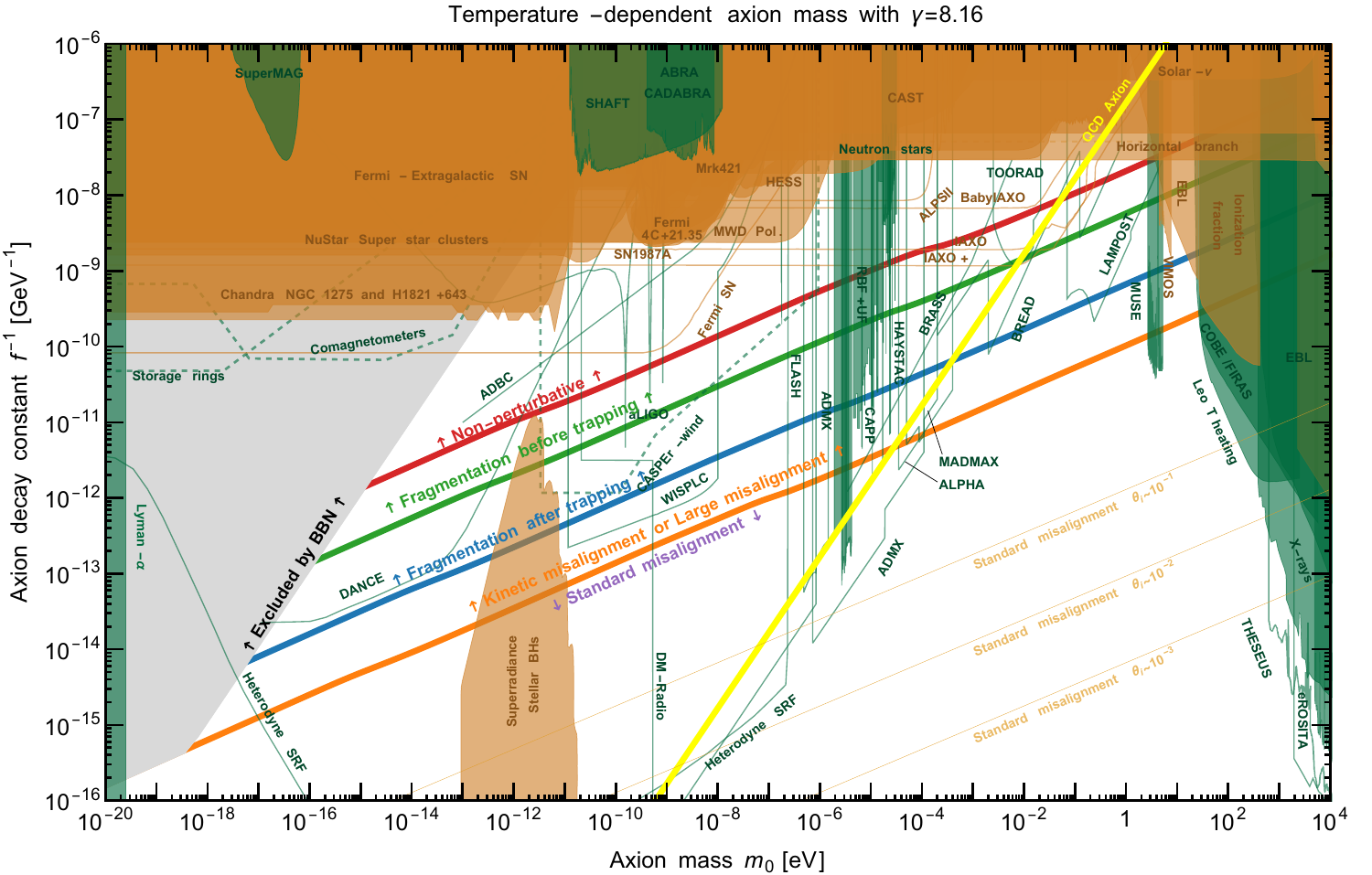}
	\caption{\it \small Parameter space for ALP dark matter. Every point in the white region can have the correct relic abundance to explain DM.  We assume a temperature-dependent axion mass with $\gamma=8$, see \eqref{eq:25} for definition, consistent with the QCD axion. The orange line separates the Standard and Kinetic Misalignment regions. The region above the orange line can also be reached without the initial kinetic energy, but by choosing the initial angle very close to the top of the potential, i.e. Large Misalignment Mechanism. Above the blue line, the fragmentation is efficient enough so that all the energy density is transferred from the homogeneous mode to the fluctuations. Above the green line, the fragmentation becomes efficient before the ALP gets trapped by the potential; see Section \ref{sec:overv-fragm-regi} for the precise definitions. Above the red line, the variance of the ALP angle becomes larger than unity before the onset of fragmentation, so our calculation cannot be trusted. In the gray region, the ALP field is rolling during BBN with a large enough kinetic energy so that it spoils the BBN predictions, see Eq.~\eqref{eq:279}. Thin lines correspond to experimental projections. The bounds/projections on the axion-photon coupling are translated into bounds on the  axion decay constant by assuming a KSVZ-like coupling given in \eqref{eq:282}.   Orange constraints apply to any ALP while the green ones assume the ALP is DM.}
	\label{fig:money-plot-g8}
\end{figure}

\begin{figure}[tbp]
	\centering
	\includegraphics[width=\textwidth]{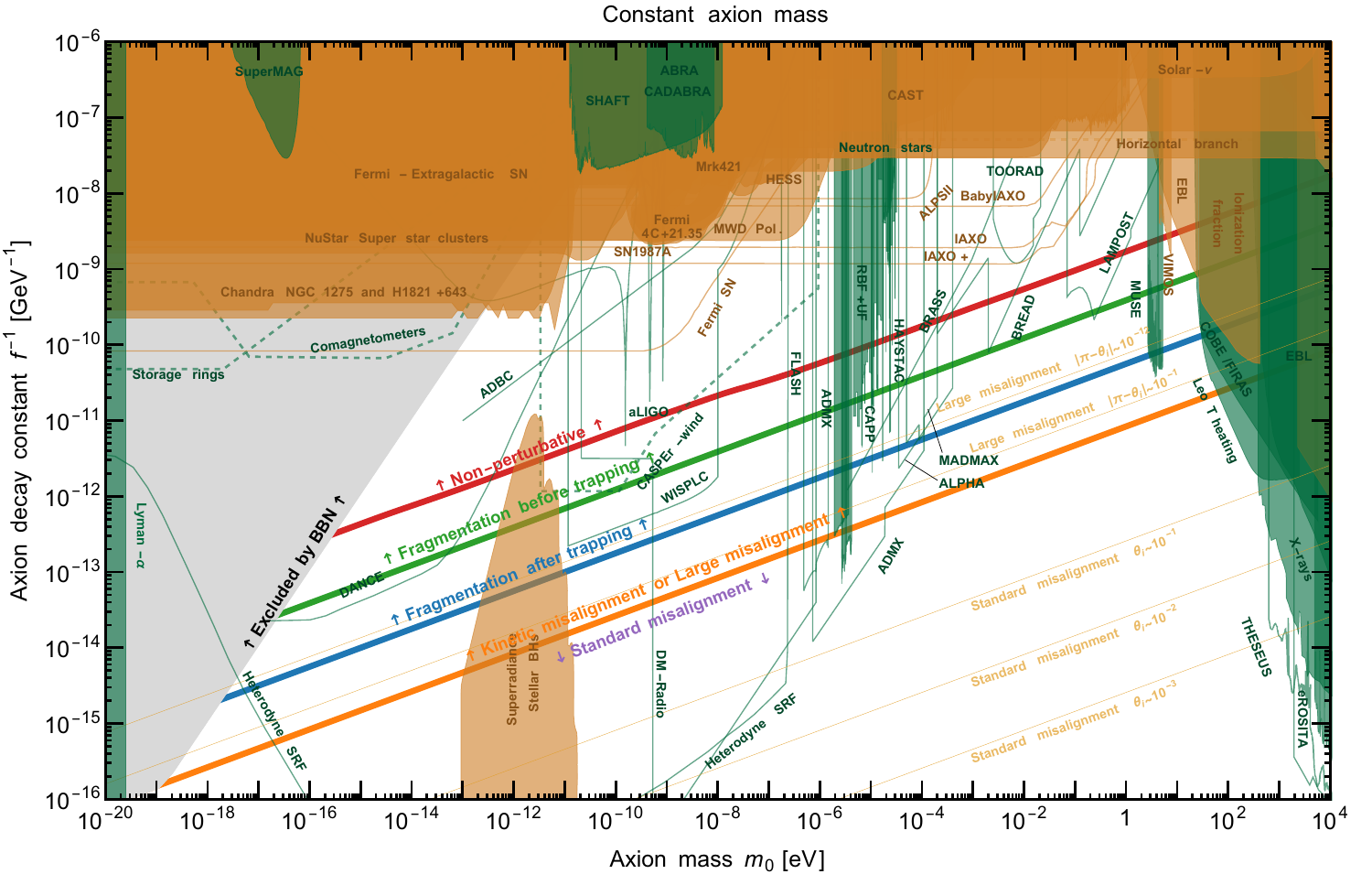}
	\caption{\it \small Same as Figure \ref{fig:money-plot-g8} but now by assuming a constant axion mass.}
	\label{fig:money-plot-g0}
\end{figure}

\subsection{ALP parameter space with existing and future experiments}
We now show the fragmentation regions of Section \ref{sec:overv-fragm-regi}, and the model constraints of Section \ref{sec:constr-alp-param} on the ALP parameter space together with a long list of experiments constraints/projections of diverse nature in Figures \ref{fig:money-plot-g8} and \ref{fig:money-plot-g0}. Current constraints are included as filled regions while projections are distinguished by having outlines only. In order to translate the bounds/projections on the axion-photon coupling $g_{\theta\gamma\gamma}$ to the bounds/projections on the axion decay constant $\decay$, we assumed a KSVZ-like coupling:
\begin{equation}
  \label{eq:282}
  g_{\theta\gamma\gamma}^{\rm KSVZ}=\frac{\alpha_{\rm EM}}{2\pi}\frac{1.92}{\decay}\approx \frac{2.23\times 10^{-3}}{f}.
\end{equation}
Figures \ref{fig:money-plot-g8} and \ref{fig:money-plot-g0} also include projections for constraints on the ALP-neutron coupling, which are indicated with dashed lines. These assume a KSVZ-like coupling of
\begin{gather}
C_{\theta n} \approx -0.02,
\end{gather}
where the ALP-neutron coupling $ C_{\theta n} $ is defined from
\begin{gather}
\mathcal{L}\supset C_{\theta n}\frac{\partial_\mu \theta}{2f_a} \bar{n}\gamma^{\mu}\gamma_5 n.
\end{gather}
For other ALP couplings, the constraints and the projections of all the experiments need to be adjusted with the exception of superradiance bounds which depend on the axion mass and the axion decay constant directly. The references for all the bounds/projections can be found in Appendix \ref{sec:experimental-surveys} and the majority of the digitized bounds are sourced from the \verb|AxionLimits| repository~\cite{AxionLimits}.

The Kinetic Misalignment Mechanism (KMM) is not the only method to get ALP dark matter with lower decay constants. Another possibility is the Large Misalignment Mechanism (LMM) where one chooses an initial angle very close to the top \cite{Zhang:2017dpp,Arvanitaki:2019rax}. The distinction between the Standard and Large misalignment is arbitrary. In this work, we will define the LMM region as the region above the orange line in Figures \ref{fig:money-plot-g8} and \ref{fig:money-plot-g0} without the initial kinetic energy. 

\section{Gravitational waves}
\label{sec:gravitational-waves}

In the case of efficient parametric resonance, the ALP field at the end of fragmentation can have significant anisotropic stress, and this stress can source gravitational waves (GW). A precise study of this process is not the topic of this paper, and will be studied in a future work. Here, we will present a very rough estimate based on the method introduced in \cite{Chatrchyan:2020pzh}. A more detailed analysis can be found in \cite{Madge:2021abk}.

\begin{figure}[tbp]
  \centering
  \includegraphics[width=\textwidth]{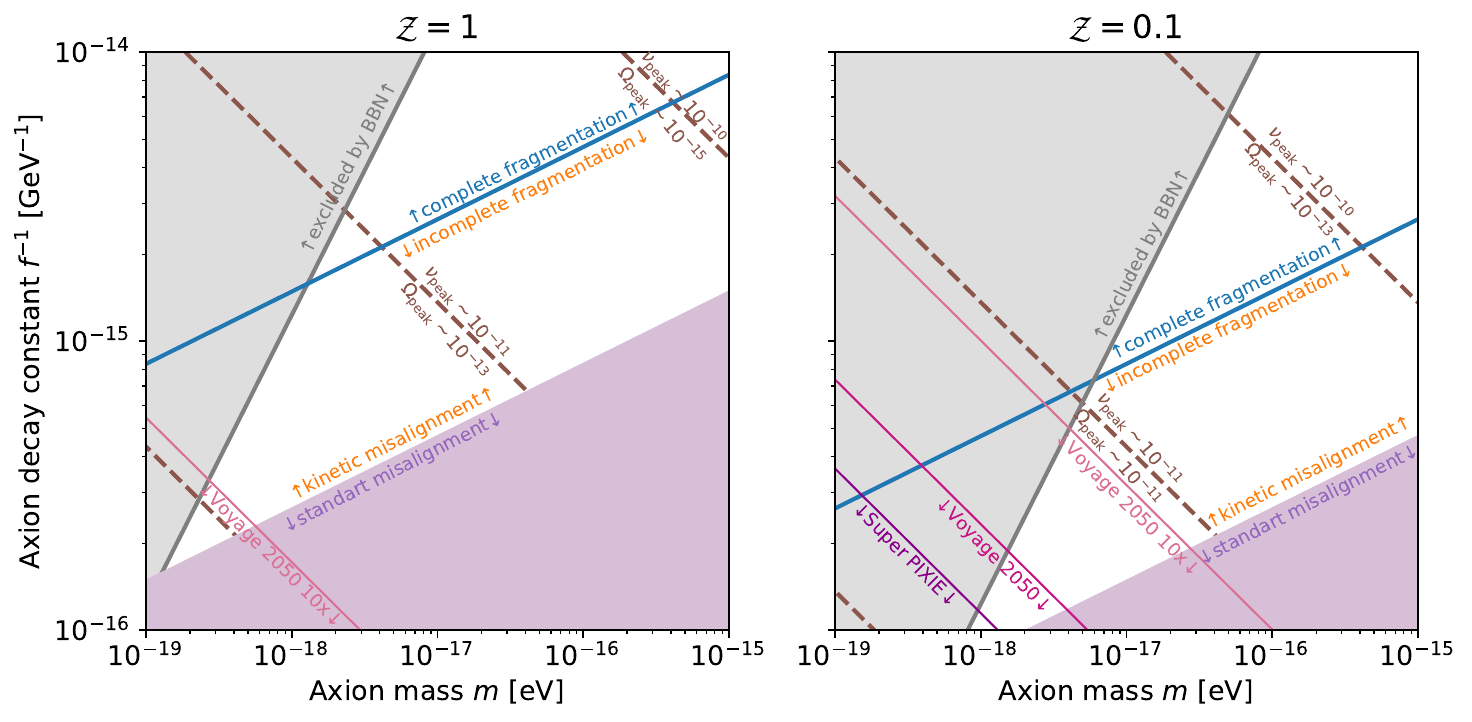}
  \caption{\it \small The peak frequency (Eq.~\ref{eq:369}) and peak amplitude (Eq.~\ref{eq:376}) contours in the ALP parameter space, and the regions that can be probed by gravitational waves (GW) induced by the fragmentation process by measuring the $\mu$-distortions in the CMB \cite{Kite:2020uix}. On the left plot we assume that the fragmentation does not dilute the relic density, while the right plot assumes a factor of $10$ dilution. Above the blue lines fragmentation is complete, and it is likely that the efficient GW production only happens in this region. The gray regions are excluded by the BBN constraints studied in Section \ref{sec:constr-alp-param}. GW prospects of kinetic fragmentation are thus severely constrained by BBN.}
  \label{fig:gravitational-waves}
\end{figure}

We start by discussing the frequency of the gravitational waves. Today's frequency is related to the comoving momentum $k$ by
\begin{equation}
  \label{eq:366}
  \nu=\frac{1}{2\pi}\frac{k}{a_{\present}}=\frac{1}{2\pi}\frac{k}{a_{\rm emit}}\frac{a_{\rm emit}}{a_{\present}},
\end{equation}
where $a_{\rm emit}$ is the scale factor at which the GW is emitted. By assuming that emission happens around trapping $a_{\rm emit}\approx a_{\trap}$ we get
\begin{equation}
  \label{eq:367}
  \nu\approx \frac{m_{\trap}}{2\pi}\kappa_{\trap}\frac{a_{\trap}}{a_0}.
\end{equation}
We can calculate $a_{\trap}/a_{\present}$ by solving \eqref{eq:24}. In this equation, $\rho_{\Theta,0}$ is the ALP energy density today assuming no fragmentation. The non-linear effects after the fragmentation can dilute the relic density by a factor of $\mathcal{O}(1\text{--}10)$ \cite{Berges:2019dgr}. Let $\mathcal{Z}$ be this suppression factor so that $\rho_{\theta,0}=\mathcal{Z}\rho_{\Theta,0}$ is the correct ALP energy density with fragmentation. By requiring that $\rho_{\theta,0}$ should match the dark matter density we get
\begin{equation}
  \label{eq:368}
  \frac{a_{\trap}}{a_{\present}}=\qty(\frac{3\pi}{8}\frac{\Omega_{\rm DM}}{\mathcal{Z}}\frac{\mpl^2H_0^2}{m_0m_{\trap}f^2})^{1/3}.
\end{equation}
We are interested to know the frequency at which the GW amplitude is peaked. Since most of the fragmented modes have the momenta $\kappa_{\trap}\sim 1$, it is likely that the GW spectrum will also be peaked at those momenta. Setting $\kappa_{\trap}\sim 1$ in \eqref{eq:367}, and using \eqref{eq:368} we obtain the peak frequency in terms of model parameters as
\begin{equation}
  \label{eq:369}
  \nu_{\rm peak}\sim 8\times 10^{-11}\,\si{\hertz}\qty(\frac{m_{\trap}}{m_0})^{2/3}\qty(\frac{m_0}{10^{-16}\,\si{\electronvolt}})^{1/3}\qty(\frac{f}{10^{14}\,\si{\giga\electronvolt}})^{-2/3}\mathcal{Z}^{-1/3}.
\end{equation}
Next we estimate the peak gravitational wave amplitude $\Omega_{\rm GW}(\nu)$. It is defined as the fraction of energy density in gravitational waves per logarithmic frequency:
\begin{equation}
  \label{eq:370}
  \frac{\rho_{\rm GW}}{\rho_{\rm crit}}=\int \dd{\ln \nu}\Omega_{\rm GW}(\nu).
\end{equation}
After the emission, the energy density in gravitational waves dilutes as radiation. Today's GW amplitude is given in terms of the amplitude at emission as
\begin{equation}
  \label{eq:371}
  \Omega_{\rm GW,0}=\frac{\rho_{\rm GW,0}}{\rho_{\rm crit,0}}=\frac{\rho_{\rm GW,emit}}{\rho_{\rm crit,emit}}\frac{\rho_{\rm GW,0}}{\rho_{\rm GW,emit}}\frac{\rho_{\rm crit,emit}}{\rho_{\rm crit,0}}=\Omega_{\rm GW,emit}\qty(\frac{a_{\rm emit}}{a_0})^4 \qty(\frac{H_{\rm emit}}{H_0})^2.
\end{equation}
Let $k_{\rm peak}$ be the comoving momentum corresponding to the peak frequency $\nu_{\rm peak}$. Ref.~\cite{Chatrchyan:2020pzh} gives the following estimate for the peak amplitude at emission:
\begin{equation}
  \label{eq:372}
  \Omega_{\rm GW,emit}^{\rm peak}\sim \frac{64\pi^2}{3\mpl^4H_{\rm emit}^2}\frac{\rho_{\theta,\rm emit}^2}{\qty(k_{\rm peak}/a_{\rm emit})^2}\frac{\alpha^2}{\beta},
\end{equation}
where $\alpha\lesssim 1$ roughly measures the fraction of the energy stored in the fluctuations, and $\beta\gtrsim 1$ is the typical logarithmic width of the spectrum of fluctuations in momentum space. We set both of them to unity for our estimates. Again we assume that the GW emission takes place at trapping. 
\begin{figure}[tbp]
  \centering
  \includegraphics[width=\textwidth]{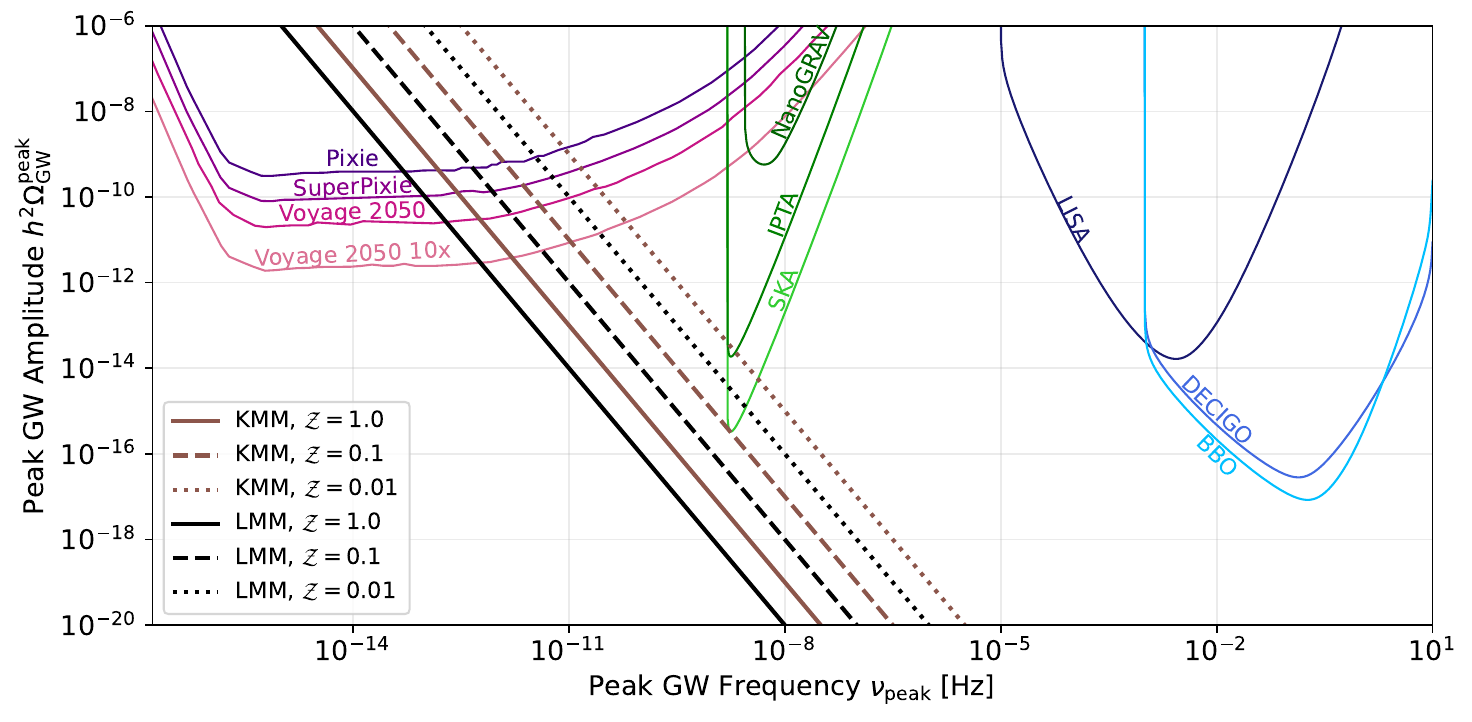}
  \caption{\it \small Peak amplitude and frequency of gravitational waves induced by axion fragmentation occurring in kinetic misalignment mechanism  (KMM) versus large misalignment mechanism (LMM) according to (Eq.~\ref{eq:376}). The lines assume constant axion mass  and should not be understood as GW spectra. The predictions are compared to the sensitivity of future experiments. The expression of $\nu_{\rm peak}$ in terms of the axion mass and decay constant is given by Eq.~\ref{eq:369}, see contours in Fig.~\ref{fig:gravitational-waves} and \ref{fig:gw-estimate-lmm}. }
  \label{fig:gravitational-waves-spectra}
\end{figure}
Then, the energy density of the ALP field at emission is
\begin{equation}
  \label{eq:373}
  \rho_{\theta,\rm emit}\approx 2 m_{\trap}^2f^2.
\end{equation}
Also, the peak momentum becomes $k_{\rm peak}=\kappa_{\trap}a_{\trap}m_{\trap}\sim a_{\trap}m_{\trap}$. Then \eqref{eq:372} is simplified to
\begin{equation}
  \label{eq:374}
  \Omega_{\rm GW,\ast}^{\rm peak}\sim \frac{256\pi^2}{3}\qty(\frac{m_{\trap}}{H_{\trap}})^2\qty(\frac{f}{\mpl})^4.
\end{equation}
Evolving this amplitude until today by using \eqref{eq:371} and \eqref{eq:368} we obtain
\begin{equation}
  \label{eq:375}
  \Omega_{\rm GW,0}^{\rm peak}\sim 1.5\times 10^{-15}\qty(\frac{m_{\trap}}{m_0})^{2/3}\qty(\frac{m_0}{10^{-16}\,\si{\electronvolt}})^{-2/3}\qty(\frac{f}{10^{14}\,\si{\giga\electronvolt}})^{4/3}\mathcal{Z}^{-4/3}.
\end{equation}
By combining this result with \eqref{eq:369} we can obtain a simple relation between the peak frequency and the peak amplitude:
\begin{equation}
  \label{eq:376}
  \Omega_{\rm GW,0}^{\rm peak}\sim 10^{-35}\qty(\frac{m_{\trap}/m_0}{\qty(\nu_{\rm peak}/\si{\hertz})\mathcal{Z}})^2.
\end{equation}
From this, we learn  that ALP models with a constant mass have  better prospects for an observable gravitational-wave signal. Secondly, the models with a lower peak frequency predict a larger GW amplitude. Finally, if there is an additional dilution in the energy density due to the fragmentation, the gravitational-wave amplitude is also enhanced. We show the contours of the peak GW frequency calculated via \eqref{eq:369}, and the peak GW amplitude calculated via \eqref{eq:376} in the upper plots of Figure \ref{fig:gravitational-waves}.

A similar estimate has also been obtained in \cite{Madge:2021abk} by a different method, but their estimate is lower than ours by a factor of $\sim 20$. This reference also presents a numerical calculation of the GW spectrum at the linearized level, and confirms that their estimate predicts the peak amplitude quite well. However, the GW amplitude can be enhanced significantly during the non-linear phase which cannot be captured by the linearized analysis \cite{Dufaux:2007pt}. Therefore, one can interpret our estimate and the estimate of \cite{Madge:2021abk} as optimistic and conservative respectively.

The power-law-integrated-sensitivity for the SKA mission \cite{Janssen:2014dka} can reach up to $h^2\Omega_{\rm GW}\sim 3\times 10^{-16}$ at frequency $\nu\sim 2\times 10^{-9}\,\si{\hertz}$ assuming a signal-to-noise of $1$ and an observation time of $20$ years \cite{Schmitz:2020syl}. So even considering a constant mass and assuming a suppression of factor of $10$, i.e. $\mathcal{Z}=0.1$, our optimistic estimate tells us that the signal is barely observable. Recently, \cite{Kite:2020uix} did point out that the gravitational waves with frequencies much smaller than the ones probed by SKA might be observable by measuring the $\mu$-distortions in the CMB via the experiments such as COBE/FIRAS \cite{1994ApJ...420..439M,Fixsen:1996nj}, PIXIE \cite{2011JCAP...07..025K}, SuperPIXIE \cite{Kogut2019CMB}, and Voyage 2050 \cite{Chluba:2019kpb}. The COBE/FIRAS experiment sets an upper limit on the $\mu$-distortions ($\mu< 9\times 10^{-5}\,95$\%CL), while the forecasted constraints for the experiments are $\mu<3\times 10^{-8}$ for PIXIE, $\mu<7.7\times 10^{-9}$ for SuperPIXIE, and $\mu<1.9\times 10^{-9}$ for Voyage 2050.

These upper bounds on the observable $\mu$-distortions can be translated into lower bounds on the observable gravitational wave signal for a given frequency \cite{Kite:2020uix}. By using these bounds, we can estimate the potentially observable regions in the ALP parameter space which we show in Figure \ref{fig:gravitational-waves}. 
The curve labeled ``Voyage 2050 10x'' assumes an upper limit of $\mu<1.9\times 10^{-10}$, see \cite{Kite:2020uix} for details. We also show the critical lines separating complete and incomplete fragmentation (blue line), as well as the region which is excluded by the BBN constrains (gray region) discussed in Section \ref{sec:constr-alp-param}. Likely, the efficient GW production does happen only in the region where the fragmentation is complete. We conclude that the BBN bound severely constrains the parameter space which can be observable by GWs. The prospects do improve if the fragmentation efficiently dilutes the relic density of ALPs, since this will increase the energy budget for the GWs without overclosing the universe. Precise estimations of the dilution factor, peak GW amplitude and the GW frequency require a dedicated lattice analysis.

\begin{figure}[tbp]
  \centering
  \includegraphics[width=\textwidth]{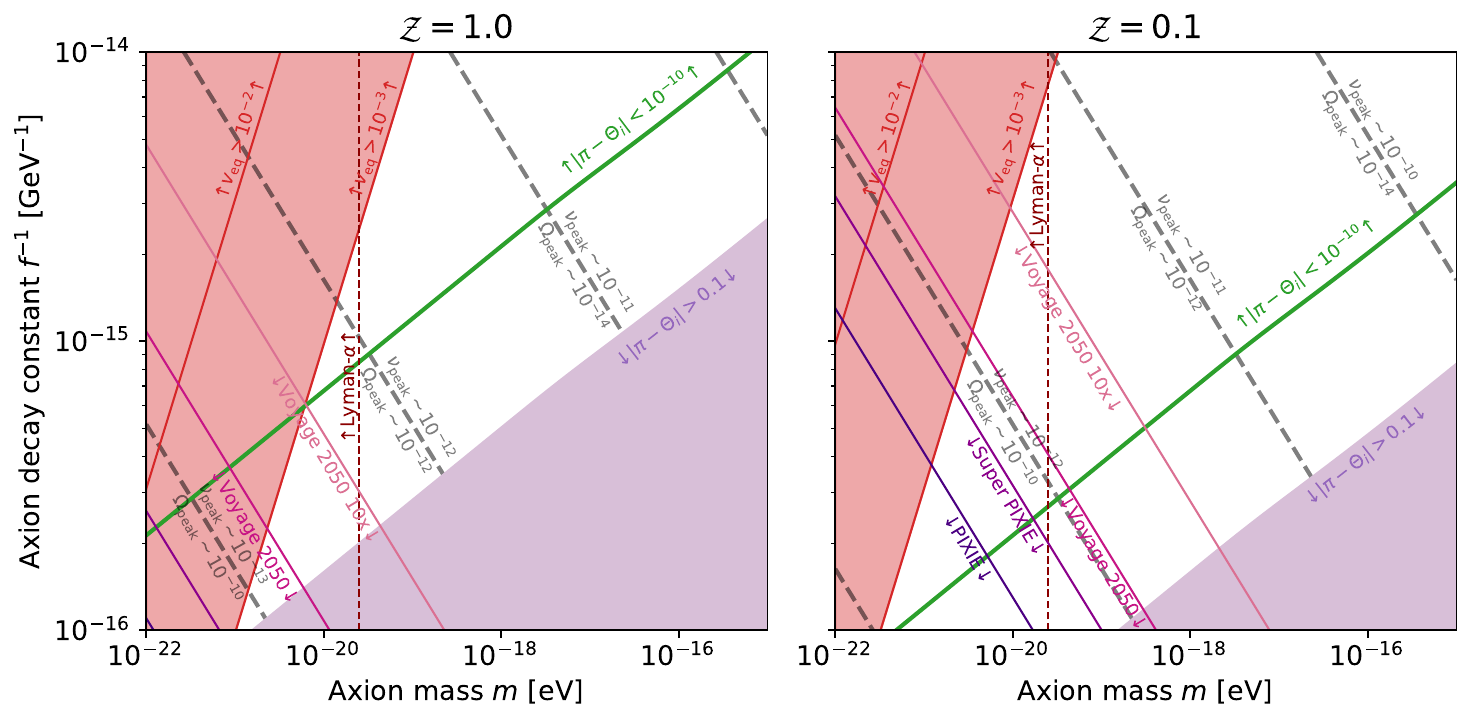}
  \caption{\it \small Analog of Figure \ref{fig:gravitational-waves} but for the case of large misalignment. Unlike the case of  kinetic misalignment  there are no BBN constraints, however the requirement that the ALPs are not too warm during the matter-radiation equality gives a similar but weaker bound. In the purple shaded region on the lower right, the initial angle is not close to the top, and the standard misalignment is at play. The green line shows the parameter space where $\abs{\pi-\Theta_i}\approx 10^{-10}$ above which the density contrast of the ALP field becomes $\mathcal{O}(1)$, and it is expected that complete fragmentation will happen above this line. It is likely that efficient GW production will happen only in this region. }
  \label{fig:gw-estimate-lmm}
\end{figure}

Since the BBN constraint is the main obstacle to get an observable gravitational wave signal, one might wonder what happens in a model which is immune to the BBN constraint. The Large Misalignment Mechanism is an example of such a model since the energy density prior to the oscillations does not redshifts as $a^{-6}$. Therefore, we did repeat the above analysis for the Large Misalignment scenario, and obtained a result which is very close to \eqref{eq:376} except an $\mathcal{O}(1)$ factor which depends very mildly on the value of $\abs{\pi-\Theta_i}$. We show the results in Figure \ref{fig:gw-estimate-lmm}. Even though the BBN constraints are absent, there is still a similar but weaker constraint that the ALPs should not be too warm at matter-radiation equality in order to be consistent with the structure formation. The green line shows approximately the critical initial angle $\abs{\pi-\Theta_i}\approx 10^{-10}$ at which the density contrast of the ALP field becomes $\mathcal{O}(1)$. Above this line, the initial angle needs to be chosen closer to the top of the potential, and we expect that complete fragmentation and the efficient GW production does happen in this region. We see that the absence of the BBN constraint opens up a sizable region which might be probed by future CMB surveys.

The ALP masses below $\sim 10^{-20}\,\si{\electronvolt}$ are constrained by various probes such as Lyman-$\alpha$ observations \cite{Rogers:2020ltq}, and the galactic rotation curves \cite{Bar:2018acw,Bar:2019bqz,Bar:2021kti}. However, it is possible that some of these contraints can be evaded as a consequence of the fragmentation \cite{Leong:2018opi}.

 We close this section by stressing another much stronger signal of GW that can arise in the Kinetic Misalignment Mechanism, and which is not related to fragmentation. While in this work we have assumed a 
radiation-dominated universe during the period when the equation of state of the axion scales as $a^{-6}$,  it is also possible that the axion energy density dominates temporarily, inducing a kination era inside the radiation era. Such a kination era enhances primordial signals of GW from inflation and cosmic strings leading to striking peak features that are observable by upcoming interferometers such as LISA and the Einstein Telescope \cite{Co:2021lkc,Gouttenoire:2021wzu,Gouttenoire:2021jhk}. The occurrence of a kination era in specific UV completions and the precise parameter space region  that lead to observable GWs are worked out in \cite{Gouttenoire:2021jhk} and will also be discussed in 
\cite{Eroncel:2025qlk}.

\section{Conclusion}
\label{sec:conclusion}

In summary, the correct ALP relic abundance to explain dark matter can be recovered naturally in the whole [axion mass $m$, decay constant $f$] plane. The region which is usually discarded because it leads to insufficient dark matter  becomes open in the scenario of kinetic misalignment, in which  oscillations are delayed due to the initial velocity of the axion field.
This was pointed out in \cite{Co:2019jts,Chang:2019tvx} and the underlying framework was extensively discussed in \cite{Co:2019wyp,Co:2020jtv,Co:2020xlh} in some specific UV completions.
This is an extremely good news for a whole generation of  experiments as it provides them with a strong motivation for ALP dark matter in the region of low values for the axion decay constant. It is in particular exciting that experiments such as ALPSII which is about to start running at DESY is  sensitive to dark matter. It is also quite remarkable that IAXO can probe the QCD axion as DM. 
What we have shown in this paper is that one can no longer describe this regime in terms of the homogeneous zero-mode. In fact, the axion field entirely fragments. We have provided a detailed analytical derivation of the phenomenon.

One main model-independent consequence is a distinct prediction for dense compact mini-clusters that is presented in detail in a companion paper \cite{Eroncel:2022efc}.
There is also a stochastic gravitational wave background generated by axion fragmentation as we discuss in our Section  \ref{sec:gravitational-waves}. However, this signal appears at extremely low frequency and is typically below the sensitivity of Pulsar Timing Arrays. On the other hand, it would be visible in CMB experiments such as Voyager.

Our results apply generically to any ALP, including the QCD axion.
One key question concerns  the motivations for an initial  axion velocity.
To address this, one should start with the complex scalar field from which the axion  originates as the angular direction.
There is a whole class of models where the axion receives a kick from the radial mode of the complex scalar field at very early times in the cosmological evolution, see \cite{Co:2019jts,Co:2019wyp,Co:2020jtv,Co:2020xlh,Gouttenoire:2021jhk}.
We will present the precise determination of the viable parameter space of these models for ALP DM from fragmentation  in a companion paper \cite{Eroncel:2022a}.
Note that in a significant region of parameter space, one can get a striking signal in gravitational waves  in these models from a kination era induced by the axion \cite{Co:2021lkc,Gouttenoire:2021wzu,Gouttenoire:2021jhk}. We do not discuss this in our section \ref{sec:gravitational-waves} which only deals with the GW signal  from the fragmentation effect, ignoring the potential signal from a kination era, as in this paper we always assume radiation domination at early times.  
 
The  main results of this paper, which are independent from any UV completion, are summarised in the following expressions and figures:

\begin{itemize}

\item The temperature $T_{\trap}$ when the field gets trapped by the barrier is given in  Eq.~\ref{eq:26}. Fig.~\ref{fig:trapping} shows the contours  of the trapping temperature.

\item The duration of the kination-like scaling underlying the kinetic misalignment mechanism is constrained by \eqref{eq:kination-bound}.

\item Fig.~\ref{fig:mastoverHast} shows the contours  of $m_*/3H_*$  at the time when the field gets trapped, which is orders of magnitude larger compared to the unit value that characterises the time when the field starts oscillating around its minimum in the standard misalignment mechanism. 

\item  The amplification factor of a given mode depends crucially on the value of $m_*/H_*$ as expressed in Eq.~\ref{eq:64}.  It is also plotted in Fig.~\ref{fig:dilution}.

\item Fig.~\ref{fig:money-plot-g8} and \ref{fig:money-plot-g0} show the contours of the different regimes of dark matter production for respectively the constant axion mass and temperature-dependent mass cases.  The expressions for these region boundaries are given in  (Eq.~\ref{eq:377}, \ref{eq:378}, \ref{eq:379}, \ref{eq:165}). 
The BBN constraint, $T_{\trap}> 20 $ keV,  translates into a bound on the size of the zero-temperature barrier, see (\ref{eq:279}).

\item All phenomenological implications of fragmentation and in particular the determination of the different regimes 
are controlled by the power spectrum (\ref{eq:150}) of the mode functions (\ref{eq:197}).

\item Figure  \ref{fig:dilution}-left indicates that the impact of fragmentation on the relic abundance prediction from kinetic misalignment is relatively weak. It typically differs by a factor of order 1, we do not expect  more than one order of magnitude effect. This can be understood as the momentum of the produced axions during fragmentation is of the same order as the axion mass (see Fig.~\ref{fig:dilution}-right), higher modes are not excited.
A precise estimate of the ${\cal O}(1)$ factor  requires a lattice calculation which  we will present in a future work. 

\item We estimated the GW signal in Fig.~\ref{fig:gravitational-waves-spectra} and the parameter space that can be probed this way in Fig.~\ref{fig:gravitational-waves}.
We showed that there are better prospects for GW from large misalignment in Fig.~\ref{fig:gw-estimate-lmm}, something which had not been mentioned in \cite{Arvanitaki:2019rax}.

\end{itemize}

There are many interesting questions that deserve further investigation.
Primarily, the precise predictions for the QCD axion DM definitely deserve better scrutiny, in particular using lattice calculations.
Another compelling aspect is the correlation with GW signatures  that can be derived once a specific UV completion
 is defined.
 Finally, the possibility to relate the halo spectrum properties to the parameters of the axion potential is an exciting opportunity in view of the upcoming observational prospects. 

\section*{Acknowledgements}
This work is supported by the Deutsche Forschungsgemeinschaft under Germany Excellence Strategy - EXC 2121 ``Quantum Universe'' - 390833306.

\appendix

\section{Detailed discussion of the parametric resonance}
\label{sec:deta-disc-param}

In this Appendix we discuss in detail the solutions of \eqref{eq:7}. If we neglect the expansion of the universe then \eqref{eq:7} becomes
\begin{equation}
  \label{eq:15}
  \ddot{\theta}_k+\qty(\frac{k^2}{a^2}+m^2\cos\Theta)\theta_k=0,
\end{equation}
where both the scale factor $a$ and the axion mass $m$ are constant. We define a dimensionless time $t_m\equiv mt$, and dimensionless momentum variable $\kappa\equiv k/am$ so that \eqref{eq:15} takes a simpler form:
\begin{equation}
  \label{eq:28}
  \theta_{\kappa}''+\qty(\kappa^2+\cos\Theta)\theta_{\kappa}=0,
\end{equation}
where primes denote derivatives with respect to $t_m$. The solutions of this equation have been studied in \cite{Greene:1998pb} in the case of oscillations after trapping, i.e. $\eps<1$. We review the method of getting the solutions, at the same time we generalize it to obtain the solutions for the rolling axion, $\eps>1$.

We start by defining a new time variable $z$ by
\begin{equation}
  \label{eq:29}
  z(t_m)\equiv \cos\Theta(t_m),
\end{equation}
where $\Theta$ is the solution of the homogeneous mode. Without expansion the energy density $\eps$ becomes a constant of motion. By taking the derivative of $z(t_m)$ by using the fact that $\eps$ is conserved we obtain
\begin{equation}
  \label{eq:30}
  \dv{z}{t_m}=\pm\sqrt{2(1-z^2)(2\eps - 1+z)},
\end{equation}
where the sign depends of the sign of $\sin\Theta$ and $\Theta'$. Plugging this result into \eqref{eq:28} gives
\begin{equation}
  \label{eq:31}
  2(1-z^2)(2\eps -1+z)\theta_{\kappa}''+\qty[1+2(1-2\eps)z-3z^2]\theta_{\kappa}'+\qty(\kappa^2+z)\theta_{\kappa}=0,
\end{equation}
where primes are now denoting derivatives with respect to $z$. Note that $z$ is not single-valued for the whole oscillation. Thus this equation can describe the solutions only for quarter of the period for $\eps <1$, and half of the period for $\eps>1$. However we can still use this equation to get the solutions in each patch, and then glue them together to get the full solution.

Let $\theta_{\pm}(z)$ denote the two linearly independent solutions to \eqref{eq:31}, where we have omitted the momentum subscript for cleaner notation. It turns out that the linear combinations $\theta_+^2$, $\theta_-^2$, and $\theta_+\theta_-$ obey a third order equation:
\begin{equation}
  \label{eq:32}
  2(z^2-1)(2\eps -1 +z)M'''+\qty[9z^2-6(1-2\eps)z-3]M''+2(z-1+2\eps-2\kappa^2)M'-2M=0.
\end{equation}
This equation has a polynomial solution given by
\begin{equation}
  \label{eq:33}
  M(z)=z-1+2\eps-2\kappa^2.
\end{equation}
The original equation we are trying to solve \eqref{eq:28} is a \emph{Hill differential equation} so according to the \emph{Floquet's theorem}, the solutions must be of the form
\begin{equation}
  \label{eq:34}
  \theta_{\pm}(t_m;\kappa)=\psi_{\pm}(t_m;\kappa)e^{\pm \mu_{\kappa} t_m}.
\end{equation}
Therefore the polynomial solution \eqref{eq:33} should correspond to $\theta_+\theta_-$ which implies
\begin{equation}
  \label{eq:37}
  \theta_+(z)\theta_-(z)=\mathcal{N}^2\qty(z-1+2\eps-2\kappa^2),
\end{equation}
where $\mathcal{N}$ is a normalization constant. To obtain the individual solutions we note that the Wronskian of the system $W\equiv \theta_+'\theta_--\theta_+\theta_-'$ obeys
\begin{equation}
  \label{eq:35}
  W'(z)=-\frac{1+2(1-2\eps)z-3z^2}{2(1-z^2)(2\eps -1 +z)}W(z)=-\dv{}{z}\ln\sqrt{(1-z^2)(2\eps-1+z)}W(z).
\end{equation}
This can easily be solved as
\begin{equation}
  \label{eq:36}
  W(z)=\theta_+'(z)\theta_-(z)-\theta_+(z)\theta_-(z)=\frac{c_{\kappa}\,\mathcal{N}^2}{(1-z^2)(2\eps-1+z)},
\end{equation}
where $c_{\kappa}$ is an integration constant coming from the integration of \eqref{eq:35} which we will determine shortly. By combining \eqref{eq:34} and \eqref{eq:36} we obtain the following differential equations for $\theta_{\pm}$:
\begin{equation}
  \label{eq:38}
  2M(z)\dv{\ln \theta_{\pm}}{z}=M'(z)\pm \frac{c_{\kappa}}{\sqrt{(1-z^2)(2\eps - 1+z)}}.
\end{equation}
The solutions are\footnote{Note that while $(1-z^2)(2\eps-1+z)$ is always non-negative, $M(z)$ does cross zero at $z=1-2\eps+2\kappa^2$. If this point lies in the range of the integral, then the integral is understood as its Cauchy principal value. }
\begin{equation}
  \label{eq:39}
  \theta_{\pm}(z)=\mathcal{N}\sqrt{\abs{M(z)}}\exp(\pm \frac{c_{\kappa}}{2}\int^z\frac{\dd{z'}}{M(z')\sqrt{(1-z'^2)(2\eps-1+z')}}).
\end{equation}
The integration constant $c_{\kappa}$ can be obtained by plugging this solution into \eqref{eq:31}. The result is
\begin{equation}
  \label{eq:40}
  c_{\kappa}^2=8\kappa^2(\eps-\kappa^2)(1-\eps+\kappa^2).
\end{equation}
These coefficients determine the instability bands in the parametric resonance. If $c_{\kappa}^2>0$, the Floquet exponent will be real, and parametric resonance happens. Otherwise they will be imaginary, and mode functions will only have oscillatory solutions. Therefore by using \eqref{eq:40} we can directly find the modes which are inside the instability bands. They are given by
\begin{align}
  \label{eq:42}
  \eps-1<\kappa^2<\eps,&\qfor \eps>1,\\
  \label{eq:105}
  0<\kappa^2<\eps,&\qfor \eps<1.
\end{align}
To find the value of the Floquet exponent we need to do a little bit more work. We work out the cases before and after trapping separately.

\paragraph{Before trapping:}

Without loss of generality we can assume that the homogeneous mode travels from $\Theta=0$ to $\Theta=2\pi$ during one period. In the first half of the period, $\Theta$ moves from $0$ to $\pi$, and $z$ decreases from $z=1$ to $z=-1$. In this patch the exponentially growing solution is $\theta_+$. By choosing the normalization factor $\mathcal{N}$ such that the solution is unity initially, its value after half a oscillation is given by
\begin{equation}
  \label{eq:41}
  \theta_{\kappa}^{(1/2)}=\sqrt{\frac{\abs{M(-1)}}{\abs{M(1)}}}\exp( \frac{c_{\kappa}}{2}\int_1^{-1}\frac{\dd{z'}}{M(z')\sqrt{(1-z'^2)(2\eps-1+z')}})
\end{equation}
In the second half of the period, $\Theta$ moves from $\pi$ to $2\pi$ while $z$ increases from $z=-1$ to $z=1$. Now the exponentially growing solution is $\theta_-$. In order to glue the solutions the normalization $\mathcal{N}$ should be chosen such that the full solution is continuous. Then we obtain the solution after a full oscillation as
\begin{equation}
  \label{eq:45}
  \begin{split}
    \theta_{\kappa}^{(1)}&=\theta_{\kappa}^{(1/2)}\sqrt{\frac{\abs{M(1)}}{\abs{M(-1)}}}\exp(- \frac{c_{\kappa}}{2}\int_{-1}^{1}\frac{\dd{z'}}{M(z')\sqrt{(1-z'^2)(2\eps-1+z')}})\\
    &=\exp(c_{\kappa}\int_{1}^{-1}\frac{\dd{z'}}{M(z')\sqrt{(1-z'^2)(2\eps-1+z')}}).
  \end{split}
\end{equation}
The Floquet exponent $\mu_{\kappa}$ can be obtained by matching this solution to \eqref{eq:34} and using the fact that $\psi_{\pm}$'s are periodic functions. We then obtain
\begin{equation}
  \label{eq:46}
  \mu_{\kappa}^{(\eps>1)}=\frac{c_{\kappa}}{\mathcal{T}_{>}}\int_{1}^{-1}\frac{\dd{z'}}{M(z')\sqrt{(1-z'^2)(2\eps-1+z')}},
\end{equation}
where $\mathcal{T}_{>}$ is the period of oscillation of the homogeneous mode in physical time $t$ when $\eps>1$. This can be derived from the conservation of energy as
\begin{equation}
  \label{eq:47}
  \mathcal{T}_{>}=\frac{2}{m\sqrt{\eps}}\ellipticK(1/\sqrt{\eps}).
\end{equation}
The integral in \eqref{eq:46} can be expressed in a more useful form by changing the integral path using the Cauchy's integral theorem:
\begin{equation}
  \label{eq:48}
  -\int_{-\infty}^{1-2\eps}\frac{\dd{z'}}{M(z')\sqrt{(1-z'^2)(2\eps-1+z')}}.
\end{equation}
By defining $\vartheta$ such that $z=-1/\sin^2\vartheta$ we obtain our final result.
\begin{equation}
  \label{eq:49}
  \mu_{\kappa}^{(\eps>1)}=\frac{c_{\kappa}}{\mathcal{T}_{>}}\int_0^{\arcsin(1/\sqrt{2\eps-1})}\frac{\dd{\vartheta}}{1+\qty(1-2\eps+2\kappa^2)\sin^2\vartheta}\frac{2\sin^2\vartheta}{\sqrt{\qty(1+\sin^2\vartheta)\qty[1+(1-2\eps)\sin^2\vartheta]}}.
\end{equation}

\paragraph{After trapping:}

Here we start the oscillation at the minimum $\Theta=0$. In the first half of the oscillation $\Theta$ travels from $0$ to $2\arcsin(\sqrt{\eps})$ while $z$ decreases from $z=1$ to $z=1-2\eps$. The growing solution is the $\theta_+$ solution. Again normalizing the mode functions to unity initial amplitude we find the solution after the first half of the oscillation as
\begin{equation}
  \label{eq:50}
  \theta_{\kappa}^{(1/2)}=\sqrt{\frac{\abs{M(1-2\eps)}}{\abs{M(1)}}}\exp( \frac{c_{\kappa}}{2}\int_1^{1-2\eps}\frac{\dd{z'}}{M(z')\sqrt{(1-z'^2)(2\eps-1+z')}}).
\end{equation}
In the second half of the oscillation $\Theta$ travels from $2\arcsin(\sqrt{\eps})$ back to $0$ while $z$ increases from $z=1-2\eps$ to $1$. Now the growing solution is $\theta_-$. So after a full period the solution is
\begin{equation}
  \label{eq:52}
  \theta_{\kappa}=\exp( c_{\kappa}\int_1^{1-2\eps}\frac{\dd{z'}}{M(z')\sqrt{(1-z'^2)(2\eps-1+z')}}).
\end{equation}
The Floquet exponent can directly be read from this result as
\begin{equation}
  \label{eq:51}
  \mu_{\kappa}^{(\eps<1)}=\frac{c_{\kappa}}{\mathcal{T}_{< }}\int_{1}^{1-2\eps}\frac{\dd{z'}}{M(z')\sqrt{(1-z'^2)(2\eps-1+z')}},
\end{equation}
where the period $\mathcal{T}_{<}$ in this case is
\begin{equation}
  \label{eq:53}
  \mathcal{T}_{<}=\frac{2}{m}\ellipticK(\sqrt{\eps}).
\end{equation}
Applying the integral transformations that we did in deriving \eqref{eq:49} yields to the result
\begin{equation}
  \label{eq:54}
  \mu_{\kappa}^{(\eps<1)}=\frac{c_{\kappa}}{\mathcal{T}_{<}}\int_0^{\pi/2}\frac{\dd{\vartheta}}{1+\qty(1-2\eps+2\kappa^2)\sin^2\vartheta}\frac{2\sin^2\vartheta}{\sqrt{\qty(1+\sin^2\vartheta)\qty[1+(1-2\eps)\sin^2\vartheta]}}.
\end{equation}

\paragraph{Final result:}

\begin{figure}[tbp]
  \centering
  \includegraphics[width=\textwidth]{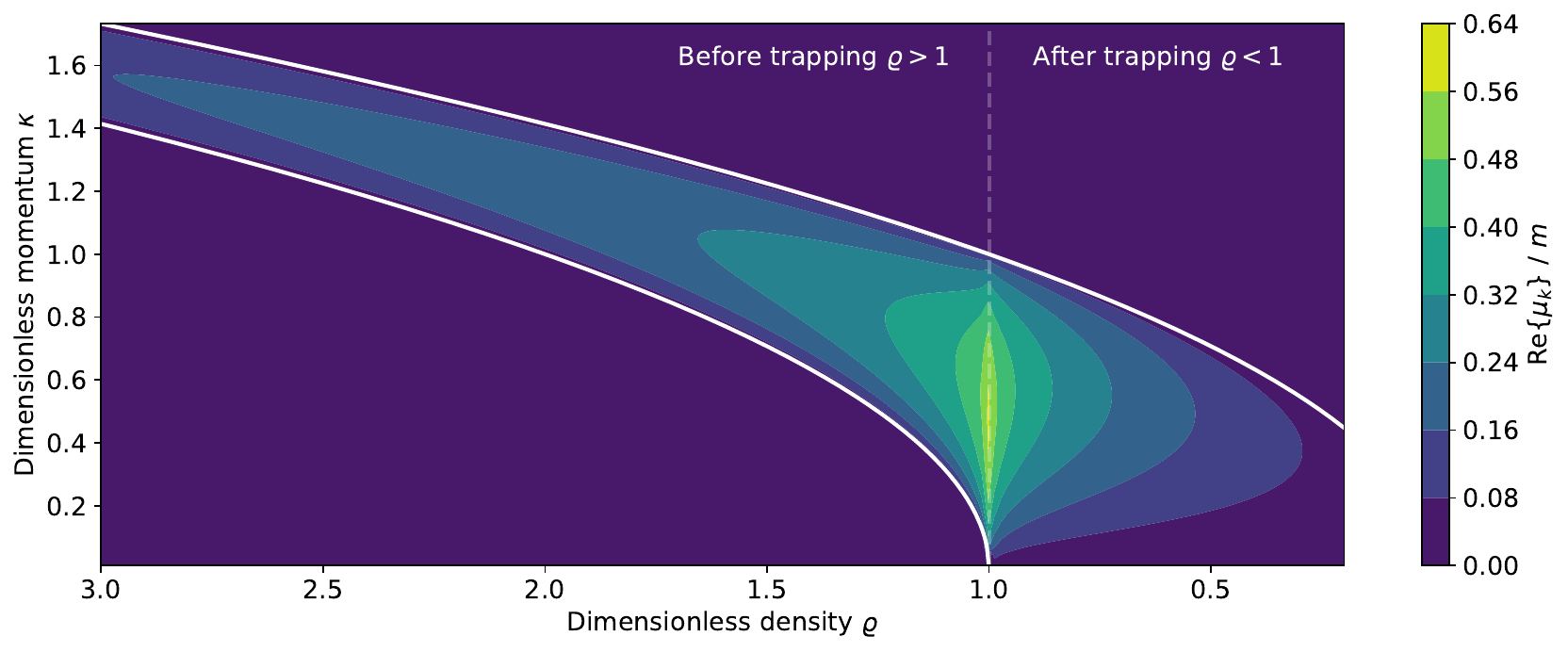}
  \caption{\it \small  Floquet exponents as a function of the dimensionless energy density $\eps$, and the dimensionless momentum $\kap$, using the analytical result (\eqref{eq:103}). The white lines denote the boundaries of the instability bands as given in (\eqref{eq:42}) and (\eqref{eq:105}). We see that the parametric resonance is most efficient around trapping $\eps \approx 1$, and for the modes $\kappa \sim 0.5$. }
  \label{fig:floquet-wo-hubble}
\end{figure}

Our final result for the Floquet exponents can be summarized as
\begin{equation}
  \label{eq:103}
  \mu_k=\sqrt{8\kap^2\qty(\eps-\kap)\qty(1-\eps+\kap^2)}\times
  \begin{cases}
    \mathcal{T}_{>}^{-1}(\eps)\,\mathcal{I}\qty(\arcsin\qty(1/\sqrt{2\eps-1})),& \eps>1\\
    \mathcal{T}_{<}^{-1}(\eps)\,\mathcal{I}\qty(\pi/2),&\eps<1
  \end{cases},
\end{equation}
where 
\begin{equation}
  \label{eq:104}
  \mathcal{I}(\varphi)=\int_0^{\varphi}\frac{\dd{\vartheta'}}{1+\qty(1-2\eps+2\kap^2)\sin^2\vartheta'}\frac{2\sin^2\vartheta'}{\sqrt{\qty(1+\sin^2\vartheta')\qty[1+(1-2\eps)\sin^2\vartheta']}}.
\end{equation}
We show a plot of the Floquet exponents together with the instability bands in Figure \ref{fig:floquet-wo-hubble}. The boundaries of the instability bands are shown in white lines. We can observe that the parametric resonance is most efficient around trapping $\eps\approx 1$, and for the modes $\kap \sim 0.5$.

\section{Calculation of the adiabatic initial conditions}
\label{sec:calc-adiab-init}

In this Appendix our primary goal is to derive the initial conditions for the mode functions given by \eqref{eq:150}. In Appendix \ref{sec:brief-revi-cosm} we briefly review the cosmological perturbation theory for completeness, and at the same time introduce our notation and conventions. We adopt the notation of \cite{maggiore2008gravitational}. The derivation of the initial conditions is presented in Appendix \ref{sec:deriv-adiab-init}.

\subsection{Brief review of cosmological perturbation theory}
\label{sec:brief-revi-cosm}

In order to study the perturbations, it is more convenient to use conformal time which we shall denote by $\eta$. We employ the \emph{Newtonian (conformal) gauge} where the perturbed metric takes the form
\begin{equation}
  \label{eq:106}
  \dd{s}^2=a^2(\eta)\qty{-\qty[1+2\Psi(\eta,\vb{x})]\dd{\eta}^2+\qty[1+2\Phi(\eta,\vb{x})]\delta_{ij}\dd{x}^i\dd{x}^j}.
\end{equation}
The perturbed stress-energy tensor can be put into the following form:
\begin{align}
  \label{eq:107}
  T^0_{\;0}&=-\overline{\rho}-\delta \rho,\\
  \label{eq:108}
  T^i_{\;0}&=-\qty[\overline{\rho}+\overline{p}]v^i,\\
  \label{eq:109}
  T^0_{\;i}&=\qty[\overline{\rho}+\overline{p}]v_i,\\
  \label{eq:110}
  T^i_{\;j}&=\qty[\overline{p}+\delta p]\delta^i_j + \qty(\partial^i\partial_j-\frac{1}{3}\delta^i_j \nabla^2)\Pi,
\end{align}
where $\rho$ and $p$ are the energy density and the pressure respectively, and quantities with an overline denote the background values. In these equations, $v^i$ is the \emph{peculiar velocity field}, and $\Pi$ is the anisotropic stress. The peculiar velocity field can be decomposed into a transverse and a longitudinal part:
\begin{equation}
  \label{eq:176}
  v^i(\eta,\vb{x})=v^{T,i}(\eta,\vb{x})+\partial^iv(\eta,\vb{x}),\quad \partial_iv^{T,i}(\eta,\vb{x})=0.
\end{equation}
Here, $v$ is the \emph{velocity potential}. Another useful quantity is the \emph{velocity divergence} defined by
\begin{equation}
  \label{eq:177}
  \vartheta\equiv \partial_iv^i=\nabla^2v.
\end{equation}
It is more usual to express the perturbations in terms of \emph{density contrast} $\delta$, and the \emph{sound speed} $c_s$ defined respectively by
\begin{equation}
  \label{eq:111}
  \delta\equiv \frac{\delta\rho}{\overline{\rho}}\qq{and}c_s^2\equiv \frac{\delta p}{\delta \rho}.
\end{equation}
One should keep in mind that both $\delta$ and $c_s^2$ are gauge-dependent quantities due to the fact that $\delta\rho$ and $\delta p$ are gauge-dependent. To overcome this, one can work with gauge-independent variables. In Newtonian gauge, the gauge-invariant generalizations of $\delta \rho$ and $\delta p$ are given by
\begin{equation}
  \label{eq:308}
  (\delta \rho)^{\rm GI}\equiv \delta \rho - 3\mathcal{H}(\overline{\rho}+\overline{p})v\qq{and}(\delta p)^{\rm GI}\equiv \delta p - 3\mathcal{H}(\overline{\rho}+\overline{p})c_{s,(a)}^2v,
\end{equation}
where $\mathcal{H}=aH=a'/a$ is the conformal Hubble parameter, and $c_{s,(a)}^2$ is the \emph{adiabatic speed of sound}:
\begin{equation}
  \label{eq:309}
  c_{s,(a)}^2\equiv \eval{\frac{\delta p}{\delta \rho}}_s = w + \frac{\overline{\rho}\dot{w}}{\dot{\overline{\rho}}},
\end{equation}
where $w=\overline{p}/\overline{\rho}$ is the equation of state. Then the gauge-invariant versions of the density contrast and the sound speed are given by
\begin{equation}
  \label{eq:310}
  \delta^{\rm GI}=\delta - 3\mathcal{H}(1+w)v\qq{and}(c_s^2)^{\rm GI}=\frac{(\delta p)^{\rm GI}}{(\delta \rho)^{\rm GI}}.
\end{equation}
For modes deep inside the horizon $\delta$ and $c_s^2$ are approximately equal to their gauge-invariant friends, so one can use them interchangebly. However one should be careful about the super-horizon modes.
\paragraph{Einstein equations:}

The linearized Einstein equations $\delta G_{\mu\nu}=8\pi G \delta T_{\mu\nu}$ can be used to relate metric perturbations to the perturbations in the stress-energy tensor. This way, we obtain four equations:
\begin{align}
  \label{eq:182}
  \nabla^2\Phi -3\mathcal{H}\qty(\Phi'-\mathcal{H}\Psi)&=-4\pi G a^2\delta \rho,\\
  \label{eq:183}
  \Phi'-\mathcal{H}\Psi&=4\pi G a^2\qty(\overline{\rho}+\overline{p})v,\\
  \label{eq:184}
  \Phi+\Psi&=-8\pi Ga^2\Pi,\\
  \label{eq:185}
  \Phi''+2\mathcal{H}\Phi'-\mathcal{H}\Psi'-(2\mathcal{H}'+\mathcal{H}^2)\Psi&=-4\pi G a^2\qty(\delta p + \frac{2}{3}\nabla^2\Pi).
\end{align}
In the rest of this section we will assume that for all fluids, the anisotropic stress does vanish $\Pi=0$\footnote{The anisotropic stress vanishes for scalar fields at linear order in perturbation theory. Photons develop anisotropic stress only during the matter era when they are sub-dominant. Only free-streaming neutrinos have non-negligible anisotropic stress, but we do neglect them in this work.}, so \eqref{eq:184} implies $\Psi=-\Phi$.

\paragraph{Perturbations of a scalar field:}

Consider an ALP field with the Lagrangian:
\begin{equation}
  \label{eq:112}
  \mathcal{L}=-\frac{\decay^2}{2}g^{\mu\nu}\partial_{\mu}\theta\partial_{\nu}\theta-V(\theta,T).
\end{equation}
where the potential is in general depends on the temperature $T$ of the universe. The corresponding stress-energy tensor is given by
\begin{equation}
  \label{eq:178}
  \decay^{-2}T_{\mu\nu}=\partial_{\mu}\theta\partial_{\nu}\theta-g_{\mu\nu}\qty(\frac{1}{2}g^{\rho\sigma}\partial_{\rho}\theta\partial_{\sigma}\theta+\frac{1}{f^2}V(\theta,T)).
\end{equation}
We now write the ALP field $\theta(\eta,\vb{x})$ as the sum of the homogenous component $\Theta(\eta)$ and fluctuations $\delta\theta(\eta,\vb{x})$ which are assumed to be small. We also denote the temperature fluctuations\footnote{The contribution of the temperature perturbations to the ALP density constrast has recently been pointed out in \cite{Sikivie:2021trt,Kitajima:2021inh}.} by $T=\overline{T}+\delta T$, where $\overline{T}$ represents the background value. By expanding the stress-energy tensor up to linear order in $\delta\theta$, $\Phi$ and $\delta T$, and matching with the expressions \eqref{eq:107}---\eqref{eq:110}, we obtain the zeroth order quantities
\begin{align}
  \label{eq:113}
  \overline{\rho}_{\theta}\equiv \rho_{\Theta}&=\frac{\decay^2}{2}\qty(\frac{\Theta'}{a})^2+V(\Theta,\overline{T})\\
  \label{eq:114}
  \overline{p}_{\theta}\equiv p_{\Theta}&=\frac{f^2}{2}\qty(\frac{\Theta'}{a})^2-V(\Theta,\overline{T}),
\end{align}
and the first order quantities
\begin{align}
  \label{eq:115}
  \delta\rho_{\theta}&=\decay^2\qty[\frac{\Theta'\delta\theta'}{a^2}+\Phi \frac{\Theta'^2}{a^2}+\frac{1}{\decay^2}\qty(\eval{\pdv{V}{\theta}}_{\Theta,\overline{T}}\delta\theta+\eval{\pdv{V}{T}}_{\Theta,\overline{T}}\delta T)]\\
  \delta p_{\theta}&=\decay^2\qty[\frac{\Theta'\delta\theta'}{a^2}+\Phi \frac{\Theta'^2}{a^2}-\frac{1}{\decay^2}\qty(\eval{\pdv{V}{\theta}}_{\Theta,\overline{T}}\delta\theta+\eval{\pdv{V}{T}}_{\Theta,\overline{T}}\delta T)]\\
  v_{\theta}&=-\frac{\decay^2\Theta'\delta\theta}{a^2\qty(\rho_{\Theta}+p_{\Theta})}=-\frac{f^2\Theta'\delta\theta}{a^2\rho_{\Theta}(1+w_{\Theta})},
\end{align}
where $w_{\Theta}$ is the equation of state of the background ALP field. Perturbing the Klein-Gordon equation up to linear order in fluctuations gives at zeroth order the evolution of the background field
\begin{equation}
  \label{eq:116}
  \Theta''+2\mathcal{H}\Theta'+\frac{a^2}{\decay^2}\eval{\pdv{V}{\theta}}_{\Theta,\overline{T}}=0,
\end{equation}
where $\mathcal{H}=aH$ is the conformal Hubble factor. At the first order, we obtain the evolution of the fluctuations:
\begin{equation}
  \label{eq:117}
  \delta\theta''+2\mathcal{H}\theta'-\nabla^2\delta\theta+\frac{a^2}{\decay^2}\eval{\pdv[2]{V}{\theta}}_{\Theta, \overline{T}}\delta\theta=2\Phi \frac{a^2}{\decay^2}\eval{\pdv{V}{\theta}}_{\Theta,\overline{T}}-4\Phi'\Theta'-\frac{a^2}{\decay^2}\eval{\pdv[2]{V}{\theta}{T}}_{\Theta,\overline{T}}\delta T.
\end{equation}
In Fourier space this becomes
\begin{equation}
  \label{eq:179}
  \theta_k''+2\mathcal{H}\theta_k'+k^2\theta_k+\frac{a^2}{\decay^2}\eval{\pdv[2]{V}{\theta}}_{\Theta, \overline{T}}\theta_k=2\Phi_k \frac{a^2}{\decay^2}\eval{\pdv{V}{\theta}}_{\Theta,\overline{T}}-4\Phi_k'\Theta'-\frac{a^2}{\decay^2}\eval{\pdv[2]{V}{\theta}{T}}_{\Theta,\overline{T}}\delta T.
\end{equation}

\paragraph{Curvature perturbations in radiation era:}

The Fourier modes of the curvature perturbations $\Phi_{\vb{k}}$ have exact solutions valid during the radiation era:
\begin{equation}
  \label{eq:119}
  \Phi_{\vb{k}}(\eta_k)=3 \Phi_{\vb{k}}(0) \qty(\frac{\sin \eta_k - \eta_k\cos
    \eta_k}{\eta_k^3}),
\end{equation}
where $\eta_k\equiv k\eta/\sqrt{3}$. The initial conditions $\Phi_{\vb{k}}(0)$ are imprinted by the inflation, and they are related to the \emph{comoving curvature perturbation} $\mathcal{R}_{\vb{k}}(0)$ via
\begin{equation}
  \label{eq:120}
  \Phi_{\vb{k}}(0)=\frac{2}{3}\mathcal{R}_{\vb{k}}(0).
\end{equation}
Since the comoving curvature perturbations are generated by the quantum fluctuations of the inflaton, they are stochastic variables. This means that we can only calculate and measure the \emph{power spectrum} of them:
\begin{equation}
  \label{eq:121}
  \dimps_{\mathcal{R}}^{\rm in}(k)=\expval{\abs{\mathcal{R}_{\vb{k}}(0)}^2}.
\end{equation}
More specifically the experiments measure the \emph{dimensionless power spectrum}:
\begin{equation}
  \label{eq:122}
  \dimlessps_{\mathcal{R}}^{\rm in}(k)=\frac{k^3}{2\pi^2}P_{\mathcal{R}}^{\rm in}(k)\equiv A_s\qty(\frac{k}{k_{\star}})^{n_s-1}.
\end{equation}
The Planck 2018 measurements (TT,TE,EE+lowE+lensing 68\%) \cite{Aghanim:2018eyx} are consistent with
\begin{equation}
\label{eq:123}
  A_s= 2.1\times 10^{-9},\quad n_s=0.9649,\quad k_{\star}=0.05\,\rm{Mpc}^{-1}.
\end{equation}
In this work we neglect the spectral tilt by setting $n_s=1$.

\subsection{Adiabatic initial conditions in Kinetic Misalignment}
\label{sec:deriv-adiab-init}

In this section, we derive the initial conditions for the axion perturbations that arise purely from adiabatic fluctuations. In other words, we neglect all other types of isocurvature fluctuations to obtain a result which is as model-independent as possible.

At early times, the axion mass can be neglected. Then the homogeneous mode equations of motion \eqref{eq:116} becomes
\begin{equation}
  \label{eq:124}
  \Theta''+2\mathcal{H}\Theta'=0\quad\Rightarrow\quad \Theta'\propto a^{-2},
\end{equation}
where $\Theta$ is the axion homogeneous mode, and $\prime$ denotes derivative with respect to conformal time. The mode function equation of motion \eqref{eq:117} takes the form
\begin{equation}
  \label{eq:125}
  \theta_k''+2\mathcal{H}\theta_k'+k^2\theta_k=-4\Phi_k'\Theta',
\end{equation}
where $\Phi_k$'s are given by \eqref{eq:119}. In radiation domination $\mathcal{H}=\eta^{-1}$. It will be more convenient to write this equation of motion in terms of $\eta_k$. Doing so we obtain
\begin{equation}
  \label{eq:126}
  \dv[2]{\theta_k}{\eta_k}+\frac{2}{\eta_k}\dv{\theta_k}{\eta_k}+3\phi_k=-4\dv{\Phi_k}{\eta_k}\dv{\Theta}{\eta_k}\equiv \mathcal{F}(\eta_k).
\end{equation}
The right-hand-side of this equation is a forcing term which will induce fluctuations even if they are zero initially. This equation has a homogeneous solution
\begin{equation}
  \label{eq:127}
  \theta_k^{\rm hom}(\eta_k)=\frac{1}{\eta_k}\qty[c_1 \cos(\sqrt{3}\eta_k)+c_2\sin(\sqrt{3}\eta_k)],
\end{equation}
and a particular solution
\begin{equation}
  \label{eq:128}
  \theta_k^{\rm par}=\theta_s(\eta_k)\int^{\eta_k}\dd{\eta_k'}\frac{\theta_c(\eta_k')\mathcal{F}(\eta_k')}{W\qty[\theta_c(\eta_k'),\theta_s(\eta_k')]}-\theta_c(\eta_k)\int^{\eta_k}\dd{\eta_k'}\frac{\theta_s(\eta_k')\mathcal{F}(\eta_k')}{W\qty[\theta_c(\eta_k'),\theta_s(\eta_k')]}.
\end{equation}
Here $\theta_c=\cos(\sqrt{3}\eta_k)/\eta_k$ and $\theta_s=\sin(\sqrt{3}\eta_k)/\eta_k$ are the homogeneous solutions, and $W$ is their Wronskian which is $\sqrt{3}/\eta_k^2$.

Let $\eta_{\ini}=\mathcal{H}_{\ini}^{-1}$ denote an arbitrary time during which the axion homogeneous mode scales as kination, i.e. $\rho_{\Theta}\propto a^{-6}$. We also define
\begin{equation}
  \label{eq:129}
  \eta_{k,\ini}\equiv \frac{k\eta_{\ini}}{\sqrt{3}}
\end{equation}
which tells whether a given mode is super- or sub-horizon at $\eta=\eta_{\ini}$. Then the forcing term $\mathcal{F}(\eta_k)$ is given by
\begin{equation}
  \label{eq:130}
  \mathcal{F}(\eta_k)=-12\Phi_k(0)\frac{\dot{\Theta}_{\ini}}{H_{\ini}} \frac{\eta_{k,\ini}}{\eta_k^2}\qty(\frac{\sin(\eta_k)}{\eta_k^2}+\frac{3\cos(\eta_k)}{\eta_k^3}-\frac{3\sin(\eta_k)}{\eta_k^4}).
\end{equation}
So the full solution becomes
\begin{equation}
  \label{eq:131}
  \theta_k(\eta_k)=\frac{1}{\eta_k}\qty{\cos(\sqrt{3}\eta_k)\qty[c_1-\mathcal{I}_s(\eta_k)+\mathcal{I}_s(\eta_{k,\ini})]+\sin\qty(\sqrt{3}\eta_k)\qty[c_2+\mathcal{I}_c(\eta_k)-\mathcal{I}_c(\eta_{k,\ini})]},
\end{equation}
where $\mathcal{I}_{c,s}$ are indefinite integrals
\begin{equation}
  \label{eq:132}
  \mathcal{I}_{c,s}(\eta_k)\equiv  \int^{\eta_k}\dd{\eta_k'}\frac{\theta_{c,s}(\eta_k')\mathcal{F}(\eta_k')}{W\qty[\theta_c(\eta_k'),\theta_s(\eta_k')]},
\end{equation}
where $\theta_c(\eta_k)=\cos(\sqrt{3}\eta_k)/\eta_k$, and $\theta_s(\eta_k)=\sin(\sqrt{3}\eta_k)/\eta_k$ are the homogeneous solutions of \eqref{eq:126}, and $W=\sqrt{3}/\eta_k^2$ is their Wronskian. Even though these integrals can be performed analytically, their full expressions are given by trigonometric integrals, and are too complicated to be informative. At large $\eta_k$, both integrals decay as $\eta_k^{-3}$. This means that in the deep sub-horizon limit, we can neglect the $\mathcal{I}_{c,s}(\eta_k)$ terms in \eqref{eq:131}, and keep only the $\mathcal{I}_{c,s}(\eta_{k,\ini})$ terms.

In order to get the full solution one also needs to specify the coefficients $c_1$ and $c_2$. These are determined by the solution of the mode functions at $\eta=\eta_{\ini}$. To calculate them, we will assume that the mods we are interested are super-horizon at $\eta_{\ini}$, which implies that $\eta_{k,\ini}\ll 1$. The assumption is justified towards the end of Section \ref{sec:initial-conditions}. Then it is possible to determine both $c_1$ and $c_2$ by using the adiabacity conditions. The adiabatic perturbations are defined such that the local state of matter at some spacetime point $(\eta,\vb{x})$ of the perturbed universe is the same as the background universe at some slightly different time. For a universe filled with multiple fluids, the adiabatic perturbations are induced by a common, local shift in time of all background quantities \cite{Baumann:2018muz}. Therefore at early times, the perturbations of all species can be described by
\begin{equation}
  \label{eq:180}
  \delta\rho_i(\eta,\vb{x})=\overline{\rho}_i(\eta+\delta\eta(\vb{x}))-\overline{\rho}(\eta)\approx \overline{\rho}_i'\delta \eta(\vb{x}).
\end{equation}
Using the continuity equation $\overline{\rho}_i'=-3\mathcal{H}(1+w_i)\overline{\rho}_i$, this equation implies
\begin{equation}
  \label{eq:181}
  \frac{\delta_i}{1+w_i}=\frac{\delta_j}{1+w_j}
\end{equation}
for all species $i$ and $j$. Let us consider the adiabatic perturbations in the rotating axion $\delta_{\theta}$, and in the radiation $\delta_{r}$. At early times, the axion equation of state is $w_{\theta}=1$, while for radiation it is $w_r=1/3$. Then, two perturbations should be related to each other by
\begin{equation}
  \label{eq:186}
  \delta_{\theta}=\frac{3}{2}\delta_r.
\end{equation}
Since all perturbations are comparable, the total density perturbation $\delta \rho \equiv \sum_i\overline{\rho}_i\delta_i$ is dominated by the species that carry the dominant energy density. Since the universe is radiation dominated at early times, we can assume $\delta \rho \approx \overline{\rho}_r\delta_{r}$. Using \eqref{eq:182} and the Friedmann equation $3\mathcal{H}^2\approx 8\pi G a^2\overline{\rho}_r$, one can show that $\delta_r\approx 2\Phi$ in the super-horizon limit. This implies that the adiabatic initial conditions for the rotating axions should have the following super-horizon limit:
\begin{equation}
  \label{eq:187}
  \delta_{\theta}\approx 3\Phi.
\end{equation}

We now present an alternative derivation of this result, which will also give us the correct initial conditions for the axion mode functions. Since adiabatic initial conditions arise due to the fluctuations in the total energy density, they should vanish in a time slicing where there is no perturbatiation in the energy density $\eval{\delta \rho}_{\rm unif}=0$\footnote{Note that this is true only for the super-horizon adiabatic modes. Once they enter the horizon they don't vanish even in the uniform-density gauge. Classification of perturbations as adiabatic or isocurvature are done by specifying their super-horizon behavior.}. This is known as the \emph{slicing of uniform energy density}. From a generic slicing, one can go this slicing via a time displacement given by \cite{Riotto:2002yw}
\begin{equation}
  \label{eq:188}
  \eta \rightarrow \eta + \xi,\quad \xi = \frac{\delta\rho}{\overline{\rho}'}.
\end{equation}
Under this time displacement, the perturbations $\delta S$ of a scalar quantity $S$ transforms as $\delta S \rightarrow \delta S - \overline{S}'\xi$. Since both the axion angle $\theta$ and the axion energy density $\rho_{\theta}$ are scalar quantities, we can write the following relations between the values of these quantities in the uniform density gauge, and the Newtonian gauge:
\begin{align}
  \label{eq:190}
  0&=\eval{\delta \theta}_{\rm unif}=\eval{\delta \theta}_{\rm new}-\Theta' \xi\\
  \label{eq:191}
  0&=\eval{\delta \rho_{\theta}}_{\rm unif}=\eval{\delta \rho_{\theta}}_{\rm new}-\rho_{\Theta}'\xi.
\end{align}
Since the dominant energy density is in radiation, we can write
\begin{equation}
  \label{eq:192}
  \xi=\frac{\delta\rho}{\overline{\rho}'}\approx \frac{\delta \rho_r}{\overline{\rho}_r'}=-\frac{\delta \rho_r}{3\mathcal{H}\overline{\rho}_r(1+w_r)}=-\frac{\delta_r}{4\mathcal{H}}\approx -\frac{\Phi}{2\mathcal{H}}.
\end{equation}
By plugging this result into \eqref{eq:190} and \eqref{eq:191}, and taking the Fourier transform we find the following super-horizon initial conditions for the axion mode functions in the Newtonian gauge:
\begin{align}
  \label{eq:194}
  \lim_{\eta_k \rightarrow 0}\theta_k(\eta_k)&\approx -\frac{1}{2}\Phi_k(0) \frac{\Theta'}{\mathcal{H}}=-\frac{1}{2}\Phi_k(0) \frac{\dot{\Theta}}{H},\\
  \label{eq:275}
  \lim_{\eta_k \rightarrow 0}\delta_{\theta,k}(\eta_k)&=2 \qty[\frac{\theta_k'(\eta)}{\Theta'(\eta)}+\Phi_k(\eta_k)]\approx 3\Phi_k(0)
\end{align}
We see that \eqref{eq:275} agrees with \eqref{eq:187}. Now we can take the full solution \eqref{eq:131}, evaluate at $\eta_k=\eta_{k,\ini}$, take the $\eta_{k,\ini}\rightarrow 0$ limit, and find the coefficients $c_1$ and $c_2$ by matching with the initial conditions given above. This procedure gives
\begin{equation}
  \label{eq:195}
  c_1=-\frac{1}{2}\eta_{k,\ini}\frac{\Theta'_i}{\mathcal{H}_i}\Phi_k(0),\quad c_2=0.
\end{equation}
The indefinite integrals \eqref{eq:132} have the following behavior in the $\eta_k\ll 1$ limit:
\begin{align}
  \label{eq:196}
  \mathcal{I}_c(\eta_k)&\approx \frac{4}{5\sqrt{3}}\Phi_k(0)\frac{\Theta'_i}{\mathcal{H}_i}\eta_k\eta_{k,\ini}+\mathcal{O}(\eta_k^3),\\
  \mathcal{I}_s(\eta_k)&\approx \frac{1}{\sqrt{3}}\Phi_k(0)\frac{\Theta'_i}{\mathcal{H}_i}\qty(2\sqrt{3}-3\ln \frac{\sqrt{3}+1}{\sqrt{3}-1})\eta_{k,i}+\mathcal{O}(\eta_k^2).
\end{align}
By using these expressions we can show that the mode functions have the following super- and sub-horizon behavior:
\begin{equation}
\boxed{
  \label{eq:197}
  \theta_k(\eta)\approx
  \begin{cases}
    -\dfrac{1}{2}\Phi_k(0)\dfrac{\Theta'}{\mathcal{H}},&\eta_k\ll 1,\\
    -\qty(\sqrt{3}\ln \dfrac{\sqrt{3}+1}{\sqrt{3}-1}-\dfrac{3}{2})\Phi_k(0)\dfrac{\Theta'}{\mathcal{H}}\cos(k \eta),&\eta_k\gg 1
  \end{cases}.}
\end{equation}
Note that we have replaced the initial values of the axion velocity and the conformal Hubble parameter with their dynamical values by using $\Theta'\propto a^{-2}$, and $\mathcal{H}\propto a^{-1}$. The numerical factor in the sub-horizon case is approximately $0.78$. Therefore we can approximate the field power spectrum at early times at both super- and sub-horizon scales by
\begin{equation}
  \label{eq:198}
  \dimps_{\theta}(k)=\abs{\theta_k}^2\approx \frac{1}{4}\abs{\Phi_k(0)}^2\qty(\frac{\Theta'}{\mathcal{H}})^2\cos^2(k\eta)\approx \frac{1}{9}\qty(\frac{2\pi^2}{k^3})A_s\qty(\frac{\Theta'}{\mathcal{H}})^2\cos^2(k\eta),
\end{equation}
where we have used \eqref{eq:120}, \eqref{eq:121}, and \eqref{eq:122}. By changing from the conformal time to physical time one obtains \eqref{eq:150}.

\section{Fragmentation before trapping: $\eps\gg 1$ limit}
\label{sec:fragm-before-trapp}

In the $\eps\gg 1$ limit, the fragmentation calculation can be simplified considerably. By averaging $\eps$ over one oscillation, we get
\begin{equation}
  \label{eq:82}
  \eps=\frac{1}{4}\frac{\dot{\Theta}^2}{m^2(t)}+\frac{1}{2},
\end{equation}
where we have omitted averages to simplify the notation. We note that $\eps\gg 1$ limit is equivalent to $\dot{\Theta}\gg 2 m$ limit. The instability band \eqref{eq:58} becomes
\begin{equation}
  \label{eq:83}
  \frac{\dot{\Theta}^2}{4}-\frac{m^2}{2}<\frac{k^2}{a^2}<\frac{\dot{\Theta}^2}{4}+\frac{m^2}{2}.
\end{equation}
The momentum mode $k$ is at the center of the instability band when $\dot{\Theta}=\dot{\Theta}_k$ where
\begin{equation}
  \label{eq:84}
  \frac{k}{a_k}=\frac{\dot{\Theta}_k}{2},\quad \dot{\Theta}=\dot{\Theta}_k\qq{when}a=a_k.
\end{equation}
The instability band will move with time due to the redshift of the modes, and due to the fact that $\dot{\Theta}$ does also change with time. After some algebra, we can show that the growth factor \eqref{eq:55} in the $\eps\gg 1$ limit can be approximated by
\begin{equation}
  \label{eq:85}
  \mu_k\approx \sqrt{\frac{m^4}{4 \dot{\Theta}^2}-\qty(\frac{k}{a}-\frac{\dot{\Theta}}{2})^2},\quad \eps\gg 1.
\end{equation}
Also for large $\eps$, the width of the instability band is very narrow so we can approximate the axion velocity and the scale factor during the amplification of a single mode as
\begin{equation}
  \label{eq:86}
  \dot{\Theta}(t)=\dot{\Theta}_k+\ddot{\Theta}_k(t-t_k)\equiv \dot{\Theta}_k+\ddot{\Theta}_k\Delta t\qq{,}a(t)=a_k\qty(1+H_k\Delta t).
\end{equation}
In other words, we did assume that the Hubble scale, barrier height, and the acceleration of the homogeneous mode remain constant during the amplification. With these approximations, and \eqref{eq:84}, the growth factor takes the form
\begin{equation}
  \label{eq:87}
  \mu_k\approx \frac{m^2}{2\dot{\Theta}_k}-\frac{m^2 \ddot{\Theta}_k}{2 f \dot{\Theta}_k^2}\Delta t +\mathcal{O}\qty((\Delta t)^2).
\end{equation}
From the expression for the instability bands \eqref{eq:83} we can see that the momentum mode $k$ is inside the instability band from $\Delta t=-t_{\rm amp}/2$ to $\Delta t=t_{\rm amp}/2$ where
\begin{equation}
  \label{eq:88}
  t_{\rm amp}=\frac{2 m^2}{\dot{\Theta}_k\abs{H_k\dot{\Theta}_k+\ddot{\Theta}_k}}.
\end{equation}
Then, the total growth factor of the momentum mode $k$ before trapping can be approximated as
\begin{equation}
  \label{eq:89}
  \ln \amp_k(t\gg t_k)=\int^{t_{\rm amp}/2}_{-t_{\rm amp}/2} \dd{\Delta t}\mu_k(\Delta t)\approx \frac{m^4}{f \dot{\Theta}_k^2\abs{H\dot{\Theta}_k+\ddot{\Theta}_k}}.
\end{equation}
This result is in fairly good aggreement with the one obtained in Equation 3.9 of \cite{Fonseca:2019ypl}. In fact, the exact result contains an additional $\pi/4$ with which the above results becomes more closer to the numerical solution. Therefore we shall include it for the rest of this section.

Even the back-reaction can be incorporated analytically in the $\eps\gg 1$ limit. In this limit $\kappa=(k/a)/m \gg 1$ which means the modes inside the instability bands are relativistic during this regime. Then we can approximate the redshift factor $A_k(t)$ as
\begin{equation}
  \label{eq:91}
  A_k(t)=\qty(\frac{\omega_k(t_{\ini})}{\omega_k(t)})^{1/2}\qty(\frac{a_{\ini}}{a})^3\approx \frac{k/a_{\ini}}{k/a}\qty(\frac{a_{\ini}}{a})^3=\qty(\frac{a_{\ini}}{a})^2.
\end{equation}
By plugging the initial power spectrum \eqref{eq:154} and \eqref{eq:91} into \eqref{eq:79} we obtain
\begin{equation}
  \label{eq:92}
  \begin{split}
    \frac{\Delta \rho_{\rm fluct}}{\Delta t}&\approx \frac{\decay^2}{2a^4}\qty(\frac{1}{3})^2A_s\frac{\dot{\Theta}_{\ini}^2a_{\ini}^2}{H_{\ini}^2}\int\dd{k}k \frac{\amp_k^2(t)-\amp_k^2(t-\Delta t)}{\Delta t}\\
    &\approx \frac{f^2}{2}\qty(\frac{4\pi}{9})^2 \qty(\frac{g_s(T_{\trap})}{10})\qty(\frac{Y T_{\trap}\mpl}{f^2})^2A_s\frac{a_{\trap}^2}{a^4}\int\dd{k}k \frac{\amp_k^2(t)-\amp_k^2(t-\Delta t)}{\Delta t},
  \end{split}
\end{equation}
where we have used \eqref{eq:144} in the second line. The dominant contribution to the integral will come from the modes around $k_{\rm cr}$ where
\begin{equation}
  \label{eq:94}
  k_{\rm cr}(t)=a(t)\frac{\dot{\Theta}(t)}{2}.
\end{equation}
Then we can approximate \eqref{eq:92} as
\begin{equation}
  \label{eq:95}
  \begin{split}
    \dv{\rho_{\rm fluct}}{t}&\approx \frac{f^2}{2 a^4}\qty(\frac{1}{3})^2A_s\frac{\dot{\Theta}_{\ini}^2a_{\ini}^2}{H_{\ini}^2}\abs{\dv{k_{\rm cr}}{t}}k_{\rm cr}\amp_{k_{\rm cr}}^2(t)\\
    &\approx \frac{\decay^2}{8}\qty(\frac{1}{3})^2A_s\frac{\dot{\Theta}_{\ini}^2}{H_{\ini}^2}\qty(\frac{a_{\ini}}{a})^2\abs{\ddot{\Theta}+H\dot{\Theta}}\dot{\Theta}\exp(\frac{\pi m^4}{2\dot{\Theta}^2\abs{H\dot{\Theta}+\ddot{\Theta}}})\\
    &\approx\qty(\frac{4\pi}{9})^2 \qty(\frac{g_s(T_{\trap})}{80})\qty(\frac{Y T_{\trap}\mpl}{\decay})^2A_s\frac{a_{\trap}^2}{a^2}\abs{\ddot{\Theta}+H\dot{\Theta}}\dot{\Theta}\exp(\frac{\pi m^4}{2\dot{\Theta}^2\abs{H\dot{\Theta}+\ddot{\Theta}}})
  \end{split}
\end{equation}
Then the ``conservation'' of energy implies that the evolution of the homogeneous mode is given by
\begin{equation}
  \label{eq:96}
  \ddot{\Theta}+3H\dot{\Theta}+\qty(\frac{4\pi}{9})^2 \qty(\frac{g_s(T_{\trap})}{80})\qty(\frac{Y T_{\trap}\mpl}{\decay^2})^2A_s\frac{a_{\trap}^2}{a^2}\abs{\ddot{\Theta}+H\dot{\Theta}}\exp(\frac{\pi m^4}{2\dot{\Theta}^2\abs{H\dot{\Theta}+\ddot{\Theta}}})\approx 0.
\end{equation}
Here the second and third terms represent the loss of energy due to Hubble expansion and back-reaction respectively. It is possible to solve for $\ddot{\Theta}$ exactly to get \cite{Fonseca:2019ypl}
\begin{equation}
  \label{eq:97}
  \ddot{\Theta}=-H\dot{\Theta}+\frac{\pi m^4
  }{2 \dot{\Theta}^2}\qty[-\frac{1}{\beta}+W_0\qty(\frac{\alpha}{\beta} e^{1/\beta})]^{-1}\equiv -H\dot{\Theta}-\frac{\pi m^4}{2\decay \dot{\Theta}^2}\frac{1}{\mathcal{W}(\alpha,\beta)},
\end{equation}
where we have defined
\begin{align}
  \label{eq:133}
  \mathcal{W}(\alpha,\beta)&\equiv -\frac{1}{\beta}+W_0\qty(\frac{\alpha}{\beta} e^{1/\beta})\\
  \label{eq:98}
  \alpha&\equiv \qty(\frac{4\pi}{9})^2 \qty(\frac{g_s(T_{\trap})}{80})\qty(\frac{Y T_{\trap}\mpl}{f^2})^2A_s\frac{a_{\trap}^2}{a^2} \\
\label{eq:90}
  \beta&\equiv \frac{4 H\dot{\Theta}^3}{\pi m^4},
\end{align}
and $W_0$ is the principal branch of the Lambert function. By substituting this solution into \eqref{eq:89} we find the growth factor as
\begin{equation}
  \label{eq:99}
  \amp_k=\eval{\exp(\frac{1}{2}\abs{\mathcal{W}(\alpha,\beta)})}_{\frac{k}{a_k}=\frac{\dot{\phi}_k}{2f}}.
\end{equation}
If $\alpha$ is small, then the function $\mathcal{W}(\alpha,\beta)$ can be approximated quite well by
\begin{equation}
  \label{eq:100}
  \mathcal{W}(\alpha,\beta)\approx
  \begin{cases}
    \beta^{-1},&\beta^{-1}<\ln \alpha^{-1} \quad\text{(weak back-reaction)} \\
    \ln\alpha^{-1},& \beta^{-1}>\ln\alpha^{-1} \quad\text{(strong back-reaction)}
  \end{cases}.
\end{equation}
This result tells us that there are two distinct regimes (weak and strong back-reaction) during the fragmentation before trapping. The fragmentation starts when $\beta\approx 1$, and $\beta$ continues to decrease during the later evolution. This means that at the beginning of fragmentation we are in the $\mathcal{W}\approx \beta^{-1}$ regime. In this case, \eqref{eq:97} reduces to
\begin{equation}
  \label{eq:101}
  \ddot{\Theta}+3H\dot{\Theta}\approx 0.
\end{equation}
This is the equation for the homogeneous mode in the absence of back-reaction. The conclusion is that in this regime the back-reaction is negligible. With some algebra, we can express the growth factor in this regime in terms of $\kappa_{\trap}$ and other model parameters. The result is
\begin{equation}
  \label{eq:102}
  \ln N_k\approx \sqrt{\frac{1}{128\pi^3}}\qty(\frac{2}{\pi})^{\gamma}\qty(\frac{m_{\trap}}{H_{\trap}})\frac{1}{\kappa_{\trap}^{11/2+\gamma}},\qq{weak back-reaction}
\end{equation}
We see that in this regime the growth factor is very small for large momentum modes, but increases quite rapidly as the instability band moves to towards smaller momentum modes.

As $\beta$ continues to decrease, at some point reaches the value $\qty(\ln \alpha^{-1})^{-1}$, provided that the axion is not trapped by that time. In this regime, the amplification factors can be approximated by
\begin{equation}
  \label{eq:93}
  \amp_k^2\approx \alpha^{-1}= \qty(\frac{9}{4\pi})^2\qty(\frac{80}{g_s(T_{\trap})})^2\qty(\frac{f^2}{Y T_{\trap}\mpl})^2 A_s^{-1}\eval{\qty(\frac{a}{a_{\trap}})^2}_{\frac{k}{a_k}=\frac{\dot{\phi}_k}{2f}}.
\end{equation}
We can observe that the amplification factor as a function of momentum is almost a flat function, rather than being an exponential in the $\mathcal{W}\approx \beta^{-1}$ regime. The reason behind this drastic change of behavior is the back-reaction of the fragmented modes on the homogeneous mode. Therefore we call the $\mathcal{W}\approx \beta^{-1}$ and $\mathcal{W}\approx \ln \alpha^{-1}$ regimes as weak and strong back-reaction regimes respectively. The transition between the two regimes happens arround
\begin{equation}
  \label{eq:155}
  \frac{\pi m^4(T)}{4H(T)\dot{\Theta}^3(T)}\approx \ln\qty[\qty(\frac{9}{4\pi})^2\qty(\frac{80}{g_s(T_{\trap})})^2\qty(\frac{f^2}{Y T_{\trap}\mpl})^2 A_s^{-1}\qty(\frac{T_{\trap}}{T})^2].
\end{equation}

Finally we do comment on how these conclusions change when one considers different initial conditions for the mode functions. The initial conditions matter only when the back-reaction becomes prominent. Therefore $\mathcal{W}\approx \beta^{-1}$ regime is not affected by the change of initial conditions. However, it will modify the parameter $\alpha$, so it will have a strong effect on the amplification factors during the strong back-reaction regime. In the case of Bunch-Davies initial conditions, $\alpha$ will be modified to
\begin{equation}
  \label{eq:156}
  \alpha = \frac{\dot{\Theta}^2}{128\pi^2\decay^2},\qq{Bunch-Davies initial conditions.}
\end{equation}
So instead of \eqref{eq:155}, the transition to the $\mathcal{W}\approx \ln \alpha^{-1}$ regime happens when
\begin{equation}
  \label{eq:157}
  \frac{\pi m^4(T)}{4H(T)\dot{\Theta}^3(T)}\approx \ln \frac{128\pi^2f^2}{\dot{\Theta}^2}\qq{Bunch-Davies initial conditions.}
\end{equation}

\section{Experimental surveys}
\label{sec:experimental-surveys}

In this Appendix, we list the references for the constraints and projections that are used in Figures \ref{fig:money-plot-g8} and \ref{fig:money-plot-g0}. The references are tabulated in Tables \ref{tab:referenceList1}  and \ref{tab:referenceList2} together with the principle behind the measurement/projection. All experiments in Table \ref{tab:referenceList1} as well as the astrophysical constraints in Table \ref{tab:referenceList2} rely on the axion-photon coupling. Neutron coupling or coupling independent constraints are labelled accordingly. Most of the digitized constraints are sourced from the \verb|AxionLimits| repository ~\cite{AxionLimits}.

Superradiance constraints can be derived from spin measurements of either stellar mass~\cite{Arvanitaki:2014wva,Baryakhtar:2020gao,Mehta:2020kwu} or supermassive black holes~\cite{Unal:2020jiy}. For supermassive BHs, these spin measurements are less reliable given the uncertainties in the measurement method as well as the infall rates of compact objects~\cite{Arvanitaki:2014wva}. To be conservative, we only show superradiance constraints from stellar mass BH's.

\newcommand{\listspacing}{\vspace{0.2cm}}

\clearpage
\begin{table}[h!]
\noindent
\footnotesize
\begin{tabular}{llll}

Experiment:	&Principle	& DM? & Ref.\\
\hline

&&&\\
{\it  Haloscope constraints} &&&\\
ABRACADABRA-10cm	&Haloscope	&DM &\cite{Salemi:2021gck}\\
ADMX				&Haloscope	&DM &\cite{ADMX:2009iij,ADMX:2018gho,ADMX:2019uok,ADMX:2018ogs,Crisosto:2019fcj,ADMX:2021nhd,Bartram:2021ysp} \\
BASE				&Haloscope (Cryogenic Penning Trap)	&DM &\cite{Devlin:2021fpq} \\
CAPP				&Haloscope	&DM&\cite{Lee:2020cfj,Jeong:2020cwz,CAPP:2020utb}\\
CAST-RADES	&Haloscope	&DM &\cite{CAST:2020rlf}\\
DANCE&Haloscope (Optical cavity polarization) &DM &\cite{Oshima:2021irp}\\
Grenoble Haloscope&Haloscope &DM &\cite{Grenet:2021vbb}\\
HAYSTAC				&Haloscope &DM&\cite{HAYSTAC:2018rwy,HAYSTAC:2020kwv} \\
ORGAN		&Haloscope 				&DM&\cite{McAllister:2017lkb}\\
QUAX		&Haloscope				&DM&\cite{Alesini:2019ajt,Alesini:2020vny}\\
RBF			&Haloscope				&DM&\cite{PhysRevLett.59.839}\\
SHAFT		&Haloscope				&DM&\cite{Gramolin:2020ict}\\
SuperMAG &Haloscope (Using terrestrial magnetic field) & DM & \cite{Arza:2021ekq} \\
UF			&Haloscope				&DM&\cite{PhysRevD.42.1297}\\
Upload		&Haloscope				&DM&\cite{Thomson:2019aht} \listspacing\\


 {\it   Haloscope projections} &&&\\
ABDC		&Haloscope	&DM &\cite{Liu:2018icu}\\
ADMX				&Haloscope	&DM &\cite{Stern:2016bbw} \\
aLIGO		&Haloscope	&DM &\cite{Nagano:2019rbw}\\
ALPHA		&Haloscope (Plasma haloscope)	&DM &\cite{Lawson:2019brd}\\
BRASS		&Haloscope	&DM &\cite{BRASSwebpage}\\
BREAD		&Haloscope (Parabolic reflector)&DM&\cite{BREAD:2021tpx}\\
DANCE		&Haloscope (Optical cavity polarization) &DM &\cite{Michimura:2019qxr}\\
DMRadio		&Haloscope (All stages: 50L, $ m^3 $ and GUT) &DM&\cite{DMRadio:2022pkf,Brouwer:2022bwo} \\
FLASH		&Haloscope (Formerly KLASH)&DM &\cite{Alesini:2019nzq,FLASHconference} \\
Heterodyne SRF&Haloscope (Superconduct. Resonant Freq.)&DM&\cite{Berlin:2020vrk,HeterodyneSRF} \\
LAMPOST		&Haloscope (Dielectric)	&DM&\cite{Baryakhtar:2018doz}\\
MADMAX		&Haloscope (Dielectric)	&DM&\cite{Beurthey:2020yuq} \\
ORGAN		&Haloscope 				&DM&\cite{McAllister:2017lkb}\\
QUAX		&Haloscope				&DM&\cite{QUAXprojection}\\
TOORAD		&Haloscope (Topological anti-ferromagnets) &DM&\cite{Marsh:2018dlj,Schutte-Engel:2021bqm} \\
WISPLC 		&Haloscope (Tunable LC circuit) &DM &\cite{Zhang:2021bpa}\listspacing\\

{\it  LSW and optics} &&&\\
ALPS	&Light-shining-through wall 				&Any&\cite{Ehret:2010mh}\\
ALPS II	&Light-shining-through wall 	(projection)&Any&\cite{Ortiz:2020tgs}\\	
CROWS	&Light-shining-through wall (microwave)	&Any&\cite{Betz:2013dza}\\
OSQAR	&Light-shining-through wall 			&Any&\cite{OSQAR:2015qdv}\\
PVLAS	&Vacuum magnetic birefringence						&Any&\cite{DellaValle:2015xxa}\listspacing\\

{\it Helioscopes} &&&\\
CAST & Helioscope	&Any&\cite{CAST:2007jps,CAST:2017uph}\\
babyIAXO & Helioscope (projection)	&Any&\cite{Shilon:2012te,Armengaud:2014gea,Irastorza:2018dyq}\\
IAXO & Helioscope (projection)	&Any&\cite{Shilon:2012te,Armengaud:2014gea,Irastorza:2018dyq}\\
IAXO+ & Helioscope (projection)	&Any&\cite{Shilon:2012te,Armengaud:2014gea,Irastorza:2018dyq}\\

\end{tabular}
\caption{List of experimental searches for axions and ALPs. The table is continued in table \ref{tab:referenceList2}. All experiments here rely on the axion-photon coupling.}
\label{tab:referenceList1}
\end{table}

\begin{table}
\noindent
\footnotesize
\begin{tabular}{llll}
Experiment:	&Principle	& DM? & Reference\\
\hline

&&&\\
  {\it Astrophysical constraints} &&&\\
  4C+21.35 & Photon-ALP oscillation on the $\gamma$-rays from blazars & Any & \cite{Li:2022jgi}\\
Breakthough Listen	&ALP $ \to$ radio $\gamma$ in neutron star magn. fields &DM &\cite{Foster:2022fxn}	\\
Bullet Cluster		& Radio signal from ALP DM decay & DM &\cite{Chan:2021gjl} \\
Chandra				&AGN X-ray prod. in cosmic magn. field    & Any &\cite{Wouters:2013hua,Marsh:2017yvc,Reynolds:2019uqt,Reynes:2021bpe}  \\
BBN + $ N_{\rm eff} $& ALP thermal relic perturbing BBN and $ N_{\rm eff} $ & Any & \cite{Depta:2020wmr} \\
Chandra MWD & X-rays from Magnetic White Dwarf ALP prod.   & Any &\cite{Dessert:2021bkv}\\
COBE/FIRAS	& CMB spectral distortions from DM relic decay 	& DM	& \cite{Bolliet:2020ofj}\\
Distance ladder & ALP $ \leftrightarrow \gamma $  perturbing luminosity distances&Any&\cite{Buen-Abad:2020zbd}\\
Fermi-LAT 	& SN ALP product. $\to$ $ \gamma $-rays in cosmic magn. field       & Any & \cite{Calore:2020tjw,Calore:2021hhn,Meyer:2020vzy}  \\
Fermi-LAT 	& AGN X-ray production $\to$ ALP in cosmic magn. field& Any & \cite{Fermi-LAT:2016nkz}  \\
Haystack Telescope& ALP DM decay $ \to$ microwave photons &DM &\cite{Blout:2000uc} \\
HAWC TeV Blazars & $ \gamma \to $ ALP $ \to $ $ \gamma $ conversion reducing $ \gamma $-ray attenuation &Any & \cite{Jacobsen:2022swa}  \\
H.E.S.S.	& AGN X-ray production $\to$ ALP in cosmic magn. field    & Any & \cite{HESS:2013udx} \\
Horizontal branch stars	& stellar metabolism and evolution & Any & \cite{Ayala:2014pea} \\
LeoT dwarf galaxy	&Heating of gas-rich dwarf galaxies by ALP decay	&DM & \cite{Wadekar:2021qae}\\
Magnetic white dwarf pol. & $ \gamma $ $ \to $ ALP conversion polarizing light from MWD stars	& Any & \cite{Dessert:2022yqq} \\
MUSE& ALP DM decay $\to$ optical photons & DM & \cite{Regis:2020fhw}\\
  Mrk 421	& Blazar $\gamma$-ray $\to$ ALP $\to$ $\gamma$-ray in cosmic magn. field & Any & \cite{Li:2020pcn}  \\
NuStar	&Stellar ALP production $\to\gamma$ in cosmic magn. fields &Any&\cite{Xiao:2020pra,Dessert:2020lil}\\
NuStar, Super star clusters& Stellar ALP production $\to\gamma$ in cosmic magn. fields & Any & \cite{Dessert:2020lil} \\
Solar neutrinos & ALP energy loss $\to$ changes in neutrino production &Any& \cite{Vinyoles:2015aba}  \\
SN1987A ALP decay &SN ALP production $\to\gamma$ decay       & Any & \cite{Jaeckel:2017tud}  \\
SN1987A gamma rays&SN ALP production $\to\gamma$ in cosmic magnetic field       & Any & \cite{Payez:2014xsa,Caputo:2021rux}  \\
SN1987A neutrinos&SN ALP luminosity less than neutrino flux  & Any & \cite{Lee:2018lcj,Caputo:2021rux} \\
Thermal relic compilation 	& Decay and BBN constraints from ALP thermal relic & Any & \cite{Cadamuro:2011fd}\\
VIMOS& Thermal relic ALP decay $ \to $ optical photons & Any &\cite{Grin:2006aw}  \\
White dwarf mass relation & Stellar ALP production perturbing WD metabolism & Any & \cite{Dolan:2021rya} \\
XMM-Newton	& Decay of ALP relic	& DM &\cite{Foster:2021ngm} \listspacing\\

{\it Astrophysical projections} &&&\\
eROSITA & X-ray signal from ALP DM decay & DM & \cite{Dekker:2021bos} \\
Fermi-LAT 	& SN ALP production $\to\gamma$ in cosmic magnetic field       & Any & \cite{Meyer:2016wrm}  \\
IAXO	&Helioscope detection of supernova axions &Any	&\cite{Ge:2020zww}	\\
THESEUS& ALP DM decay $\to$ x-ray photons& DM & \cite{Thorpe-Morgan:2020rwc}  \listspacing\\

{\it Neutron coupling:} &     &         &            \\
CASPEr-wind				& NMR from oscillating EDM	(projection)								& DM & \cite{JacksonKimball:2017elr,Centers:2019dyn} \\
CASPEr-ZULF-Comag.& NMR from oscillating EDM											& DM &\cite{Wu:2019exd,Centers:2019dyn} \\
CASPEr-ZULF-Sidechain	& NMR (constraint \& projection)			& DM &\cite{Garcon:2019inh,Centers:2019dyn}\\
NASDUCK 				& ALP DM perturbing atomic spins 						& DM &\cite{Bloch:2021vnn} \\
nEDM					& Spin-precession in ultracold neutrons and Hg		& DM &\cite{Abel:2017rtm,Centers:2019dyn} \\
K-3He                   & Comagnetometer				     		& DM & \cite{Vasilakis:2008yn} \\
Old comagnetometers     & New analysis of old comagnetometers		& DM &\cite{Bloch:2019lcy}   	\\
Future comagnetometers  & Comagnetometers							& DM &\cite{Bloch:2019lcy}	\\
SNO					    & Solar ALP flux from deuterium dissociation	& Any & \cite{Bhusal:2020bvx}	\\
Proton storage ring	    & EDM signature	 from ALP DM				& DM & \cite{Graham:2020kai}	\\
Neutron Star Cooling    & ALP production modifies cooling rate			& Any & \cite{Buschmann:2021juv}\\
SN1987 Cooling		    & ALP production modifies cooling rate			& Any & \cite{Carenza:2019pxu} \listspacing\\

{\it Coupling independent:} &     &         &        \\		
Black hole spin         & Superradiance for stellar mass black holes                      					& Any & \cite{Arvanitaki:2014wva,Baryakhtar:2020gao,Mehta:2020kwu}  \\
Lyman$ -\alpha $         & Modification of small-scale structure                       					& DM & \cite{Rogers:2020ltq}  \\

\end{tabular}
\caption{List of experimental searches for axions and ALPs.}
\label{tab:referenceList2}
\end{table}

\clearpage

\bibliographystyle{JHEP}
\bibliography{kinetic-frag}
\end{document}